\documentclass[a4paper,11pt]{article}
\usepackage{ILD}
\RequirePackage{rotating}
\usepackage{subcaption}
\usepackage{float}
\usepackage[affil-it]{authblk}
\usepackage{booktabs}
\usepackage{amssymb}
\usepackage{wasysym}
\usepackage{siunitx}
\usepackage{lineno}
\usepackage[backend=biber,style=numeric-comp,sorting=none,mcite=true,
doi=false,subentry]{biblatex}
\usepackage{hyperref}
\usepackage{pgfpages}
\usepackage{fancybox,graphicx}
\usepackage{graphbox}
\usepackage{amsfonts}
\usepackage{amsmath}
\usepackage{relsize}
\usepackage{slashed}
\usepackage[normalem]{ulem}
\usepackage{wrapfig,rotating}
\usepackage{tikz}
\usepackage{pgfpages}
\usepackage{fancybox,graphicx}
\usepackage{epstopdf}

\AppendGraphicsExtensions{.eps.gz}
\usepackage{times}
\usepackage[T1]{fontenc}

\usepackage{lineno}

\newcommand{\alc}[2]{\multicolumn{1}{#1}{#2}}
\newcommand{\tento}[1]{\mbox{$\cdot 10^{#1}$}}

\title{The SUSY reach of Higgs Factories in the most challenging scenario:
  scalar $\tau$-leptons with lowest cross section and small mass differences}
\ildphys{2026}{001}
\localrep{DESY-26-039\\}
\date{\today}
\addauthor{M.T. N{\'u}{\~n}ez Pardo de Vera}
{\institute{1}}
\addauthor{M. Berggren}
{\institute{1}}
\addauthor{J. List}
{\institute{1}}

\addinstitute{1}{  Deutsches Elektronen-Synchrotron DESY,\\
                      Notkestr. 85, 22607 Hamburg, Germany}

\abstract{
The direct pair-production of the superpartner of the $\tau$-lepton, the $\widetilde{\tau}$,  is one
of the most interesting channels to search for SUSY in. First of all, the $\widetilde{\tau}$ is
likely to be the lightest of the scalar leptons. Secondly the
signature of $\widetilde{\tau}$ pair production is one of the experimentally most difficult
ones, thereby constituting the ``worst'' possible scenario for SUSY searches.
The current limit on $\widetilde{\tau}$ production in the general MSSM comes from analyses
performed at LEP,
suffering from the limited energy and luminosity of this facility and the need to trigger. Limits obtained at the LHC do extend to higher masses, but they are only valid under strong assumptions.
Future electron-positron colliders will be powerful facilities for SUSY searches, offering advantages
with respect to previous $e^+e^-$ colliders as well as to hadron machines.
In order to quantify the capabilities of future $e^+e^-$ colliders, the ``worst-case'' scenario for $\widetilde{\tau}$ searches  has been studied, taking
into account the effect of the $\widetilde{\tau}$ mixing on both $\widetilde{\tau}$ production cross section and on detection efficiency. To evaluate the latter, the ILD detector concept, originally developed for the International Linear Collider (ILC), and the ILC beam conditions at a centre-of-mass energy of $500$\,GeV have been used for detailed simulations, including for the first time the effect of bunch-crossings containing no hard $e^+e^-$ interaction, but only low-$\it{P_{T}}$ hadrons from $\gamma\gamma$ interactions and $e^+e^-$ pairs from beamstrahlung. 
Still, the obtained exclusion and discovery reaches extend nearly up to the 
kinematic limit even in the worst-case scenario. This remains true also when the  $\widetilde{\tau}$ and the lightest SUSY particle are quite close in mass.
The results of the detailed simulation study are extrapolated to centre-of-mass energies, integrated luminosities and beam polarisations of other proposed Higgs factory projects and discussed in view of their respective experimental environments, in particular addressing the case of FCCee.

}
\begin{document}
\titlepage

\section{Introduction\label{sec:intro}}
The most important question in particle physics today is the validity of the Standard
Model of particle physics~\cite{Glashow:1961tr,Weinberg:1967tq,Salam:1959zz,Englert:1964et,Higgs:1964pj,Guralnik:1964eu}
(the SM) that describes the strong, weak, and electromagnetic interactions.
The existence of the intermediate vector bosons, the Z and the W,
as well that of the Higgs boson were predicted by the SM, and
all of these were subsequently discovered at CERN~\cite{GargamelleNeutrino:1974khg,UA1:1983crd,UA2:1983tsx,UA1:1983mne,UA2:1983mlz,CMS:2012qbp,ATLAS:2012yve}.
The SM has been tested in precision experiments both at
and around the Z resonance - at LEP and SLC - and 
in the measurement of complex processes
at the Tevatron and LHC involving the production of single and multiple W and Z bosons, the Higgs, 
and top quarks.

While the Standard Model so far has passed every test, it cannot be the ultimate theory of nature.
From the cosmological perspective, the SM  neither does contain the dark matter and dark energy that appears 
to make up 95\% of the energy content of the universe, nor does it explain the fact that the visible universe is made up of matter but no antimatter.
Though the SM successfully predicts the existence and
general properties of the Higgs boson, essentially all these properties
are
determined by parameters adjusted by hand.
Moreover,  quantum corrections to the Higgs mass rapidly exceed
the mass of the Higgs itself if there is no cut-off due to new phenomena occurring
at a scale of 1\,TeV or lower.
These contributions require ever more incredulous
fine-tunings as the excluded energy scale of new physics would be much higher, e.g.\ the GUT scale.
  
If such Beyond the Standard Model (BSM) phenomena indeed are present at or around  1\,TeV, they will influence the properties
of the Higgs, the intermediate vector bosons, and the fermions (in particular the top quark).
It is therefore paramount to attempt to study these known SM particles in the finest
detail,
but also to attempt to search for new states appearing at such scales.
It has become a consensus in the community of high energy physicists that the next accelerator to be
constructed should be a Higgs factory, suited to carry out such a program.

Direct signs of BSM physics are commonly studied along two main theoretical paths.
One path is to study well-defined
but incomplete theories designed to elucidate some of the above-mentioned problems
of the SM, without the expectation of full self-consistency.
This encompasses so called 
simplified models and portal models.

The other path consists of models of BSM physics which {\it are} self-consistent to high-energy scales.
These models also addresses the questions above, and among them are models of compositnes, leptoquarks,
extra dimensions, and, most notable, Supersymmetry (SUSY). 
  SUSY continues to be one of the most promising candidates for new physics, and is the topic of this paper. SUSY could explain or
  at least give some hint at solutions to some of the above-mentioned problems of the Standard Model, {\it viz.} the fine-tuning
  problem, the nature of Dark Matter  or the possible theory-experiment discrepancy of the muon magnetic moment.

The outline of the paper is the following: In Sec.~\ref{sec:susy}, we will present the details of
the SUSY model we will explore, and in Sec.~\ref{sec:ILC} we discuss the experimentation at  Higgs factories in general, but with
some emphasis on the International Linear Collider (ILC), since we rely on ILD Monte-Carlo data simulated based on its beam conditions.
In Sec.~\ref{sec:stausearches}, we argue that the case of the SUSY partner of the $\tau$ lepton, the  $\widetilde{\tau}$,
is of particular interest, and present the current status of searches for this particle.
In Sec.~\ref{sec:environment}, the general conditions of the study are presented, by discussing
the expected signal signature, background sources, detector properties, and software tools used.
In Sec.~\ref{sec:analysis} the details of the analysis procedure are given,
and finally the expected results are given in Sec.~\ref{sec:limits},
both from the detailed study at ILC, and as recasts of the results to
other proposed Higgs factories. In Sec.~\ref{sec:concl}, we
summarise and conclude.

\section{SUSY\label{sec:susy}}
  
   Supersymmetry (SUSY)
     ~\cite{Gervais:1971ji,Golfand:1971iw,Wess:1974tw} (for reviews, see \cite{Martin:1997ns,Nilles:1983ge,Haber:1984rc,Barbieri:1982eh})
  is a symmetry of
  space-time relating fermions and bosons. For every SM particle it introduces a superpartner with the
  same quantum numbers except for the spin. The spin of the SUSY particle differs by half a unit from the value of its SM partner.
  These superpartners are collectively denoted sparticles.
  Supersymmetry requires that the Lagrangian is analytic, meaning that the option that the Higgs field and its
  conjugate have different couplings is not allowed, which in turn means that the SM Higgs sector must be extended to
  a two-doublet model to explain the difference in Yukawa couplings of up- and down-type fermions.
  This results in an extended Higgs sector of the SM, with two neutral CP-even states, one CP odd state, and two charged Higgs bosons.
  The scalar partners of the SM fermions are called sfermions, and the fermionic partners of the SM bosons are called
  bosinos.
  A SUSY model that only includes these new particles is denoted the Minimal Supersymmetric extension to Standard Model, MSSM for short.
  There are several extensions to the MSSM proposed, such as the nMSSM~\cite{Ellwanger:1996gw}, in which a further scalar is included, or GMSB~\cite{Giudice:1998bp}, in which
  the SUSY partner of the graviton, the gravitino, plays an important role.
  
  The colourless bosinos will mix to form the physical states,
  four neutral states that are majorana fermions, called the neutralinos,
  and two charged states, the charginos.
  For the sfermions, it should be noted that there will be one partner to each of the weak hyper-charge
  states of the fermions (the L and R states).
  Since the sfermions are scalars, these partners are not mass-degenerate, unlike the SM fermionic partners, for
  which chiral symmetry imposes degeneracy.
  The L and R sfermions might mix due to the interaction with the super-Higgs field, an interaction that involves
  both the sfermion and its SM partner.
  The mixing is therefore expected to be more pronounced for the partners to the heavier fermions, i.e.\ the third
  family ones.
  
  A new parity, R-parity~\cite{Farrar:1978xj}, is commonly introduced in SUSY, which has a crucial impact in SUSY phenomenology.
  R-parity takes an even value for SM particles and odd value for the SUSY ones.
  Multiplicative R-parity conservation\footnote{The introduction and conservation of
  this symmetry is inspired by flavour physics constraints since the most general SUSY Lagrangian induces
  flavour-changing neutral interactions that are avoided imposing R-parity conservation.}, assumed in
  most of the SUSY models, implies that the SUSY particles are always created in
  pairs and that the lightest SUSY particle (the LSP) is stable and,  when cosmological constraints are taken into account,
  also electrically neutral and colourless~\cite{Ellis:1983}.

  An important feature in the study of SUSY is the fundamental principle stating that 
  the couplings of particles and sparticles are related by symmetry.
   This allows to know the cross sections
  for SUSY pair production solely from the centre-of-mass energy the colliding elementary particles,
  the masses of the involved SUSY
  particles, and possibly a mixing-angle.
  Likewise, the properties of the decays of the produced sparticles are determined by the parameters
  of the SUSY model-point.

  To study SUSY experimentally, enough energy must be available to pair-produce sparticles.
  This yields certain advantages for lepton colliders, like LEP, and other ones for  hadron colliders, such as the LHC.
  At lepton colliders the full beam-energy is available for each interaction, and the colliding particles
  (the electron and the positron) are elementary ones.
  The sparticle-pair
  would be produced alone, with a well-known energy and momentum , and practically no background.
  Since the initial energy is know, the production cross-section is given by the assumed masses
  of the sparticles.
  The kinematics of the sparticle decays are also known from the initial e$^+$e$^-$ state and the
  assumed SUSY model point.
  However, leptons are harder to accelerate to high energies than protons.
  So, at hadron colliders, even though only a small fraction of the beam-energy is carried by the
  interacting partons, potentially higher SUSY masses can be probed.
  Hadron colliders also have a certain advantage in that strong production of sparticles
  is available. However, this only applies to coloured sparticles (the squarks and the gluino),
  which are likely to be the most massive ones, and of course comes with the draw-back that
  strong production also occurs for the SM processes, yielding background levels many orders of
  magnitude higher than at lepton colliders.
  A further disadvantage of hadron colliders comes again from the fact that only a small fraction of the
  beam-energy is carried by the interacting partons: not only is this fraction small, but it is also
  un-known on the event-by-event basis, and
  the exact kinematics of the sparticle decays as well as the production cross-section will obtain
  uncertainties from this lack of knowledge.
  This leads to a need to include further, arbitrary, constraints on the model-point under study.
  
  The consequence of these differences is that limits reported by lepton colliders
  are  valid for any value of the model parameters not explicitly shown in the exclusion plots,
  but do not reach
  as far in mass, 
  while those reported by hadron colliders reach further in mass but are only valid if many constraints
  on the model parameters are fulfilled.
  This is particularly true in
  the search of the lighter colour-neutral SUSY states, such as sleptons (the SUSY partners of the leptons),
  charginos or neutralinos.
  Hadron colliders do not benefit from a strong production for such colourless states,
  and they also suffer from the effects of higher backgrounds and the unknown initial state.
 
\section{Experimental Environment at Higgs Factories\label{sec:ILC}}
A variety of electron-positron colliders has been proposed with the main
motivation to study the Higgs boson in unprecedented detail.
Among them, the International Linear Collider (ILC)~\cite{ILCInternationalDevelopmentTeam:2022izu, Behnke:2013xla, ILC:2013jhg, Adolphsen:2013jya, Adolphsen:2013kya, Behnke:2013lya} can operate at 
energies of $250-500$\,GeV and with upgrade capability to $1$\,TeV.
The Cool Copper Collider (C$^3$)~\cite{Vernieri:2022fae}
targets a similar range of energies,
while the Compact Linear Collider (CLIC)~\cite{Brunner:2022usy, CLICdp:2018cto, Linssen:2012hp} reaches even up to $3$\,TeV.
The Linear Collider Facility (LCF) recently proposed for CERN~\cite{LinearCollider:2025lya} assumes major runs at $250$\,GeV and $550$\,GeV with higher integrated luminosities than ILC.
Circular colliders, in particular the Circular Electron Positron Collider
(CEPC)~\cite{Gao:2022lew, CEPCPhysicsStudyGroup:2022uwl, CEPCStudyGroup:2018rmc, CEPCStudyGroup:2018ghi}
and the $e^+e^-$ option of the Future Circular Collider (FCCee)~\cite{Bernardi:2022hny, FCC:2018byv, FCC:2018evy,FCC:2025uan},
put more emphasis on lower energies, but can be upgraded to $365$\,GeV. 
All $e^+e^-$ colliders share the advantages of a well defined initial
state and a clean and reconstructable final state.
At linear colliders, SUSY searches would profit in addition from a well
defined spin configuration in the initial state from longitudinally polarised beams:
The ILC foresees high electron and positron beam polarisations of ($\pm$80$\%$,$\mp$30$\%$)\footnote{
  We introduce the following notation for beam polarisations,
$\mathcal{P}\equiv( \mathcal{P}_{e^-}, \mathcal{P}_{e^+} )$,
and define the pure beam polarisations as
$\mathcal{P}_{LR}\equiv(-1, +1)$ and $\mathcal{P}_{RL}\equiv(+1, -1)$.
The nominal beam polarisations for the ILC are defined as
$\mathcal{P}_{-+}\equiv(-0.8, +0.3)$, $\mathcal{P}_{+-}\equiv(+0.8, -0.3)$,
$\mathcal{P}_{--}\equiv(-0.8, -0.3)$ and $\mathcal{P}_{++}\equiv(+0.8, +0.3)$.}.
The LCF assumes the same polarisations at $250$\,GeV, but a twice higher degree of positron polarisation at $550$\,GeV.

The CLIC and C3 baselines only foresee the electron beam to be polarised,
but consider positron polarisation as an option.
Very recently, also CEPC is studying the possibility to maintain electron
polarisation from the source through the booster ring into the collider ring.
FCCee is not considering longitudinal beam polarisation.

At all  electron-positron colliders,
the main electron-positron process is accompanied by parasitic $\gamma\gamma$ interactions
between virtual photons produced by the Weizs{\"a}cker-Williams process~\cite{Zolotorev:2000}.
These interactions produce hadrons with low
transverse momentum (low-$\it{P_{T}}$ hadrons).
The Weizs{\"a}cker-Williams process is mainly
mediated by vector meson dominance (VDM), i.e.\ the photon fluctuating to a neutral
vector meson (usually the $\rho^0$), and thus best described as
$\rho^0 - \rho^0$ scattering (i.e.\ in a process mediated by the strong
interaction).
In~\cite{Sasikumar:2020qxa} it is shown that this process predominately produces
charged pions and $\rho^0$'s (which decay with a branching ratio close
to 100~\% to $\pi^+ \pi^-$), while the presence of $\gamma$'s from $\pi^0$
or $\eta$ decays is quite rare.
Under ILC-500 conditions the cross-section for this process is
380~nb, and
the expected number of low-$\it{P_{T}}$-hadron events from $\gamma\gamma$ interactions
is three orders of magnitude higher than any other SM process,
and nine orders of magnitude higher than the  production rate of typical
SUSY processes.
Hence, it will be required to have a rejection factor stronger
than $\mathcal{O}(10^{-9})$ to make them negligible.

Among the strongest advantages of the proposed circular colliders is their
tremendous luminosity, in particular at the $Z$ pole.
This, however, comes at a price: the final focus magnets are required to be
close to the interaction point, sticking far inside the detectors.
Thus, the detectors can cover only angles down to about $50$\,mrad,
while linear collider detectors are typically hermetic down to $6\,$mrad.
Furthermore, unlike linear collider detectors, and due to the much
higher collision-rate, the circular collider detectors
might not be able to run trigger-less, which is a clear disadvantage when
searching for unexpected signatures.

Linear colliders have a lower collision frequencies which
enables the trigger-less operation, but also leads to a
rather high luminosity per bunch crossing.
This high  luminosity per bunch crossing might create a few pile-up $\gamma\gamma$ events.
The probability that the above-mentioned low-$\it{P_{T}}$ hadron production
can occur during
the same bunch-crossing as a signal event cannot be neglected
at linear colliders.
In addition,
under the conditions at such machines,
large numbers of real photons are created by beamstrahlung.
These will contribute to the total $\gamma\gamma$ interaction-rate,
but they can also incoherently interact with the strong field inside the bunches,
creating electron-positron pairs.
For instance at ILC with $\sqrt{s}=500$\,GeV, the number of electron-positron
pairs per bunch crossing from beamstrahlung is of the order of $10^5$.
However, as a consequence of their very low transverse momentum,
they are mainly curled up in
the magnetic field of the detector, and do not reach the tracking system; 
only $\mathcal{O}(10)$  are expected to leave signals in the tracking system
of the detector in each bunch crossing.
It should be noted that at CLIC at 3\,TeV, the fields inside the bunches becomes so
high that the Schwinger limit is reached,
and also {\it coherent} pair-production becomes active,
and must be taken into account.
The ensemble of all these physics processes with cross-sections so high that they create activity in
every or nearly every bunch crossing is summarised by the term ``overlay''.

If these processes occur during the same bunch crossing as a hard e$^+$e$^-$ process,
they are referred to as overlay-on-physics, if they occur
alone, they are referred to as 
overlay-only events.
The low-$\it{P_{T}}$ hadron part of the overlay-only events occur e
qually frequently at any machine,
while overlay-on-physics can be neglected at the circular colliders.
Among the linear collider options,
CLIC, where the detector needs to integrate over many bunch-crossings,
suffers from a substantial amount of overlay-on-physics events,
which can however be reduced significantly by timing information.
At ILC and C$^3$, much less pile-up is expected.
However, it is not completely absent -
e.g.\ at ILC with $\sqrt{s}=500$\,GeV an  average of 1.05
low-$\it{P_{T}}$-hadron  events per bunch crossing would be expected~\cite{ILDConceptGroup:2020sfq}. Thus, even at the same centre-of-mass energy, circular and linear colliders have different advantages and disadvantages for SUSY searches, and we will address the question whether the overlay-on-physics at linear colliders or the lack of hermeticity of circular collider detectors is the larger challenge for SUSY searches.

\section{Searches for the $\widetilde{\tau}$ \label{sec:stausearches}}
  For evaluating the power of SUSY searches at future facilities,
  it is beneficial to focus on the lightest particle in the SUSY spectrum that could be accessible. Since
  the cosmological constraints require a neutral and colourless LSP, the next-to-lightest SUSY
  particle, the NLSP,  would usually be the first one to be detected. The NLSP can only decay to the LSP and
  the SM partner of the NLSP\footnote{In this paper we do not consider the case that the LSP-NLSP mass-difference is smaller than the mass of the SM partner, such that only three- or four-body decays mediated by the SM partner would be allowed, often leading to displaced signatures and thus requiring a different kind of analysis.}.
  This already makes the NLSP production special: heavier states might well decay in
  cascades, and thus have signatures that depend strongly on the model.
  Furthermore, there is only a finite set of sparticles that could be the NLSP, so a systematic search
  for each possible case is feasible.
  This also means that one can a priori estimate which will be the most difficult case, namely the
  NLSP that combines small production cross section with a difficult experimental signature.
  Also, the requirement that the model should be as challenging as possible to detect singles out case of the MSSM with conserved R-parity:
If R-parity is broken, or the gravitino is
  the LSP, the lightest bosino would decay
  in the detector, which LEP experience shows gives stronger limits than in the R-parity conserving case.
  This is true both for prompt and for displaced decays~\cite{DELPHI:2003noy}. 

  The super-partner of the ${\tau}$ - the stau slepton (the $\widetilde{\tau}$) - satisfies both the conditions above: due to mixing, the production cross section
  might be quite small, and because of the fast and partly invisible decay of its SM partner,
  the experimental signature is the least clear one.
  The searches for the $\widetilde{\tau}$ can be affected by overlay-only events since the low-$\it{P_{T}}$ hadrons
  from $\gamma\gamma$ interactions are kinematically very close to
the visible part of the signal events for small mass differences.

  Therefore, studies  of $\widetilde{\tau}$ production
  might be seen as the way to determine the guaranteed discovery or exclusion reach for SUSY: any other NLSP
  would be easier to find.
  In addition, the search of a light $\widetilde{\tau}$ is also theoretically motivated: SUSY models with a light
  $\widetilde{\tau}$ can accommodate the observed relic density, by enhancing 
  the $\widetilde{\tau}$-neutralino co-annihilation process~\cite{Ellis:1998},
  if the mass difference between the  $\widetilde{\tau}$ and the neutralino is not more than $\sim$\,10\,GeV.
 The  $\widetilde{\tau}$ might also play an important role in determining the SUSY
contribution to the muon g-2 anomaly. The study described in~\cite{Endo:2022qnm}
shows that, if this contribution comes
mainly from the bino, and all the sleptons and the lightest neutralino
are within the ILC reach, measurement of $\widetilde{\tau}$ parameters could
allow the determination of the SUSY contribution to this anomaly
with a precision of ~1\,\%.

\subsection{Properties  of the $\widetilde{\tau}$\label{sec:thestau}}
\begin{figure}[htbp]
    \centering
      \includegraphics [width=.4\textwidth]{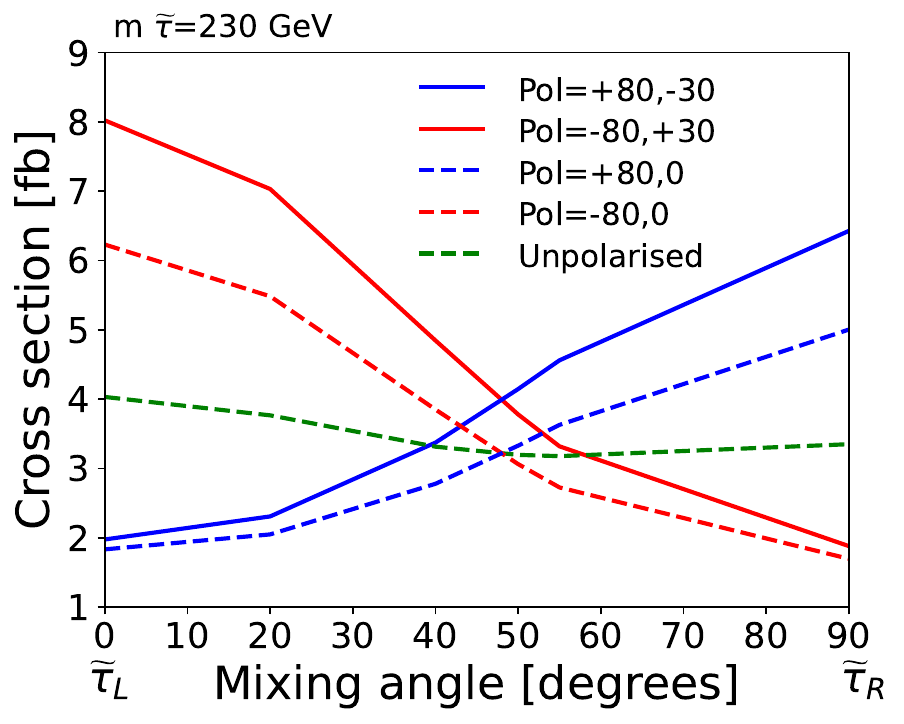}
    \caption{
      Cross section for $\widetilde{\tau}$ pair production as a function of the $\widetilde{\tau}$ mixing
      angle for five different settings of the beam polarisations. Flipping of the polarisation signs inverts the dependence on the mixing angle. For unpolarised beams, the dependence vanishes nearly, indicating loss of information for determining the mixing angle. 
             }
    \label{fig:crosssections}
  \end{figure}

For the $\widetilde{\tau}$, like for any other fermion or sfermion, there
are two weak hyper-charge states, $\widetilde{\tau}_{R}$ and $\widetilde{\tau}_{L}$,
and for the $\widetilde{\tau}$, being a scalar,
  there is no reason to
  expect that  $\widetilde{\tau}_{R}$ and $\widetilde{\tau}_{L}$ would have the same mass, as explained in Sec.~\ref{sec:susy}.
  Mixing between the weak hyper-charge states
  yields the physical states, and, as also discussed in Sec.~\ref{sec:susy},
  the mixing is expected to be more prominent for the super-partners of the third generation fermions.
  As a consequence of the mixing, the lightest
  $\widetilde{\tau}$, $\widetilde{\tau}_{1}$, would most likely  be the lightest slepton, due to the seesaw
  mechanism: the mass of the lightest physical (mixed) state would be smaller than the mass of
  any un-mixed weak hyper-charge state. The cross section of the $\widetilde{\tau}$ also differs from the
  one of the ${\tau}$, not only due to phase-space limitations - the $\widetilde{\tau}$ being more massive
  than the ${\tau}$ - but also due to the mixing.
  In $e^+e^{-}$ colliders, assuming R-parity conservation, the $\widetilde{\tau}$ will be pair-produced, with
  contribution of the s-channel only, via $Z^{0}$/$\gamma$ exchange. The strength of 
  the  $Z^{0}$/$\gamma$ $\widetilde{\tau}$ $\widetilde{\tau}$ coupling depends on 
  the $\widetilde{\tau}$ mixing, reaching its minimum value when
  the coupling $\widetilde{\tau}_{1}$ $\widetilde{\tau}_{1}$ $Z^{0}$ vanishes.
The dependence of the cross section on the $\widetilde{\tau}$ mixing angle is
shown in Fig.~\ref{fig:crosssections}, for the two beam-polarisations considered at
ILC. 
Also shown is its dependence for unpolarised beams, which
shows a minimum at 53$^\circ$, and the cross sections if only the electron beam is polarised.

\begin{figure}[htbp]
    \centering
    \subcaptionbox{}{\includegraphics [scale=1.0]{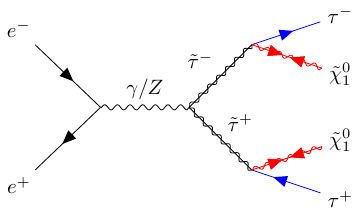}}

    \subcaptionbox{}{\includegraphics [scale=1.0]{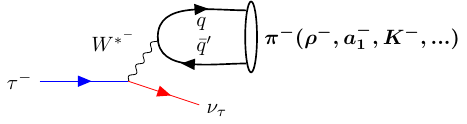}}
    \subcaptionbox{}{\includegraphics [scale=1.0]{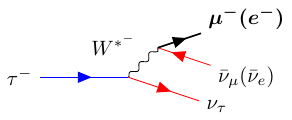}}
    \caption{(a) $\widetilde{\tau}$ pair production and decay to $\tau$ and $\widetilde{\chi}^0_1$; (b) hadronic $\tau$ decay; (c) leptonic $\tau$ decay.
             Blue lines indicates the $\tau$, red lines indicate undetectable final state particles, while thick black lines indicate the
             detectable final state particles.}
    \label{fig:signdiagrams}
  \end{figure}

  Assuming that the $\widetilde{\tau}$ is the NLSP, R-parity conservation implies that the $\widetilde{\tau}$
  will decay to a ${\tau}$ and an LSP (assuming mass differences above the mass of the ${\tau}$, as is done
  in this study).
  The LSP, as already mentioned, is stable and at most weakly interacting, hence it will leave the detector
  without being detected. 
  The ${\tau}$, with a proper lifetime of 2.9$\tento{-13}$ s, will decay
  before leaving any signal in the detectors. The only activity in the signal events is therefore
  the decay products of the two ${\tau}$'s, but
  some of these decay products are
  undetectable neutrinos.
  This makes the identification 
  more difficult than the decay of selectrons to electrons or of smuons to muons.
  In addition,
  since the decay products are only partially detectable, that kinematic signatures get blurred.
  This less clear experimental signature in combination with the potentially low cross section makes 
  the search of  $\widetilde{\tau}$ the worst case.
  Figure~\ref{fig:signdiagrams}(a) shows the diagram of the $\widetilde{\tau}$ production and decay, while
  Figs.~\ref{fig:signdiagrams}(b) and (c) show the diagrams of the subsequent $\tau$ decays, in the hadronic
  and leptonic modes, respectively.
 In hadronic $\tau$ decays,  it is quite common that the charged particles are
accompanied by close-by photons:
the dominating decay-modes $\tau \rightarrow \rho^\pm \nu_\tau$ and  $\tau \rightarrow a^\pm_1 \nu_\tau$
always (in the $\rho$ case), or often (in the $a_1$ case) contain $\pi^0$'s in the subsequent
decays of the mesons. Also in the electronic decay mode, it is common to observe such close-by
photons, in this case due to brems-strahlung when the electron transverses the detector material.
The visible part of the $\tau$ decay products will be refered to as a ``jet'' in the following,
even if it only contains a single charged particle.

  Since
  the $\widetilde{\tau}$ decay is, for $\Delta M$ $>$ $M_{\tau}$, a two body decay, it is possible to
  determine the maximum and minimum momentum of each of the  decay products as
  a function of the $\widetilde{\tau}$ mass, the mass of LSP
  and the centre-of-mass energy of the collider. The minimum momentum
  can not easily be observed due to the presence of neutrinos in the $\tau$ decay, with
  the corresponding decrease of observable momentum.
  It can be seen as an inflection point in the spectrum of the visible $\tau$ decay products,
  different for different decay modes,
  but will in most cases be hidden by background, and distorted by the event selection. 
  The upper edge, on the other hand, is readily observable, as the end-point of the spectrum of
  the decay products, an end-point that does not depend on the decay mode, and is robust with respect to cuts.
  The expression for the maximum
  jet momentum is given by~\cite{Berggren:2015ar}:

  \begin{equation}
    P_{max} = \frac{\sqrt{s}}{4}\left(1-\left ( \frac{M_{LSP}}{M_{\widetilde{\tau}}} \right )^2\right)\left(1+\sqrt{1-\frac{4M_{\widetilde{\tau}}~^2}{s}}\right)
    \label{eq:pmax}
  \end{equation}

\begin{figure}[htbp]
    \centering
       \includegraphics [width=.4\textwidth, trim=0 -0.8cm 0 0]{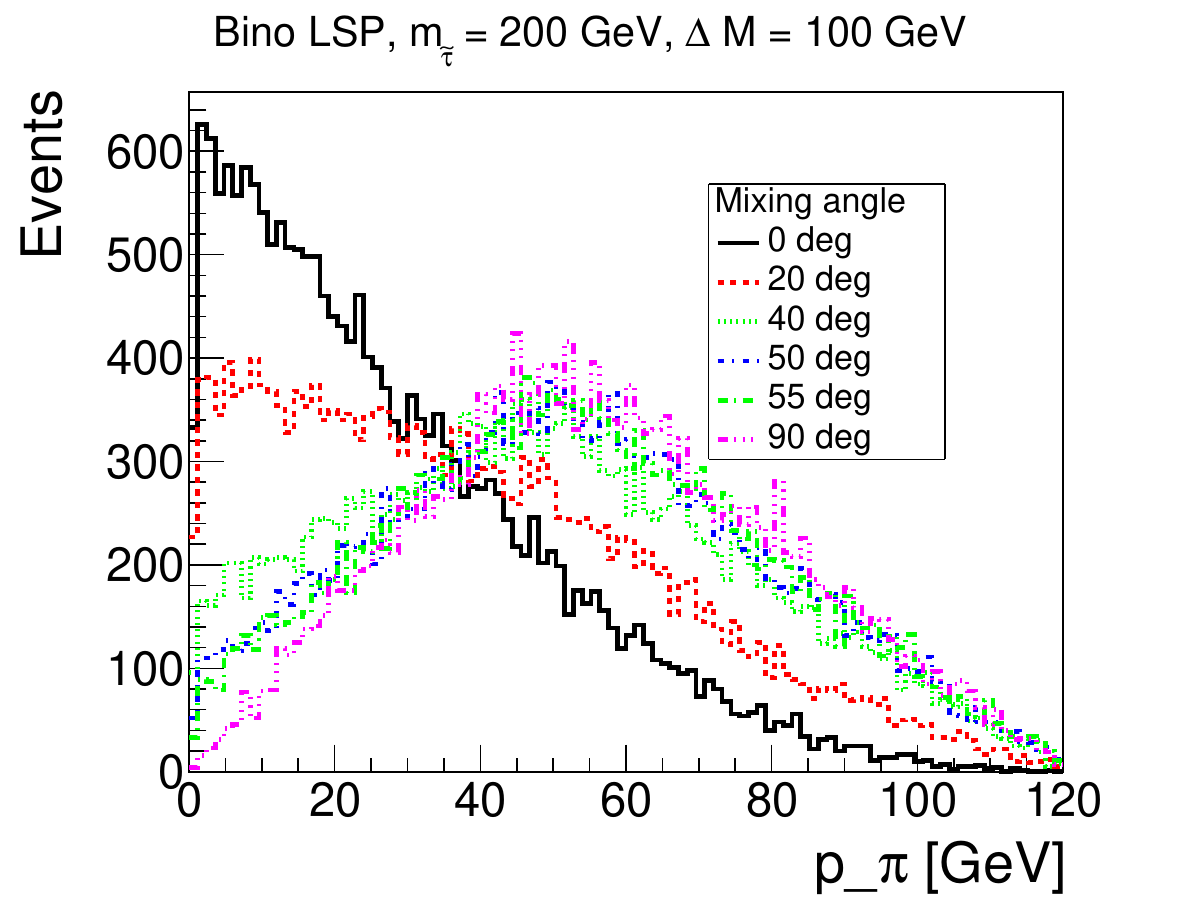}
    \caption{
      Momentum distribution of the pions coming from ${\tau}$ decays for different $\widetilde{\tau}$ mixing angles.
      The neutralino was taken to be pure bino.
      }
    \label{fig:pispec}
  \end{figure}

The polarisation of the $\tau$ is also important, since it influences the momentum 
distribution of the ${\tau}$-decay products,
and hence the signal efficiency: if the $\tau$ is pre-dominantly left-handed, and
since there are no right-handed neutrinos in the standard model, the invisible
neutrino will tend to align with the direction of the $\tau$ and take a larger 
fraction of the momentum, and consequently the visible system will take less.
This is illustrated in Fig.~\ref{fig:pispec} which shows the momentum distribution 
of the pions coming from ${\tau}$ decays (in the single pion decay mode) for different
$\widetilde{\tau}$ mixing angles and a bino LSP.
The ${\tau}$ polarisation from the $\widetilde{\tau}$ decay depends not only on the $\widetilde{\tau}$ mixing but also on the
neutralino mixing. This is because of the different behaviour of the interaction between higgsinos and
binos with the $\widetilde{\tau}$. The higgsino (like the Higgs boson) carries weak hyper-charge, while the bino
(like the $Z$ or the photon) does not. Therefore, with a higgsino LSP, the $\widetilde{\tau}$  produces 
a ${\tau}$ with the opposite chirality with respect to that of the $\widetilde{\tau}$, while with a bino LSP, 
the produced ${\tau}$ has the same chirality.
Since the lowest efficiency is expected for left-handed $\tau$'s, we
evaluate the efficiency for the assumption that the neutralino is a pure bino for
$\widetilde{\tau}$ mixings below 45$^\circ$ (i.e.\ for a $\widetilde{\tau}$ more left
than right), and for a pure higgsino LSP for mixings above  45$^\circ$.
This assures that the most difficult case is studied at any mixing.

  \subsubsection{Signal characteristics\label{sec:signalchar}}
  From the properties of  $\widetilde{\tau}$ production and decay explained above, one can conclude that the signal
  will be characterised by a number of properties necessarily present:
  the  signal events must have two ${\tau}$ candidates and a large missing energy, not only due to the
  invisible LSPs but also due to the neutrinos from both ${\tau}$ decays. There should only be small additional activity in the event, meaning that the total seen multiplicity is low.
  
  Furthermore, since the $\widetilde{\tau}$'s are scalars
  and hence isotropically produced, these events have a large fraction of the detected activity in the
  central region of the detector. 
  The $\widetilde{\tau}$'s must also be rather heavy, so they will not have a large boost in the lab-frame.
  An LSP close in mass to the  $\widetilde{\tau}$ will constitute the experimentaly most difficult
  case, and is the one we focus on. 
  In this case, also the LSP is quite heavy, and 
  the direction of the $\widetilde{\tau}$ does not  strongly correlate to that of
  the visible $\tau$ after the  decay.
  As a consequence, the two $\tau$ candidates are expected to go in directions quite independent
  of each other resulting in events with un-balanced total and transverse momentum, 
  large angles between the 
  two $\tau$-candidate directions and zero forward-backward asymmetry.
  These properties are  however not necessarily present in any event - the two $\tau$'s could accidentally happen
  to be back to back, for example.

\begin{figure}[htbp]
    \centering
    \includegraphics [width=.8\textwidth]{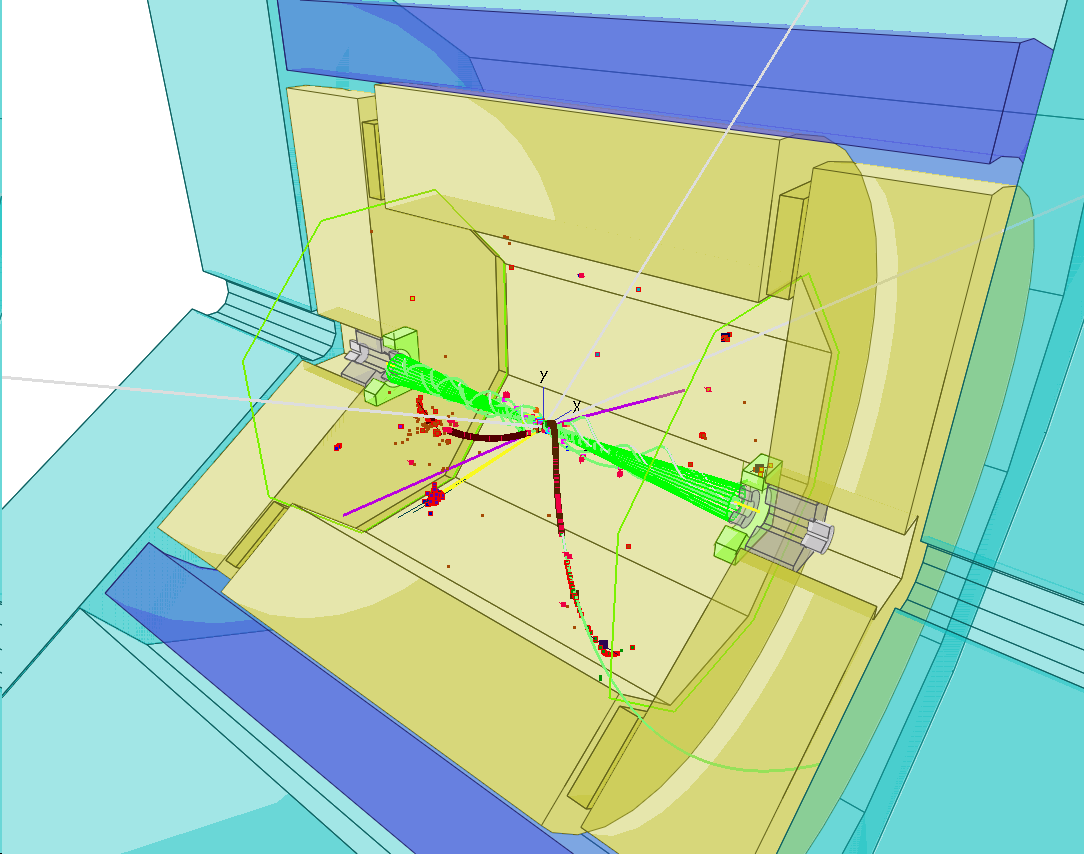}
    \caption{
      A $\widetilde{\tau}$ event at ILC operating at $\sqrt{s}$ = 500~GeV, fully simulated in the ILD detector.
      $M_{\widetilde{\tau}}$ = 230~GeV, and $\Delta M =$~10~GeV.
      The properties of this event are discussed in detail in the text (Sec.~\ref{sec:signalchar}).
      }
    \label{fig:evdisp}
  \end{figure}

  Most of these properties are well illustrated by the fully simulated event shown in Fig.~\ref{fig:evdisp}.
  In this event, the $\widetilde{\tau}$ mass is 230\,GeV, and the LSP mass is 220\,GeV (i.e.\ $\Delta M =$~10~GeV).
  The  $\widetilde{\tau}$ mixing angle is 55$^\circ$, and
  the detector shown is ILD, see Sec.~\ref{sec:environment} for details.
  The nominal $\sqrt{s}$ is 500\,GeV, 
 and the sources of beam-background discussed in Sec.~\ref{sec:ILC} are included.
  One can see two charged tracks, a muon (downwards, sightly to the right), and
  a hadron (straight to the left). The muon is easily identified by the minimum ionising signal in the
  electromagnetic calorimeter (the part of the track after the break in the string of hits), 
and
  the hadron is also easily identifiable by the typical hadronic shower-shape in the calorimeter.
  In addition, a prominent signal is visible in the calorimeter below the hadron track. No charged track
  points in this direction. Thanks to the high granularity of the  electromagnetic calorimeter of ILD, this
  deposit has been reconstructed as two separate electromagnetic clusters,
  i.e.\ as two photons.
  
  In addition to the measured quantities, also the true information of invisible partcles are shown in the 
  event display, for information:
  The three neutrinos in the event are indicated by the grey lines, and the two LSPs by the magenta ones.
  Already from the event display, without any detailed reconstruction, one can see that
  this event has a huge missing energy: From the large curvature of the two tracks, and the small size of the
  photon clusters, the seen energy is not more than some 10\,GeV. Indeed, the full reconstruction yields
  $E_{\mu}= 1.54 (1.54)$~GeV, $E_{hadron} = 3.60 (3.63)$~GeV and $E_{\gamma 's} = 5.83 (6.33)$ and $ 1.78 (1.44)$~GeV;
  the true values are given within the parenthesis. Hence, the total reconstructed energy is only 12.75\,GeV.
  The topology of the event is well compatible with a $\widetilde{\tau}$ pair-production, with both
  $\widetilde{\tau}$'s decaying to $\tau \widetilde{\chi}^0_1$, followed by decays of the $\tau$'s.
  One  $\tau$ decays in the chain $\tau \rightarrow \rho \nu_\tau \rightarrow \pi \pi^0 \nu_\tau \rightarrow \pi \gamma \gamma \nu_\tau$, and the other
  one directly $\tau  \rightarrow \mu \nu_\tau \nu_\mu$.
  This is indeed the correct interpretation. 
  The total multiplicity of the event is also low: in addition to the $\widetilde{\tau}$ decays, there are three overlay-on-physics tracks reconstructed,
  with energies of a few hundred MeV; none of them has a transverse momentum above 140 MeV.
  One can also see that the event is unbalanced in seen momentum and transverse momentum, that the two $\tau$ candidates are far from being
  back-to-back, and that almost all activity is in the barrel region of the detector.

\subsection{Limits on $\widetilde{\tau}$ production at LEP, LHC and previous ILC studies\label{sec:otherlimits}}
According to the PDG~\cite{ParticleDataGroup:2024cfk},  
the most solid limit on the $\widetilde{\tau}$ mass comes from the DELPHI experiment at LEP~\cite{Abdallah:2003xe},
and the PDG cites 81.9\,GeV as the limit, a limit valid for any  $\widetilde{\tau}$ mixing if $\Delta M > 15$\,GeV. 
In the same publication, DELPHI also presents  an analysis targeted at low mass differences and excludes a $\widetilde{\tau}$ 
with mass below 26.3\,GeV, for any mixing, and any mass difference larger than the $\tau$ mass~\cite{Abdallah:2003xe}.
The LEP SUSY working group also combined results from all four LEP 
experiments~\cite{LEPSUSYWG/04-01.1}.
By doing so,  they could set a somewhat higher limit with minimum value that ranges from 87 to 93\,GeV depending on
the mass difference between the $\widetilde{\tau}$ and
the neutralino, not smaller than 7\,GeV.
These limits
are valid for any mixing 
and any value of the model-parameters, other than the two masses.
However, there was no combination done of the low mass difference case, $\Delta M$ $<$ 7\,GeV.

At the LHC, ATLAS and CMS have determined limits on the $\widetilde{\tau}_{R}$ mass, analysing data from Run 1 and
  Run 2~\cite{ATLAS:2024fub,CMS:2022syk}.
  ATLAS  excludes masses from 100 to 330\,GeV for a mass-less LSP,
  and CMS from 150 to  240\,GeV.
  Neither of the collaborations can exclude masses just above the LEPII limits at any LSP mass.
  The pure $\widetilde{\tau}_{R}$ pair production can be considered the closest case to the physical $\widetilde{\tau}$,
  since it is likely to be the lightest of the two  weak hyper-charge states and is the one with the lower cross 
  section. However, also for $pp$ collisions, it can not be considered the worst case, since the dependence of the
  mixing on the cross section and kinematics of the decay products has to be taken into account.
  
  The future HL-LHC should provide an improvement on the $\widetilde{\tau}$ limits provided by ATLAS and
  CMS, not only because of an increase of the luminosity but also because of an expected gain in sensitivity
  to direct $\widetilde{\tau}$ production again due to the use of different analysis methods.
  Simulation studies have already been performed in both
  experiments~\cite{ATLAS:2018diz,CMS:2018imu}. Upper limits for $\widetilde{\tau}$ masses
  are indeed increased by about 300\,GeV, but they suffer from the same constraints as the previous
  studies. ATLAS achieves projected exclusion limits for pure $\widetilde{\tau}_{R}$ pair production for higher mass values than
  the ones from the last experimental analysis, but still would have no potential for discovery.
  The exclusion reach for $\widetilde{\tau}$ would not cover co-annihilation scenarios, i.e.\ scenarios with  $\Delta M \lesssim$ 10\,GeV.
  It can also be noted that if the heavy higgses (the scalar $H$ or the pseudo-scalar $A$) are light enough to be observed
  at HL-LHC, their decay to $\widetilde{\tau}$-pairs could at certain regions in the parameter space have a large branching
  factor, and could become a  $\widetilde{\tau}$ discovery channel \cite{Arganda:2021qgi}. At the relevant regions in parameter-space, the  $\widetilde{\tau}$
  would, however, not be the NLSP. 

   Searches for the $\widetilde{\tau}$ at the ILC have been also performed in previous studies~\cite{Berggren:2013vna}.
  In this fast simulation study, they assumed an integrated luminosity of only 500\,fb$^{-1}$ at $\sqrt{s}=500$\,GeV and only used the $\mathcal{P}_{+-}$
  data sample,
  contrary to the present study, where 
  all four polarisation samples are used
  and the full ILC 500 integrated luminosity was assumed, corresponding to a data-set more than six times larger.
  The previous study was aimed at scanning the entire $M_{LSP} - M_{NLSP}$ plane, and doing so for several different
  NLSP candidates.
  Quite generic cuts were therefore used, and were optimised ``on-the-fly'' at different points. 
  More specifically, the limits presented in that study do not have a dedicated
  analysis for low mass differences between the $\widetilde{\tau}$ and the LSP.
   Even so, it was found that the exclusion limit goes up to 240\,GeV with a discovery potential
  up to 230\,GeV for large mass differences.

  Another study of slepton production at ILC and,  in general, future e$^+$e$^-$ colliders 
         can be found in~\cite{Baum:2020gjj}.  They
could
 demonstrate
         that these colliders would be able to probe most of the kinematically
         accessible parameter space with only a few days of data, with the
         capability of discovering/excluding new physics that would evade
         detection at the LHC.
However,  this work used  simplified parameterised detector
response, and as such would not be reliable at the more difficult parts of
parameter space, i.e.\ low mass-differences and/or close to the kinematic limit.

\section{Environment of the study\label{sec:environment}}
The study described in this paper was performed on fully simulated $\widetilde{\tau}$ events as well as all
SM backgrounds at a centre-of-mass energy of 500\,GeV, according to the beam conditions expected at ILC.
The standard ILC running scenario~\cite{Barklow:2015tja} was followed,
which assumes that an integrated luminosity of 1.6\,ab$^{-1}$ at $\sqrt{s}=500$\,GeV
for each of the beam polarisations $\mathcal{P}_{+-}$ and  $\mathcal{P}_{-+}$ should be collected.
In addition, also 0.4\,ab$^{-1}$ will be collected with each of the $\mathcal{P}_{--}$ and $\mathcal{P}_{++}$ polarisations,
yielding a total integrated luminosity of 4\,ab$^{-1}$.

The International Large Detector (ILD) concept~\cite{ILD:2025yhd, ILDConceptGroup:2020sfq, Behnke:2013lya}  
in its ILC version is used in the Geant4-based detector simulation implemented in key4hep~\cite{Carceller:2025ydg, Key4hep:2021rub}.
The main tracker of ILD is a large TPC, offering excellent pattern recognition and 
particle identification capabilities 
as well as very good momentum resolution, with a minimal material budget.
Inside the TPC, closest to the interaction point, a silicon pixel vertex detector allows to reach impact-parameter
measurement precision of 5\,$\mu$m, and outside the TPC, a large silicon strip detector helps to further enhance the
momentum resolution down to $\sigma(1/P_T)$ = 2 $\cdot 10^{-5}$. 
The highly granular electromagnetic and hadron calorimeters
are both placed inside the 3.5\,T superconducting solenoid, and the return yoke is instrumented to detect muons.
The low angle region is of utmost importance for this analysis. Here, 
a set of discs of silicon detectors allows to reconstruct charged tracks down to 7$^\circ$ from the beam axis.
Below this angle, the forward calorimeters are placed: The luminosity monitor, LumiCal, behind it the low angle hadron
calorimeter, LHCal, which assures that also hadrons can be detected to the lowest angles. 
In the very forward region the BeamCal is placed,
mounted directly on the beam-pipe. The holes in the BeamCal for the beam-pipes are the only uninstrumented part of the system.
Seen from the interaction point, these acceptance holes extends to an angle to the beam of 6\,mrad. 

A corresponding description of ILD at FCCee, featuring in particular the final focus quadrupoles and masks extending into the tracking volume, is currently being implemented, but is not yet available for massive simulation and reconstruction of events for physics analyses. Therefore, we will assess the impact of the FCCee-compatible machine-detector interface later in a simplified manner, disregarding any calorimeter hits below the start of the FCCee LumiCal acceptance 
at 50\,mrad from the beam axis (see \cite{FCC:2018evy}).

  The study assumes R-parity conservation and a 100$\%$ decay of the $\widetilde{\tau}$ to ${\tau}$
  and the lightest neutralino ($\chi^0_1$), the LSP in this case.
  In order to select the worst scenario, the $\widetilde{\tau}$ mixing angle was set to 53 degrees,
  and the LSP was assumed to be a pure higgsino
  as this corresponds to lowest sensitivity when all polarisation samples are combined - a combination of production cross section,
  selection efficiency and level of selected backgrounds.
  This choice will be discussed in detail later, in Sec.~\ref{sec:limitcalc}.
  
\subsection{ILC Conditions and tools used\label{sec:tools}}
The generated background event samples were those used for the Interim Design Report of
ILD (the  ``IDR'')~\cite{ILDConceptGroup:2020sfq,Berggren:2021sju}. 
The event generator {\tt Whizard}~v1.95~\cite{Kilian:2007gr} was used, and the samples
contain {\it all} standard model processes with up to six fermions in
the final state. 
For treating ${\tau}$ decays, {\tt Whizard} was interfaced to {\tt Tauola}~\cite{Jadach:1990mz}.
  {\tt Tauola} simulates the ${\tau}$ decays taking into account the ${\tau}$ polarisation in the products.
  Beam-spectra and the amount of photons in the beams were simulated with {\tt GuineaPig}~\cite{Schulte:1999tx}.
  Detector simulation and reconstruction were done on the Grid using {\tt DDSim}~\cite{Petric:2017psf} 
  and {\tt Marlin}~\cite{Gaede:2006pj}.
  The Grid production was done by the ILD production team using the {\tt Dirac}~\cite{Tsaregorodtsev:2008zz} system.
  
  The {\tt SGV} fast detector simulation~\cite{Berggren:2012ar,SGV}, adapted to ILD,
  was used for detector simulation and event
  reconstruction for signal events.
  As is shown in~\cite{Berggren:2012ar,Berggren:2025fpw}, SGV agrees excellently with estimates from full detector simulation,
  in particular for the measured properties of charged particles.
  The events were generated
  with {\tt Whizard}~v2.8.5, also interfaced to {\tt Tauola} for
  correct treatment of the ${\tau}$ decays.
  The same beam-spectrum as was used for the fully simulated background samples was also used for the signal sample.
  Both the signal and background samples were analysed using the tools
  included in {\tt SGV}.
  Signal points were specified in SUSY Les Houches accord files~\cite{Allanach:2008qq}.
  The $\widetilde{\tau}$ masses ranged from 100 to 249\,GeV: 100, 150, then from 200 in intervals of 5\,GeV (200, 205,
  210, ..., 235) and from 240 to 249 in intervals of 1\,GeV. In both cases $\Delta M$ was taken from 2 to 10\,GeV in 1\,GeV steps. Signal samples with higher $\Delta M$ values, 12, 34, 56, 78, 100, 122 and 124\,GeV, were also generated for each of the above mentioned $\widetilde{\tau}$ masses
  (except at 100 and 150\,GeV) in order to cross-check previous studies.
  At all points, the $\widetilde{\tau}$ mixing angle was 53 degrees,
  and the LSP was a pure higgsino.
  In addition, at some selected points events were also generated with other mixing angles,
  and other LSP composition; this was in order to verify that the selected working-point indeed
  was the worst possible one, see Sec.~\ref{sec:limitcalc}.
  For all of these signal points, 10000 events were generated for each of the
  beam-polarisations $\mathcal{P}_{+-}$ and  $\mathcal{P}_{-+}$.
  With appropriate reweighting, these samples were also used to evaluate the signals expected in
  the smaller $\mathcal{P}_{--}$ and $\mathcal{P}_{++}$ samples.

  As discussed in Sec.~\ref{sec:ILC}, spurious events are expected to be present in each bunch crossing. At the ILC
  with $\sqrt{s}=500$\,GeV an average of 1.05 low-$\it{P_{T}}$-hadron events 
  from $\gamma\gamma$ interactions 
  is expected per bunch crossing. 
  A number ($\mathcal{O}(10)$ ) of  electron-positron pairs from beam-beam interactions is also expected to reach the
  tracking system of the detector in each bunch crossing.
  The $\gamma\gamma$ interactions were generated either with {\tt Pythia} 6.422~\cite{Sjostrand:2006za} (if  $M_{\gamma\gamma}$ > 2\,GeV)
  or a dedicated generator~\cite{Chen:1993dba} for $\gamma\gamma$ interactions (otherwise).
  The electron-positron pairs from beam-beam interactions were generated with
  {\tt GuineaPig}. A pool of such events were created, and random events picked from
  the pool were added to each physics event during full simulation.
  The {\tt SGV} simulation did not implement this mechanism, instead reconstructed
  objects coming from overlay were extracted from random background events and
  overlaid on the signal events at analysis level\footnote{To also study the situation at
    circular colliders, at some selected points the study was repeated without
    overlay-on-physics (both signal and background), as this effect is expected to
    be absent at circular colliders, see Sec.~\ref{sec:ILC}.}.

The study of overlay-only events was focused on two mass differences
between the $\widetilde{\tau}$ and the LSP, one low and one moderate,
since the mass difference is highly relevant to the effect of these events.
The strong rejection factor needed, stronger than $\mathcal{O}(10^{-9})$ (see Sec.~\ref{sec:ILC}),
implies that  it would be needed to simulate $\mathcal{O}(10^{10})$ overlay-only events to
estimate with confidence such a strong rejection factor.
This is beyond the reach of the computing resources currently thinkable for the analysis (albeit it {\it would}
be possible with the resource a future running experiment should have).
This problem is solved by
identifying sets of {\it independent} cuts;
the
total rejection factor was computed as the product of the factors obtained with either of these sets.
This will be further discussed below, in Sec.~\ref{sec:varycuts}.

 Well identifying $\tau$'s is obviously a key requirement for this analysis.
  To do this, we use the DELPHI $\tau$ finder~\cite{Abdallah:2003xe}.
  This algorithm was particularly developed to identify $\tau$'s in low multiplicity
  events. 
  It iteratively builds $\tau$ candidates from all possible combinations of charged tracks, starting with single tracks,
  then in each iteration adding tracks to existing combinations, always retaining the combination yielding the
  lowest mass, and requiring that the mass is below 2\,GeV. Characteristics of $\tau$ decays are also taken into account,
  so that e.g.\ an identified muon is not allowed to be grouped with another track. The charged track grouping is terminated
  when no more groups with mass below 2\,GeV can be made. Neutrals are then added to the found candidates, as long as the mass
  remains below 2\,GeV - any neutrals that can not be grouped are left as belonging to the ``rest-of-event''
  class.

  \begin{figure}[htbp]
    \centering
    \subcaptionbox{}{\includegraphics [scale=1.0]{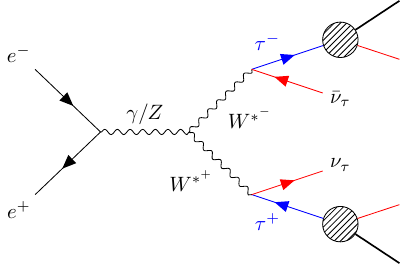}}
    \subcaptionbox{}{\includegraphics [scale=1.0]{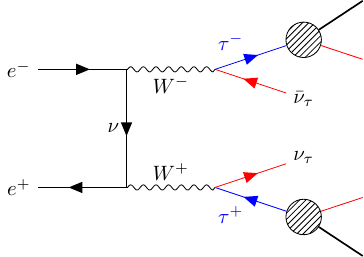}}

    \subcaptionbox{}{\includegraphics [scale=1.0]{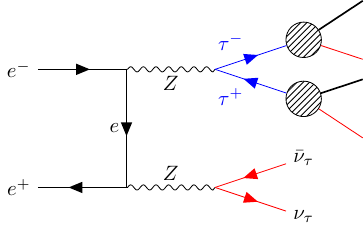}}
    \caption{Main ``irreducible'' background sources: 
(a) s-channel $WW$ production with both $W$'s decaying to $\tau$ and $\nu$; 
(b) t-channel $WW$ production with both $W$'s decaying to $\tau$ and $\nu$; 
(c) t-channel $ZZ$ production with one $Z$ decaying to $\tau$'s, the other to $\nu$'s. 
The same colour-coding is as in Fig.~\ref{fig:signdiagrams} is used, and the various $\tau$ decays modes (cf.\ Figs.~\ref{fig:signdiagrams}(b) and (c))
are collectively represented by the hatched circles.
    \label{fig:irredbckdiagrams}}
  \end{figure}

\subsection{Main background sources\label{sec:mainbck}}
  The main sources of background, given the generic signal topology, i.e.\ two $\tau$'s and an unseen
  recoil system, are SM processes with real or fake missing energy. They can be classified into
  ``irreducible'' and ``almost irreducible'' sources. The first are events which do contain two $\tau$'s
  and real missing energy from un-detected neutrinos. The main contribution to this group are $ZZ$ events
  with one $Z$ decaying to two neutrinos and the other to two $\tau$'s, and fully leptonic $WW$ events,
  where both the $W$'s decays to $\tau \nu_\tau$.
  The dominating diagrams are shown in Fig.~\ref{fig:irredbckdiagrams}.
  Also $ZWW$ and $ZZZ$ events decaying to two $\tau$'s and four neutrinos constitute a minor contribution to
  the irreducible background, 
  and are included in the simulated background samples.
\begin{figure}[htbp]
    \centering
    \subcaptionbox{}{\includegraphics [scale=1.0]{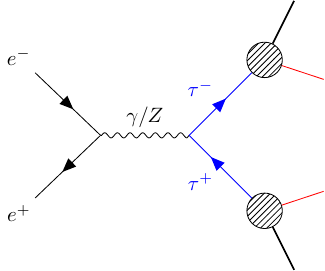}}
    \subcaptionbox{}{\includegraphics [scale=1.0]{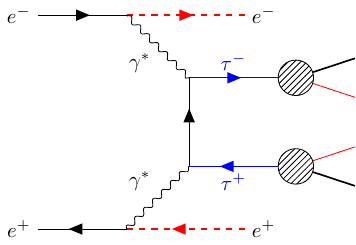}}
    \caption{Main background sources: (a) $\tau$ pair production and decay; 
      (b)  $\tau$ pair production 
      through the multi-peripheral $\gamma\gamma$ process. 
             In (b), the thick dashed red lines indicate that while the outgoing
             electrons or positrons are in principle detectable, this diagram contributes to the background only in the case that they are
             deflected so little that they escape detection by remaining inside the beam-pipe, and hence contribute to (fake) missing energy.}
    \label{fig:redbckdiagrams}
  \end{figure}

  The second category of backgrounds are those which do not contain two $\tau$'s and neutrinos, but after reconstruction look very similar.
  They are processes with two soft $\tau$-jets, with two leptons other than $\tau$'s together with true
  missing energy or two $\tau$'s accompanied by fake missing energy.
  The main sources for events with true missing energy in this group are on one hand $\tau$
  pair production, with the $\tau$'s decaying such that most energy goes to the neutrinos, (cf.\ Fig.~\ref{fig:redbckdiagrams}(a)), 
  and on the other hand, $ZZ$ or $WW$ decaying to two neutrinos and at least one lepton other than $\tau$'s, i.e.\ the
  type of processes shown in Fig.~\ref{fig:irredbckdiagrams}, but with one, or both, $\tau$'s replaced by muons or electrons.
  The background with fake missing energy comes mainly from $\tau$ pair
  production with Initial State Radiation (ISR) at very low angles, events with two $\tau$'s and
  two very low angle electrons (below the acceptance of the detector) in the final state (cf.\ Fig.~\ref{fig:redbckdiagrams}(b)) and events 
  where two $\tau$'s are produced
  by a $\gamma\gamma$ interaction and not by an $e^+e^-$ one  ; in the latter case
  energy is not actually missing, but the assumption that the initial energy is
  the energy of the incoming electron and positron is wrong.

\section{Analysis\label{sec:analysis}}
  Taking into account the signal signature and the main background sources,
  different cuts have been designed in order to separate the signal from
  the background. We will discuss them in the following by separating them into preselection cuts(Sec.~\ref{sec:cuts_1}), model-point dependent cuts (Sec.~\ref{sec:cuts_2}), cuts against non-$\tau$ background (Sec.~\ref{sec:cuts_3}) and, finally, into cuts varying by $\Delta M $ (Sec.~\ref{sec:varycuts}).
  
   Tables~\ref{tab:cutsflow_dm34} and~\ref{tab:cutsflow_dm10} show the cutflow at two model-points in the order the cuts are applied in the discussion that follows.
  The numbers correspond to the data-set with polarisation $\mathcal{P}_{+-}$. 
  In addition the final numbers after all cuts for all four polarisations are given in 
  Tables~\ref{tab:finalcounts_pols_dm34} and~\ref{tab:finalcounts_pols_dm10}.
  All figures in this chapter will show background and signal distributions for various quantities to
  which cuts are applied at the point where the cut on the quantitiy shown will be the next to be made,
  i.e.\ for events passing all previous cuts.
  
\subsection{Preselection cuts\label{sec:cuts_1}}
  The study was focused on small differences between the $\widetilde{\tau}$ and LSP
  masses,
  $M_{\tau}$ $<$ $\Delta M$ $\le$ 10\,GeV. This class of signal events are quite similar to
  high cross section events of the type illustrated in
  Fig.~\ref{fig:redbckdiagrams}(b) - known as multi-peripheral
  $\gamma\gamma$ events - with one important difference, namely that 
  in the  $\gamma\gamma$ events the final state also includes the beam-remnant electrons and positrons.
  These are scattered by some low angle, so that demanding absence of the observation of a large individual 
  shower in the calorimeter closest to the beam pipe
  {\color{black}(the BeamCal)} strongly reduces this source of background, and was required
  before applying the following cuts. 
  
  Cuts are then applied on properties that the signal often {\it does not} have,
  while some backgrounds {\it do have}:
  The  {\color{black}multiplicity} of the event can
  be constrained taking into account that the visible part of the $\widetilde{\tau}$ decays comes only from
  the decays of the two $\tau$'s and maybe an ISR photon. 
  For that reason
  the number of charged particles in the event is required to be between 2 and 10.
  This cut
  removes practically all hadronic backgrounds.
$WW$ events with each of the $W$'s decaying to a lepton and a neutrino are
   {\color{black}highly forward-backward asymmetric}; they can be effectively removed by requiring the sum of the product of the
  charge and the cosine of the polar angle, $ q \cos{ \theta_{jet}}$, of the two most energetic jets to be above -1.0,
  see Fig.~\ref{fig:qcos_emiss}(a).
  Jets are, at this stage, formed by the generic JADE algorithm~\cite{Bartel:1983jade}~\cite{Bartel:1986jade} 
with its limiting distance parameter, $y_{cut}$, set to 0.02.
  A more specific $\tau$ clustering will be used later. 
  $ZZ$ events with one $Z$ decaying to two neutrinos and the second one to a electron or muon
  pair are highly suppressed  {\color{black}demanding  a visible mass ($M_{vis}$) more
  than 4\,GeV from the $Z$ mass}.

 \begin{figure}[htbp]
    \centering
\includegraphics [scale=1.0]{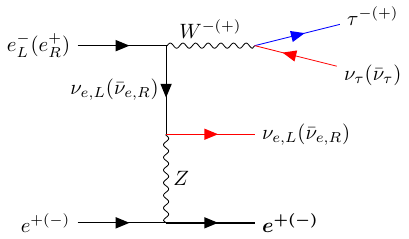}
    \caption{
     Single $W$ production with  
              the $W$ decaying to $\tau$ and $\nu$.
      The labels indicate the only beam-polarisations for which this process is possible:
             either left-handed electrons or right-handed positrons. 
    \label{fig:singleWdiagram}}
  \end{figure}

\subsection{Model-point dependent cuts\label{sec:cuts_2}}  
The cuts mentioned in the previous section do not depend on
  the properties of the signal model point. However, all following cuts applied will depend on the model-point considered,
  and so will depend on either, or both, of $M_{\widetilde{\tau}}$ and $M_{LSP}$.

  The next group of cuts contains those that correspond to the properties that the
  $\widetilde{\tau}$ events {\it must} have. Since the two LSPs from the
  $\widetilde{\tau}$ decays are invisible to the detector, signal events
  must have a {\color{black}missing energy}, $E_{miss}$, greater than $2 \times M_{LSP}$.
    \begin{figure}[htbp]
    \centering
   \subcaptionbox{}{\includegraphics [width=0.45\textwidth]{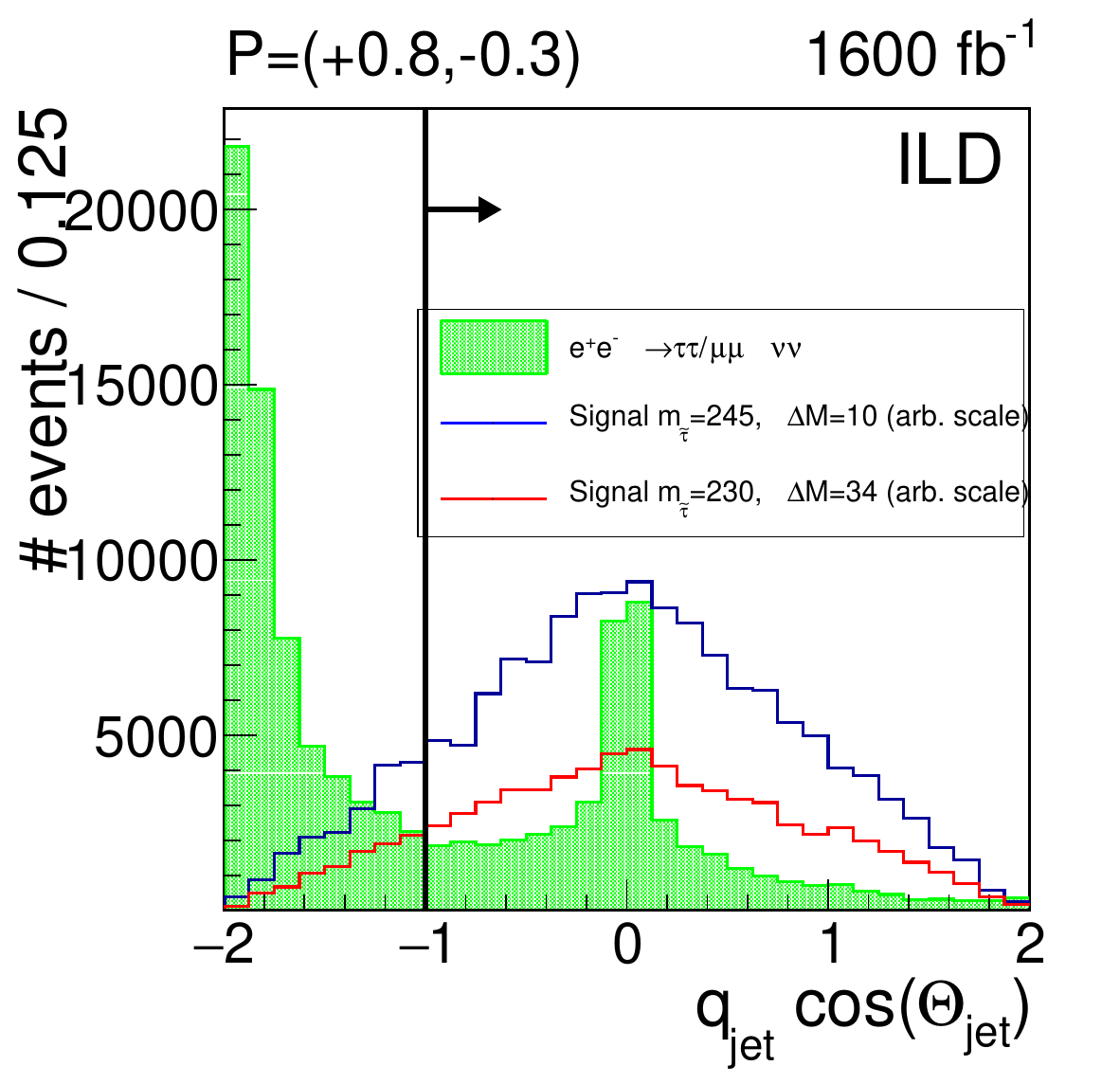}}
      \subcaptionbox{}{ \includegraphics [width=0.45\textwidth]{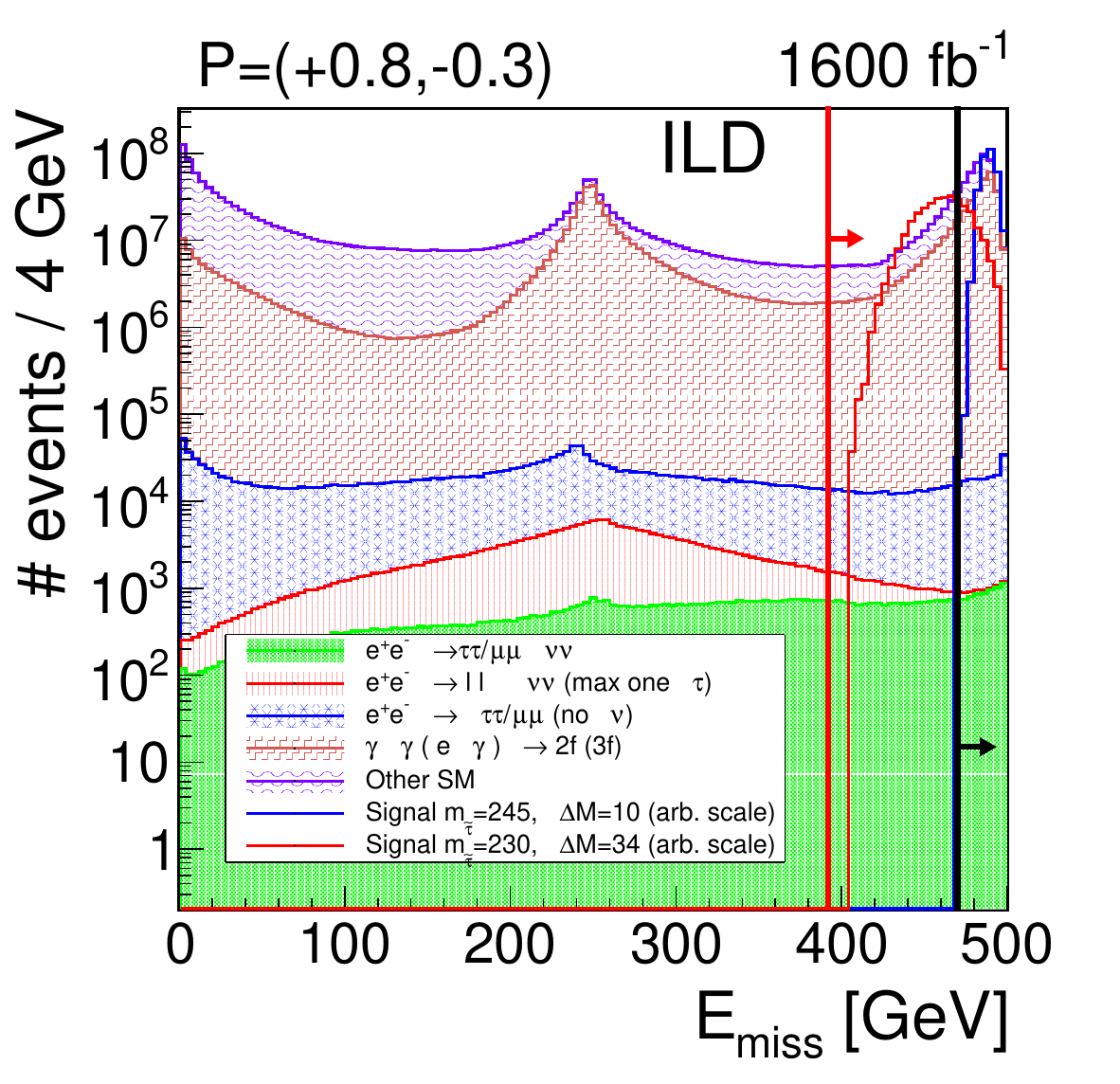}}
   \caption{Distributions for background and signal (the latter on an arbitrary scale).
     (a) : $q_{jet} \cos{\theta_{jet}}$ after the cut on  the number of charged particles. 
    Only the irreducible $e^+ e^- \rightarrow l \nu l \nu$
    is shown,
     since this background is what this cut is designed to reduce. 
     The background peak at -2 is mainly populated by  the highly asymmetric $W^+W^- \rightarrow l \nu l \nu$,
     while the peak at 0 is the  $ZZ \rightarrow l \nu l \nu$ contribution.
     (b) The missing energy after previous cuts have been applied (cf.\ Tabs.~\ref{tab:cutsflow_dm34} and ~\ref{tab:cutsflow_dm10}).
     The arrows indicate the region where events are accepted in at the model points indicated by the colour of the lines.}
    \label{fig:qcos_emiss}
  \end{figure}
  Likewise, the {\color{black}$M_{vis}$} can not be greater than $E_{CMS} - 2 \times M_{LSP}$.
  Therefore, events should fulfil $M_{vis} < E_{CMS} - 2 \times M_{LSP}$ and
  $E_{miss} > 2 \times M_{LSP}$ (cf.\ Fig.~\ref{fig:qcos_emiss} (b))
  to be considered further.
    Events with the higher jet momentum {\color{black}($p_{jet~high}$)} above the limit given by Eq.~\ref{eq:pmax} are excluded from the further analysis at the
    given model point. 

\begin{figure}[htbp]
    \centering
 \subcaptionbox{}{ \includegraphics [width=0.45\textwidth]{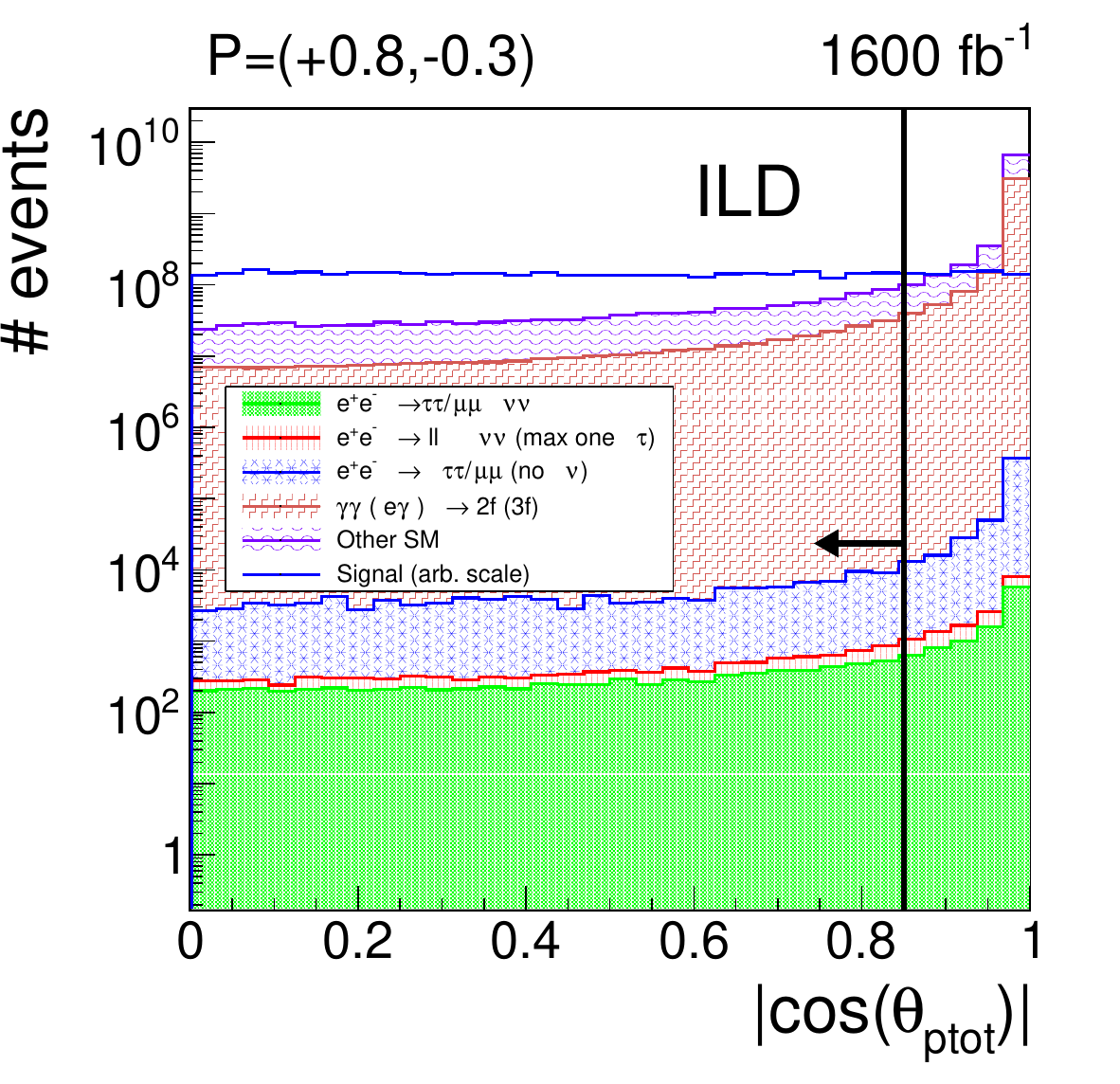}}
 \subcaptionbox{}{ \includegraphics [width=0.45\textwidth]{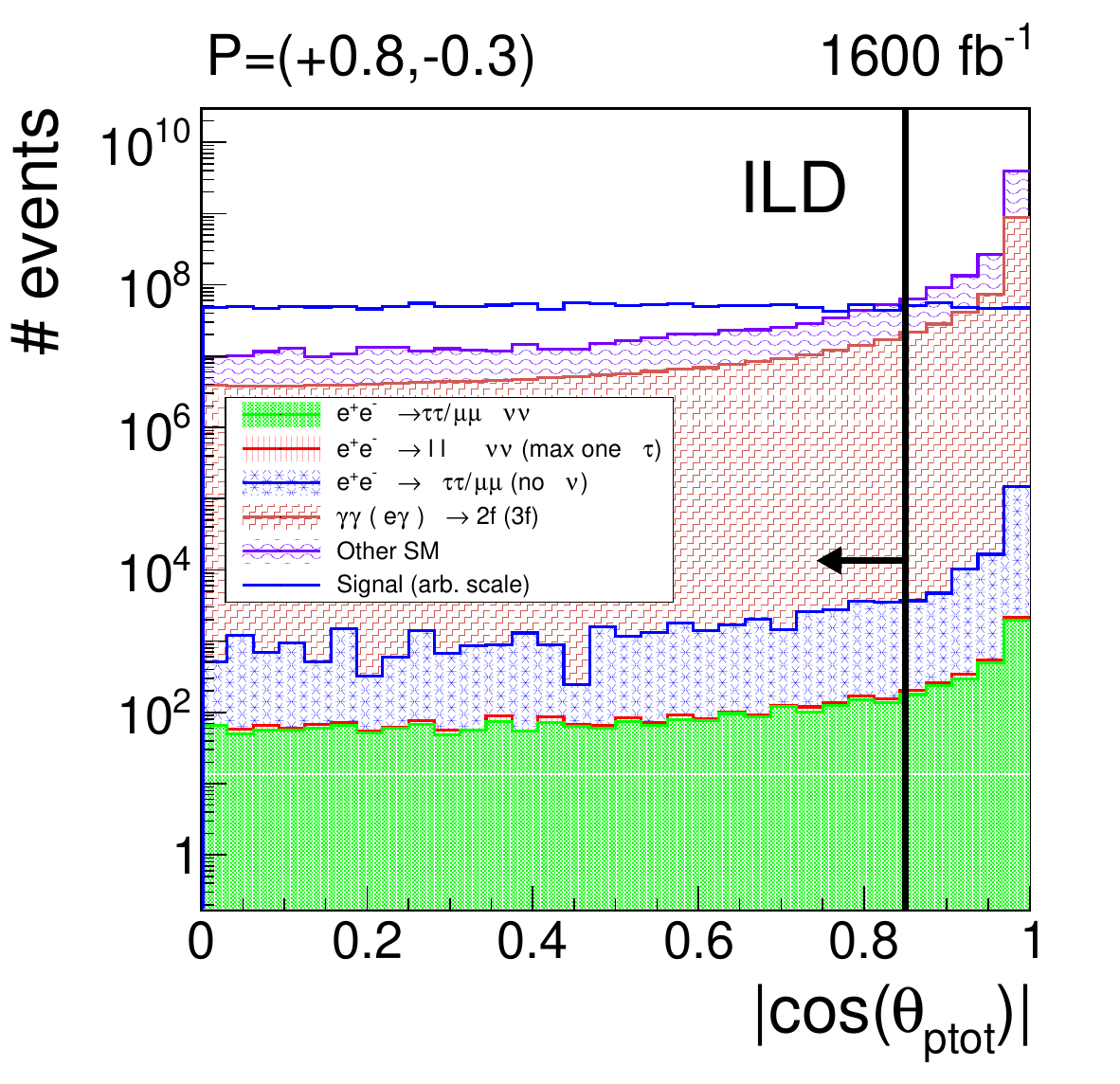}}
   \caption{
Distributions for  the polar angle of the total seen momentum
     (a)  M$_{\widetilde{\tau}}$ = 230\,GeV $\Delta M$ = 34\,GeV
     and (b)   M$_{\widetilde{\tau}}$ = 245\,GeV $\Delta M$ = 10\,GeV.
The two signals are on arbitrary scale,
     and all previous cuts have been applied (cf.\ Tables~\ref{tab:cutsflow_dm34} and~\ref{tab:cutsflow_dm10}). 
     The arrows indicate the region where events are accepted.}
     \label{fig:costh_ptott_dm34_dm10}
  \end{figure}

As already pointed out, the  $\widetilde{\tau}$'s are
  scalars, and therefore isotropically produced, while the backgrounds are
  either fermions or vector bosons, and tend to be produced at small angles to the beam
  axis.
The {\color{black}total seen momentum}, $\sum \bar{p}$, therefore tends to be in an almost random direction
  in signal events, while it tends to point close to the beam-axis for most
  backgrounds. Therefore a cut on the direction of the total momentum, $|\cos(\theta_{\sum \bar{p}})|<0.85$, was
  imposed, see Fig.~\ref{fig:costh_ptott_dm34_dm10}.

\subsubsection{Cuts against non-$\tau$ background\label{sec:cuts_3}}  
To {\color{black}select the two-$\tau$ topology}, the total event should have charge 0 or $\pm$ 1, 
and the DELPHI $\tau$ finder should have found exactly two $|q|=1$~$\tau$ candidates, with opposite charges,
and made by either one or three charged particles.
  These two candidates could be accompanied by further particles,
  that are not compatible with the requirements imposed by the DELPHI $\tau$-finder
  to be considered as $\tau$-jets.

  A further set of conditions on the jets
  is devoted to the reduction of background from sources with
  leptons not from $\tau$ decays.
  The background of single $W$,
  with the $W$ decaying to $\tau$ and neutrino, see Fig.~\ref{fig:singleWdiagram},
  can be reduced by a cut depending on beam-polarisation:
  this process can only occur if at least one of the beams has the
  correct polarisation.
  Since the degree of polarisation of the positron beam is lower than that of the electron beam,
  this background is more likely to yield an electron as the beam-remnant for $\mathcal{P}_{+-}$, a
  positron for $\mathcal{P}_{-+}$. We therefore remove events with an {\color{black}electron(positron) $\tau$ candidate in the
   $\mathcal{P}_{+-}$ ( $\mathcal{P}_{-+}$)} samples.
This background
  together with the background from $WW \rightarrow e\nu_{e} \mu\nu_{\mu}$ and
  from $\gamma\gamma$ events with a beam-remnant deflected to larger
  angles is further reduced by rejecting those events in which the {\color{black}most energetic
  jet consists of a single electron}. 
  The two charged jets were also required to
  not be made by {\color{black}single leptons with the same flavour}.
  These selections reduce the signal efficiency to about 30$\%$ but with
  a reduction of the background by a factor of 50 to 80, depending on the
  region of the SUSY parameter space.

\subsection{Cuts varying by $\Delta M $\label{sec:varycuts}}
A further group of cuts is based on those properties that the
  $\widetilde{\tau}$-events {\it might} have, but will {\it rarely} be
  present in background events.
In this case, the scalar nature of  the $\widetilde{\tau}$
allows to impose cuts requiring events to have {\color{black}high missing transverse
  momentum ($P_{T miss}$)}, and {\color{black}large acoplanarity} $\theta_{acop}$.
  One type of background still present at this stage is
  from $e^+ e^- \rightarrow \tau \tau$, with the {\it visible} part of the decay of one of the $\tau$'s in the direction of its parent,
  while the other $\tau$ decays with the {\it invisible} $\nu$ closely aligned with the direction of its parent.
  These events fake the signal topology,
  having both a large missing transverse momentum and high acoplanarity.
  To reduce this source of background,
  a cut on
   the variable $\rho$ is imposed.
   This variable is  calculated by
first projecting the two jets on the x-y
  (transverse) plane, and calculating the thrust axis (the axis that maximises the sum over the jets of the absolute values
of their momenta projected unto the axis). $\rho$
 is then the sum of transverse momenta of the jets (in the plane) with respect to the
  thrust axis, and can be calculated as
   \begin{equation}
   \rho = \frac{2 || \bar{p}_{xy_1}\times   \bar{p}_{xy_2}||}{||   \bar{p}_{xy_1} \pm   \bar{p}_{xy_2}||} =
 \frac{2  p_{T_1}  p_{T_2} \sin{\theta_{acop}}}{\sqrt{ p^2_{T_1} + p^2_{T_2} + 2 p_{T_1}  p_{T_2} | \cos {\theta_{acop}}|}}
    \label{eq:rho}
  \end{equation}
  where $\bar{p}_{xy_i}$ is the two-dimensional momentum of jet $i$ projected into the x-y
  plane, and  $p_{T_i}= ||\bar{p}_{xy_i}||$ is the transverse momentum of jet $i$.
In the first expression, the positive sign should be taken if $\theta_{acop}<\pi/2$, the negative othervise.
  The $\tau \tau$ events surviving all previous cuts must have low values of $\rho$,
  in contrast to signal events, 
\subsubsection{Cuts for $\Delta M \ge 11$\,GeV\label{sec:highdmcuts}}
 
\begin{figure}[htbp]
    \centering
    \subcaptionbox{}{\includegraphics [width=0.32\textwidth, trim=0 1cm 0 0]{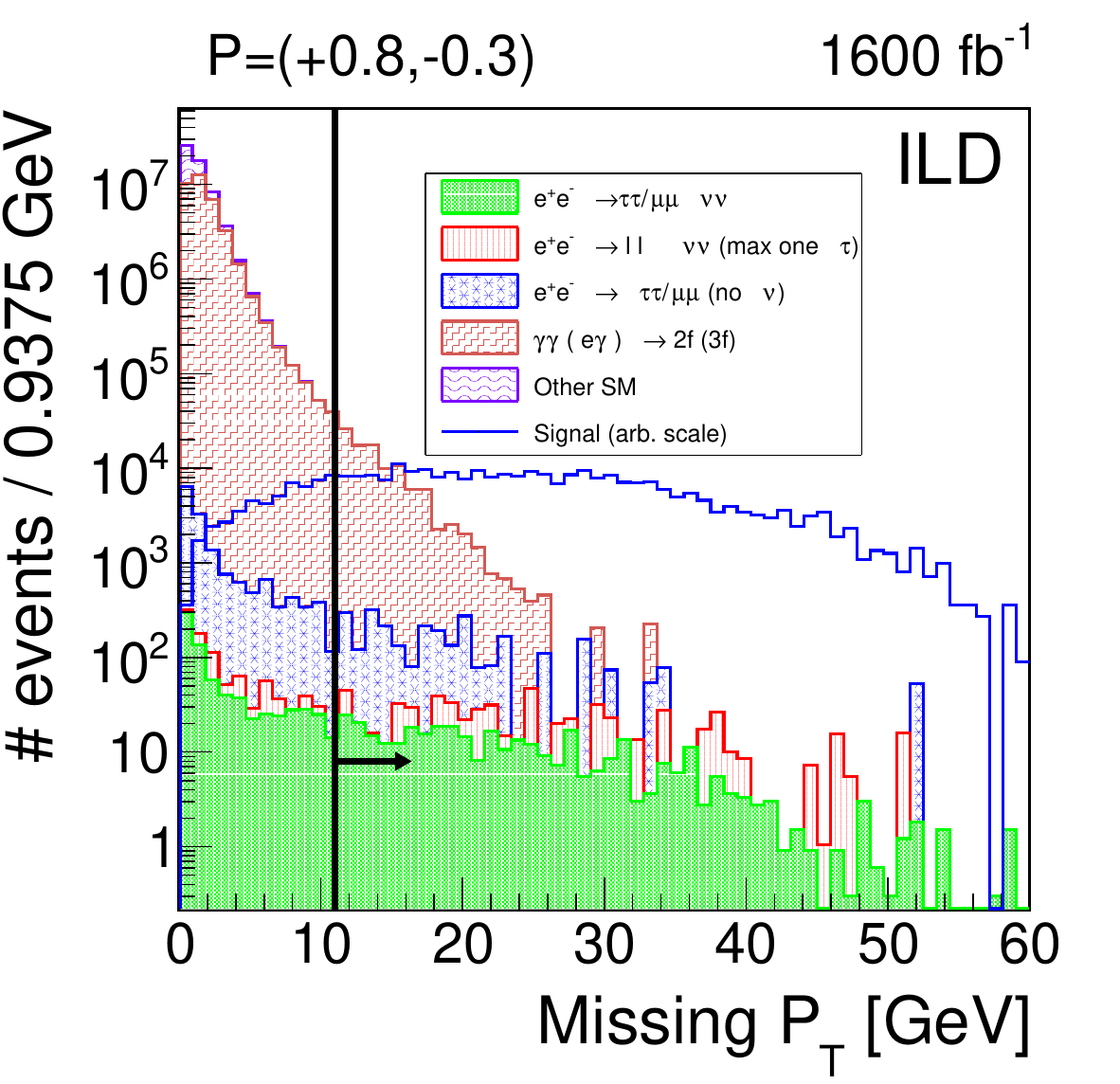}}
    \subcaptionbox{}{\includegraphics [width=0.32\textwidth, trim=0 1cm 0 0]{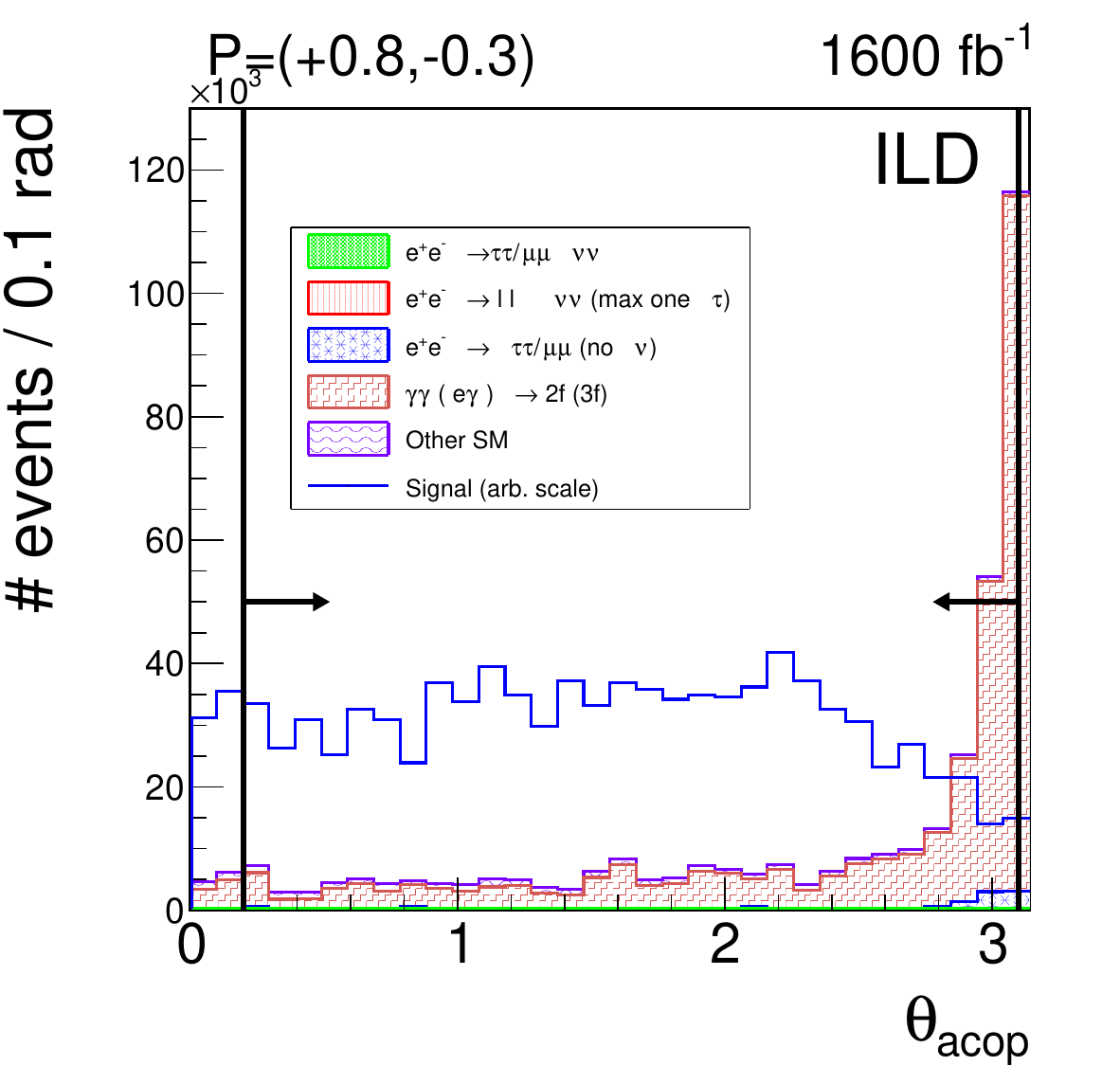}}
    \subcaptionbox{}{\includegraphics [width=0.32\textwidth, trim=0 1cm 0 0]{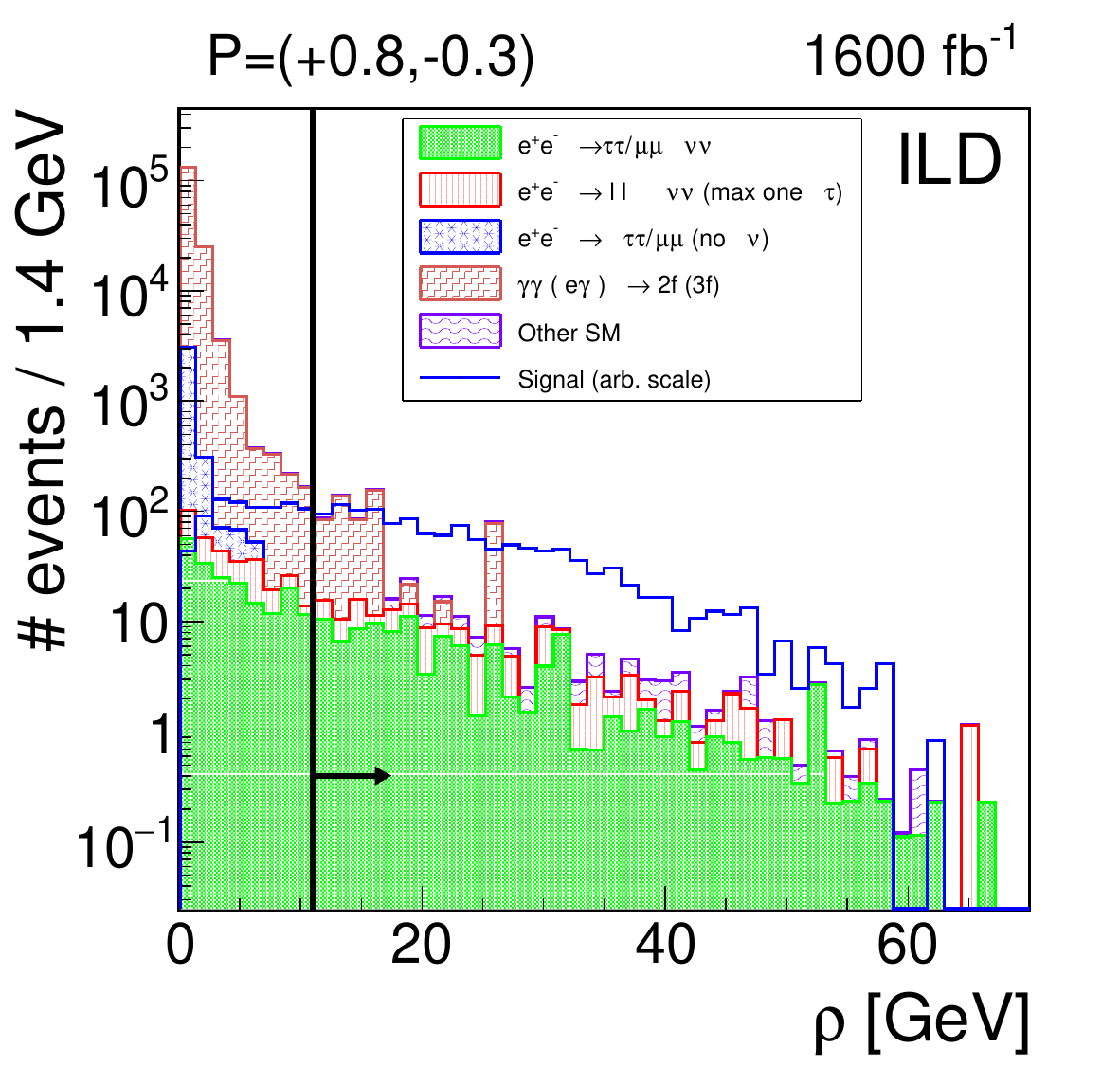}}
   \caption{
     Distributions for   M$_{\widetilde{\tau}}$ = 230\,GeV $\Delta M$ = 34\,GeV of 
     (a) the missing transverse momentum,
     (b)  the acoplanarity angle between the two jets 
     and (c) the variable $\rho$, described in the text.
The signals are on arbitrary scale,
and all previous cuts have been applied (cf.\ Table~\ref{tab:cutsflow_dm34}). The arrows indicate the region where events are accepted.}
     \label{fig:ptmiss_thetaacop_rho_dm34}
  \end{figure}

Missing transverse momentum, large acoplanarity and $\rho$ are generally more powerful as discriminators at larger mass differences.
  The distributions for these two quantities are shown in Fig.~\ref{fig:ptmiss_thetaacop_rho_dm34}(a), (b) and (c)
  for a typical model-points as they are after applying all previous cuts,
  applicable to the model-point in question.
  It is requested that  $P_{T miss} > $ 11\,GeV and that  $\theta_{acop}$ is in the interval 0.2 to 3.1 radians,
and  that $\rho$ > 11\,GeV.
   \begin{figure}[htbp]
    \centering
    \subcaptionbox{}{\includegraphics [width=0.45\textwidth]{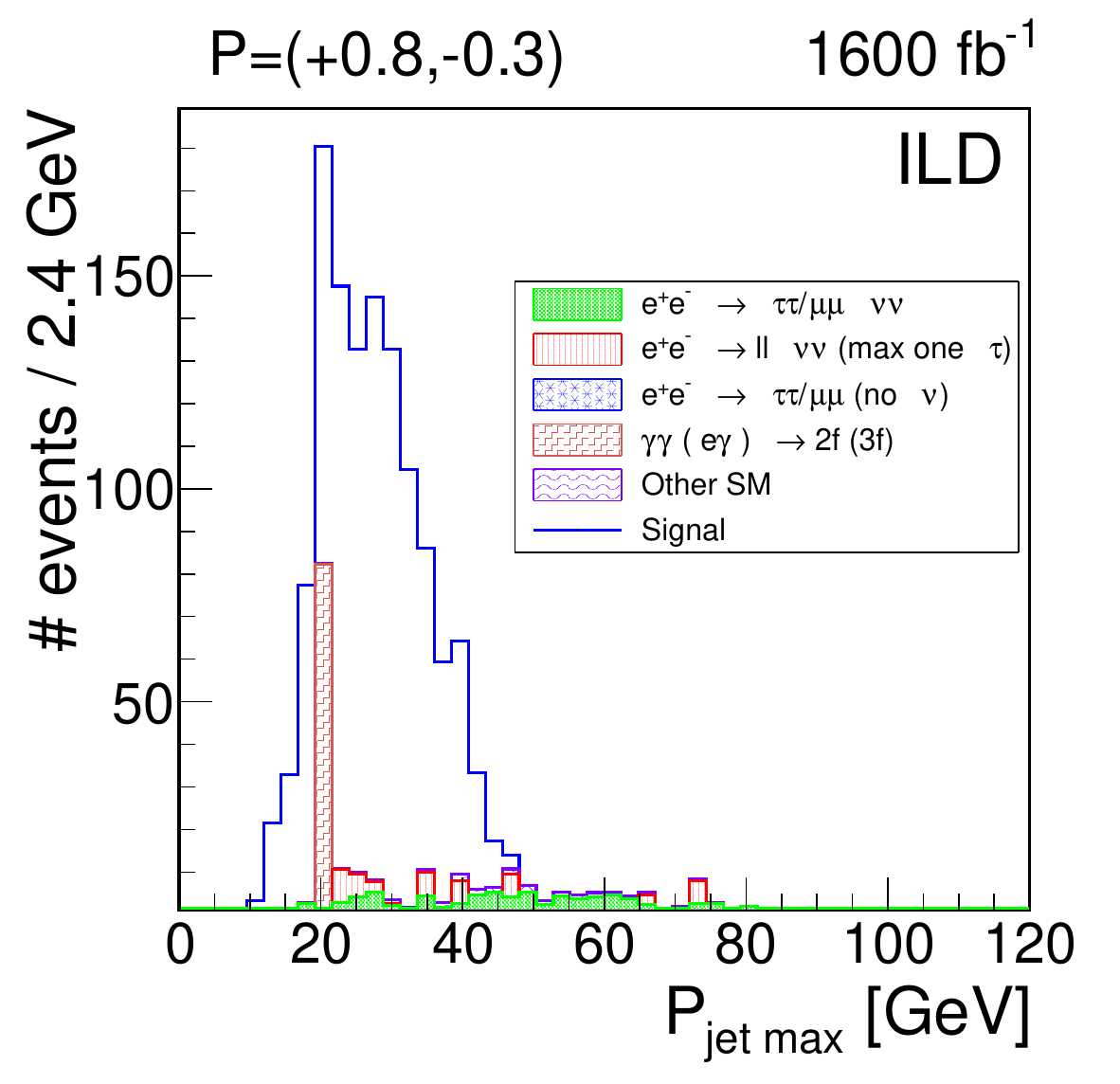}}
    \caption{
     Distribution of the higher of the two jet momenta after all cuts (except the one on $p_{jet~max}$ itself)
     corresponding to the model point M$_{\widetilde{\tau}}$ = 230\,GeV, $\Delta M$ = 34\,GeV.
     The maximum $P_{jet}$ possible at this model point is 47.63\,GeV (Eq.~\ref{eq:pmax}), and is indeed the endpoint of
     the signal distribution.
     (Contrary to other figures in this section, the signal is stacked on top of the background,
     and corresponds to the same integrated luminosity as the background. The contribution
     from $\gamma\gamma$ events at around 20\,GeV is in fact one single simulated event.)}
    \label{fig:pjetmaxend_dm34}
  \end{figure}
Cutting in these properties has a certain
  cost in efficiency but improves the signal-to-background ratio.

  As a final cut it is noted that {\color{black}sizeable energy
  detected at low angles} to the beam is rare in signal events, but common in many backgrounds.
  Events with more than 2\,GeV detected at angles lower than 20
  degrees to the beam axis are therefore rejected.
  
  After applying these cuts the main sources of remaining background are $WW$ events with each
  $W$ decaying to $\tau\nu$ and events with four fermions in the final state coming from
  $\gamma\gamma$ interactions, mostly $\tau\tau$ events.
  As an illustration of the expected signal after all cuts, Fig.~\ref{fig:pjetmaxend_dm34} shows the
  distribution of $p_{jet~max}$ for the model-point
  with  M$_{\widetilde{\tau}}$ = 230\,GeV and $\Delta M$ = 34\,GeV, when all described cuts except the one
  on $p_{jet~max}$ itself has been applied. In this figure, the signal and background
  distributions correspond to the same 1600 fb$^{-1}$ of integrated luminosity.
 
\begin{table}[htbp]
    \centering
    \caption{Cut-flow table for the large  $\Delta M$ case.
      The first column shows the cut definition, and the following columns shows the
      number of events passing all cuts up to, and including, the current row. The model point is M$_{\widetilde{\tau}}$ = 230\,GeV $\Delta M$ = 34\,GeV,
      and the numbers correspond to an integrated luminosity of 1.6\,ab$^{-1}$ at 500\,GeV for
    polarisation $\mathcal{P}_{+-}$.}
    \label{tab:cutsflow_dm34}
    \begin{tabular}{lS[table-format=4.1]S[table-format=4.1]S[table-format=4.1]S[table-format=4.1]S[table-format=4.1]S[table-format=4.1]}
      \hline\hline
      \addlinespace[2pt]
      Cut &\alc{c}{Signal}& \alc{c}{$e^+e^- \rightarrow $}     &\alc{c}{$e^+e^- \rightarrow $}  &\alc{c}{$e^+e^- \rightarrow $} &\alc{c}{$\gamma\gamma (\gamma e) \rightarrow $}&\alc{c}{All other}  \\[2pt]
          &               & \alc{c}{$\tau\tau/\mu\mu ~\nu\nu$} &\alc{c}{$ l l ~\nu\nu$}        & \alc{c}{$\tau\tau/\mu\mu $}   &\alc{c}{2(3) fermions}                         &                    \\[2pt]
          &               &                                    &\alc{c}{($\ge 1~l$ not $\tau$)}& \alc{c}{(no $\nu$'s)}         &                                               &                    \\[2pt]
      \hline
      \addlinespace[1pt]
      no cut                             &  8292.8  &  134100 &  533900 & \alc{l}{4.377 \tento{6}}  &\alc{l}{2197 \tento{6}}  & \alc{l}{2616 \tento{6}}\\
      BeamCal veto                       &  8284.5  &  133966 &  532800 & \alc{l}{4.368 \tento{6}}  &\alc{l}{1933 \tento{6}}  & \alc{l}{2603 \tento{6}}\\
      2 $\le N_{charged} \le$ 10           &  8218.0  &  108900 &  403100 & \alc{l}{3.542 \tento{6}}  &\alc{l}{910.4 \tento{6}}  & \alc{l}{1720 \tento{6} }\\
      $ q \cos{ \theta_{jet}} > -1$       &  7050.5  &   70720 &  216800 & \alc{l}{2.671 \tento{6}}  & \alc{l}{800.6 \tento{6}} & \alc{l}{1500 \tento{6} }\\
      $|M_{vis} - M_Z| > 4$               &  7029.0  &   66820 &  207200 & \alc{l}{2.552  \tento{6}} &\alc{l}{795.2 \tento{6}}& \alc{l}{1500 \tento{6} }\\
      $E_{miss} >2 M_{LSP}$                &  6950.2  &   19620 &   14570 &                 365493.0  &\alc{l}{654.4 \tento{6}}    & \alc{l}{1399 \tento{6} }\\
      $M_{vis} < \sqrt{s} - 2 M_{LSP}$    &  6772.7   &   17990 &   12490 &                 334886.7  & \alc{l}{653.9 \tento{6}}   & \alc{l}{1398 \tento{6} }\\
     $P_{jet} < P_{max}$                 &   6205.5  &    11539 &    5947 &                 209270.9   & \alc{l}{641.8 \tento{6}}   &  \alc{l}{1187\tento{6} }\\
      $|\cos(\theta_{\sum \bar{p}})|<0.85$ &  5338.1   &    4269 &    2567 &                 47236.2  &\alc{l}{71.31\tento{6} }     & \alc{l}{61.5\tento{6} } \\
      $\tau$ decay modes                &  2608.1   &    1109 &    561.1 &                16678.8  & \alc{l}{36.01\tento{6} }    & \alc{l}{22.6\tento{6} }   \\
      $P_{T miss}$ > 11\,GeV               & 2367.6   &     774.1 &   452.9 &                10123.1   &            211417            &                     28251 \\
      $\theta_{acop} \in$  [0.2, 3.1]     &  2176.9  &     493.8 &   318.5 &                6076.9   & 160121 &                     6942 \\
      $\rho$ >  11 \,GeV                       &   1138.6  &    110.4 &    65.5 &                    0.0    &  489.4 &                    45.2 \\
      $E_{<20^\circ} < 2$~GeV              &   729.8 &    74.6 &    48.5  &                    0.0     &  0.0 &                   28.6 \\

      \hline
     \hline
      \addlinespace[2pt]
      \addlinespace[2pt]
    \end{tabular}
  \end{table}
 
\begin{table}[htbp]
    \centering
    \caption{Final counts for the large  $\Delta M$ case for all polarisation cases,
      with integrated luminosities as defined by the standard ILC running scenario. The model point is M$_{\widetilde{\tau}}$ = 230\,GeV $\Delta M$ = 34\,GeV.
    For reference, the counts
      assuming unpolarised beams is also given.}

    \label{tab:finalcounts_pols_dm34}
    \begin{tabular}{lcS[table-format=4.1]S[table-format=4.1]S[table-format=4.1]S[table-format=4.1]S[table-format=4.1]S[table-format=4.1]}
      \hline\hline
      \addlinespace[2pt]
      Beam&Integrated &\alc{c}{Signal}& \alc{c}{$e^+e^- \rightarrow $}     &\alc{c}{$e^+e^- \rightarrow $}  &\alc{c}{$e^+e^- \rightarrow $} &\alc{c}{$\gamma\gamma (\gamma e) \rightarrow $}&\alc{c}{All other}  \\[2pt]
      polarisation    &Luminosity&               & \alc{c}{$\tau\tau/\mu\mu ~\nu\nu$} &\alc{c}{$ l l ~\nu\nu$}        & \alc{c}{$\tau\tau/\mu\mu $}   &\alc{c}{2(3) fermions}                         &                    \\[2pt]
          &   [ab$^{-1}$]            &                                    &\alc{c}{($\ge 1~l$ not $\tau$)}& \alc{c}{(no $\nu$'s)}         &                                               &                    \\[2pt]
      \hline
      \addlinespace[1pt] 
      $\mathcal{P}_{+-}$  &1.6& 729.8 &    74.6   &     48.5  &                     0.0 &     0.0               &         28.6 \\
      $\mathcal{P}_{-+}$  &1.6& 526.5  &  654.9   &     540.1 &                     0.0 &     0.0               &         28.6  \\
      $\mathcal{P}_{++}$  &0.4& 107.8  &   22.9    &      17.2 &                     0.0 &     0.0               &          4.4  \\
      $\mathcal{P}_{--}$  &0.4& 84.7 &   88.9    &      73.0 &                     0.0 &     0.0               &          4.4  \\
     \hline
      Unpolarised        &4.0& 1266  &  735.4   &     593.4 &                     0.0 &     0.0               &         61.8   \\

      \hline
      \addlinespace[2pt]
      \addlinespace[2pt]
    \end{tabular}
  \end{table}

  \subsubsection{Cuts for $\Delta M <  11$\,GeV \label{sec:lowdmcuts}}
\begin{figure}[htbp]
    \centering
      \subcaptionbox{}{ \includegraphics [width=0.32\textwidth]{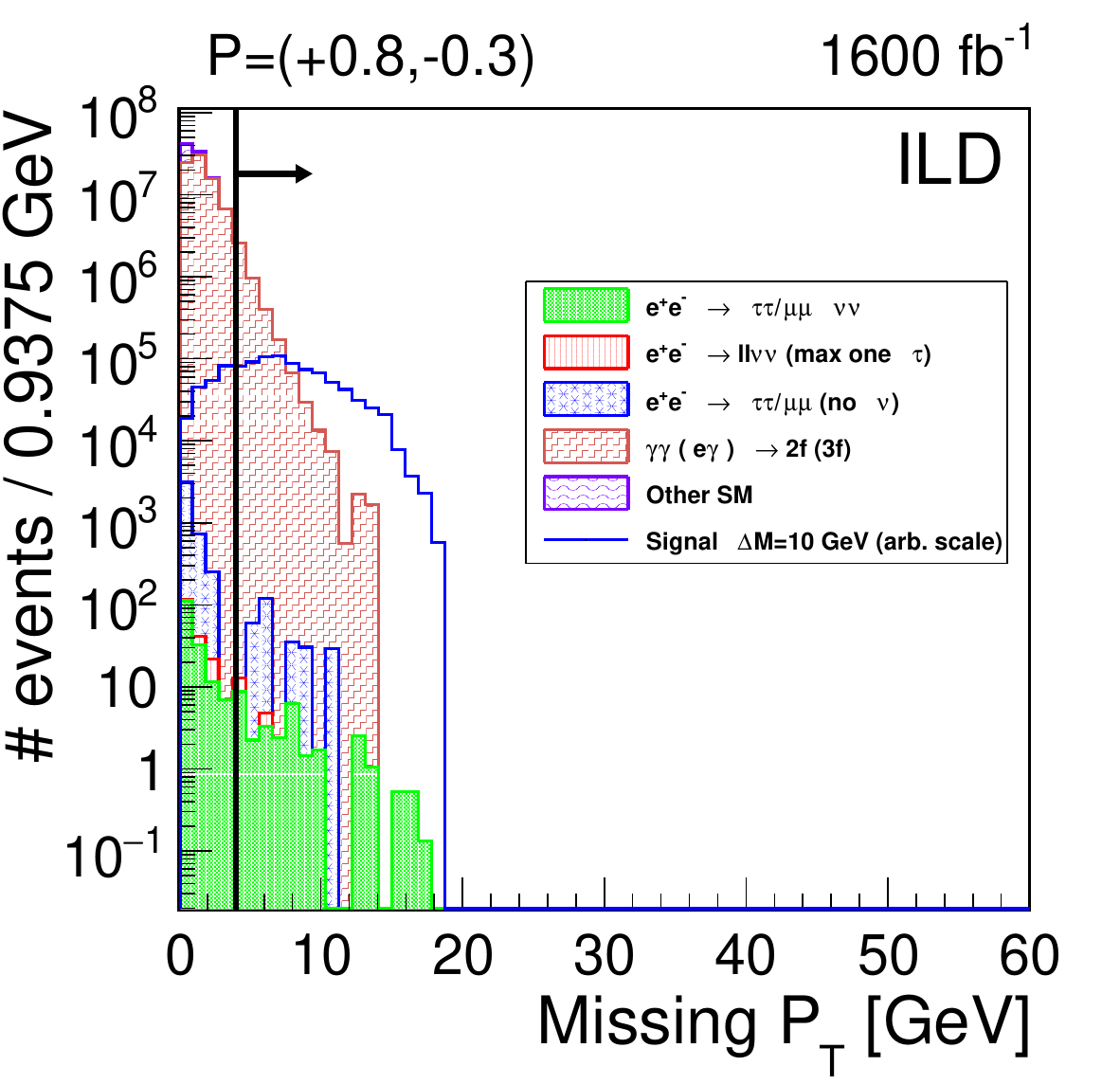}}
    \subcaptionbox{}{ \includegraphics [width=0.32\textwidth]{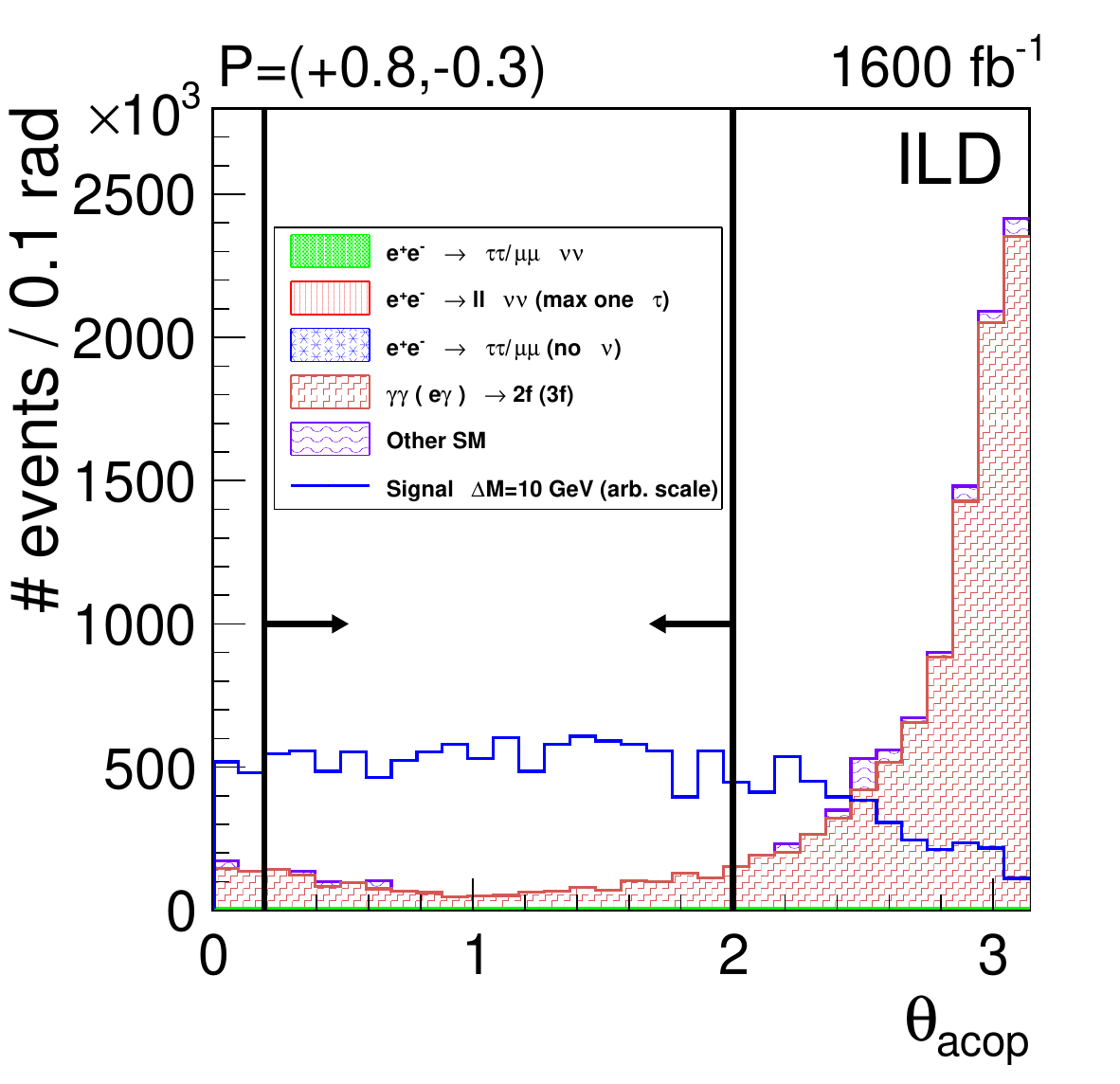}}
    \subcaptionbox{}{\includegraphics [width=0.32\textwidth]{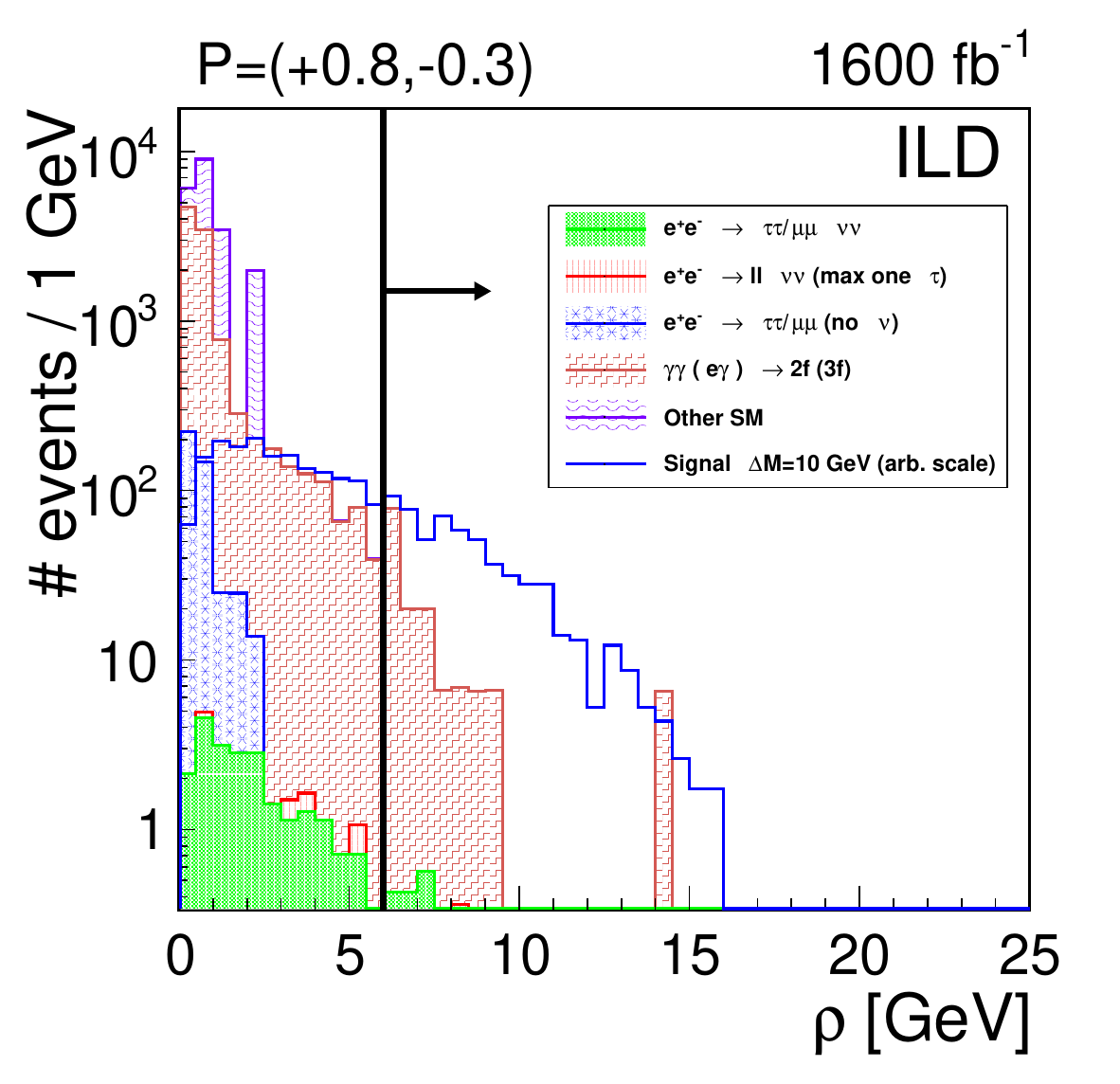}}

    \subcaptionbox{}{ \includegraphics [width=0.32\textwidth]{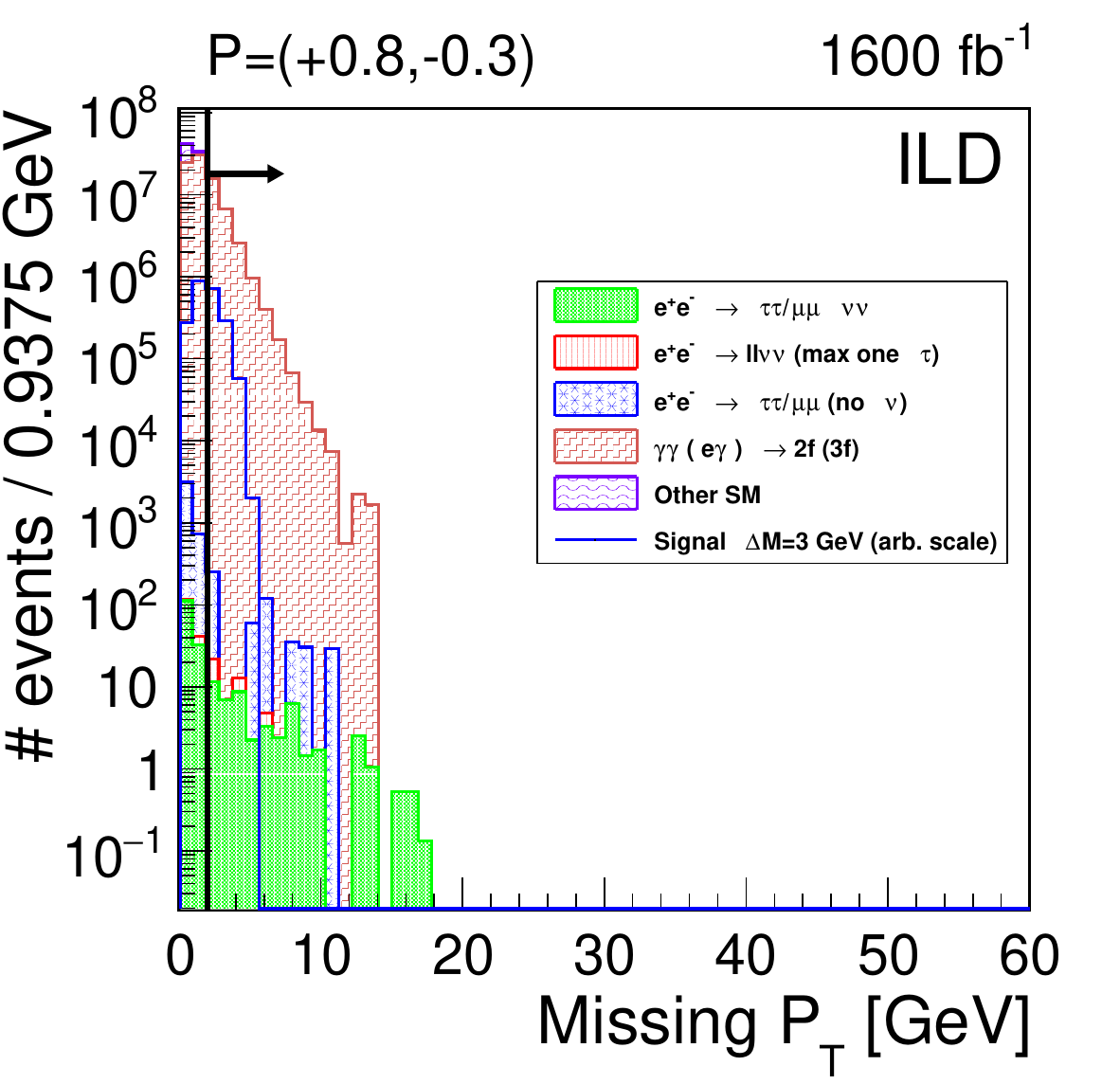}}
    \subcaptionbox{}{ \includegraphics [width=0.32\textwidth]{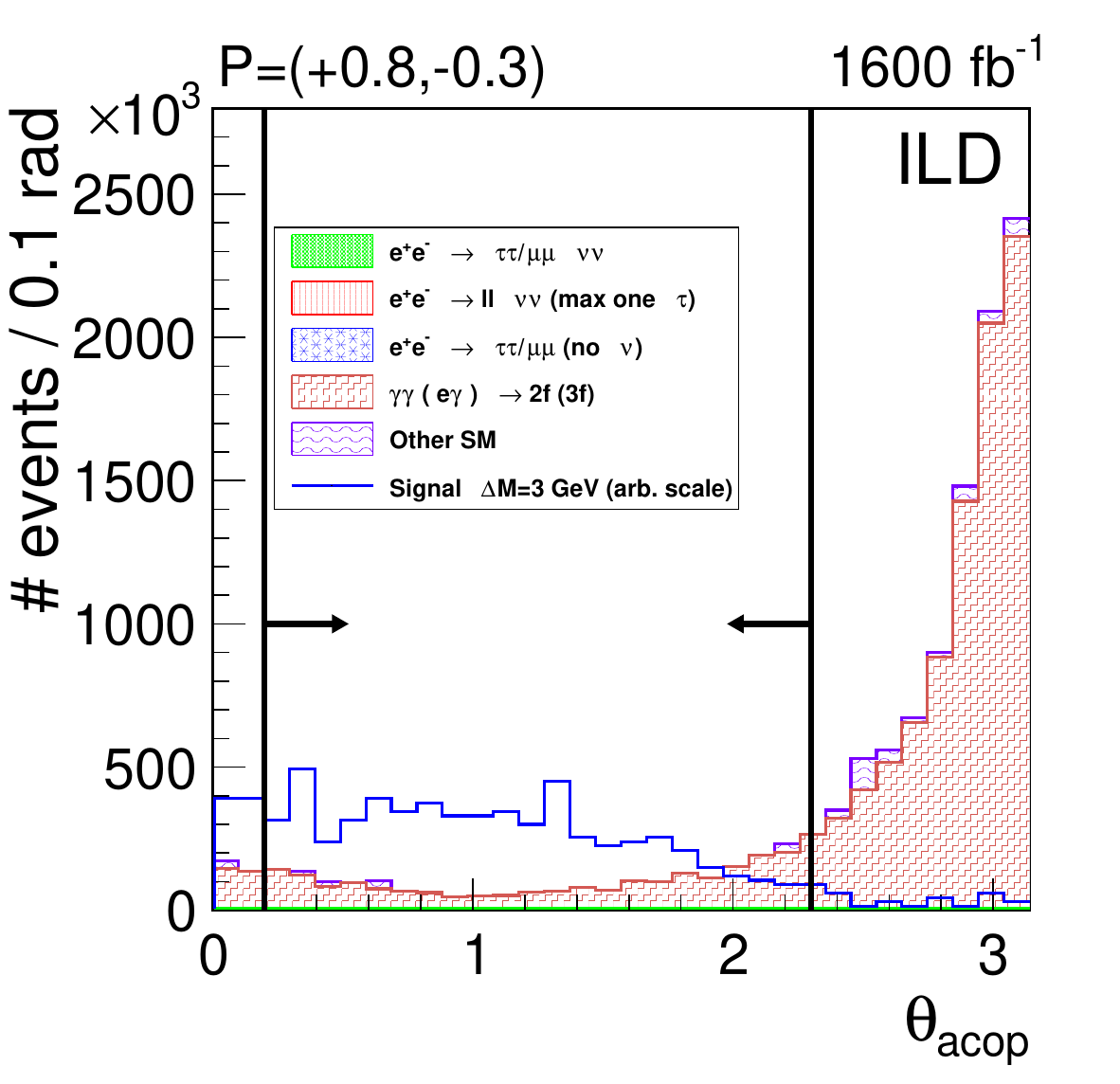}}
    \subcaptionbox{}{\includegraphics [width=0.32\textwidth]{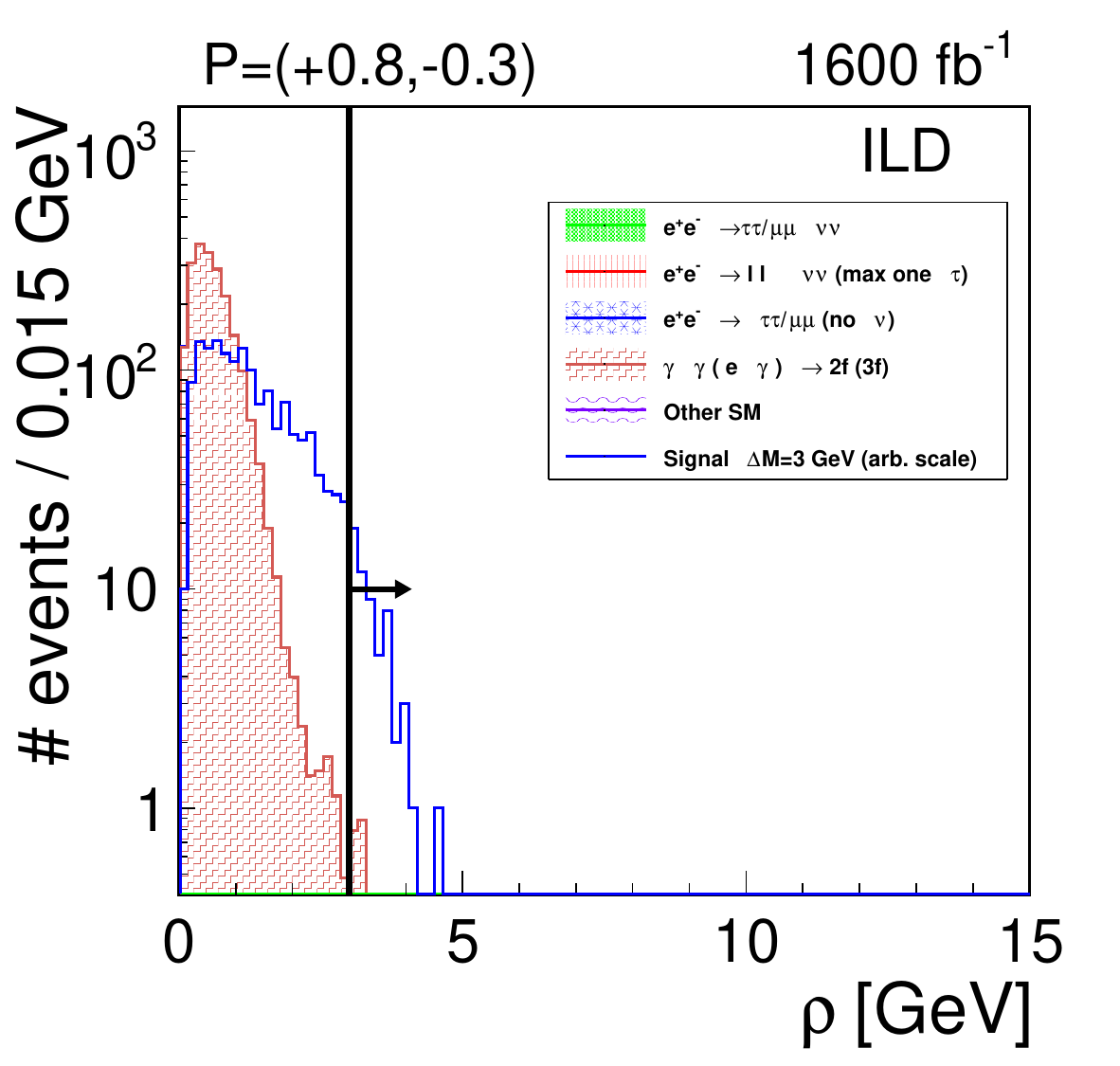}}
   \caption{
     Distributions for  M$_{\widetilde{\tau}}$ = 245\,GeV $\Delta M$ = 10\,GeV (first line) and  $\Delta M$ = 3\,GeV
     (second line) of 
     (a) and  (d): the missing transverse momentum;
     (b) and (e): the acoplanarity angle between the two jets;
     (c) and (f): the variable $\rho$, described in the text.
The signal is on arbitrary scale,
and all previous cuts have been applied (cf.\,Table~\ref{tab:cutsflow_dm10}).  The arrows indicate the region where events are accepted.}
    \label{fig:ptmiss_thetaacop_rho_dm10_dm3}
  \end{figure}

  For mass differences less than 11\,GeV, down to  the mass of the $\tau$ the kinematics of the signal events is very close to that of the $\gamma\gamma$ background
  events\footnote{For mass differences
    below the mass of the $\tau$ the lifetime of the $\widetilde{\tau}$ increases rapidly and
    the study has to be done based on a signature of long-lived particles travelling through the
    detector.}, which seriously impedes on the possibility of detecting any signal events.
  Down to $\Delta M$ = 3\,GeV, cuts on  $P_{T miss}$,  $\rho$ and  $\theta_{acop}$ are still sufficient, 
  as can be seen in Figs.~\ref{fig:ptmiss_thetaacop_rho_dm10_dm3} for
  the model points   with M$_{\widetilde{\tau}}$ = 245\,GeV and with $\Delta M$ = 10\,GeV and $\Delta M$ = 3\,GeV.
    However, it was found that the optimal cuts on these  quantities needs to be softened  compared to those applied
    at higher mass-differences, and to  depend on the mass
difference: For  $\Delta M$ from 10\,GeV down to 3\,GeV the cuts on $P_{T miss}$,
and  $\theta_{acop}$ are those given in Table~\ref{tab:cutsflow_dm10}.
For  $\Delta M$ from 10\,GeV down to  6\,GeV, the requirement on $\rho$
is that it should exceed 6 \,GeV.
At  $\Delta M$ =  5 \,GeV it should be above 5 \,GeV and at  $\Delta M$ =  4 \,GeV, above  4 \,GeV. 
For $\Delta M$ = 3\,GeV, the cuts are $P_{T miss} > $ 2\,GeV,
$\rho > $ 3\,GeV and $ \theta_{acop} \in [0.2 , 2.3 ]$.
For $\Delta M$
equal to 2\,GeV and lower, it was required that  $P_{T miss}$ > 0.1\,GeV,
and that  $ \theta_{acop} \in [0.2 , 3.1 ]$.
At this lowest mass difference, the  $\rho$ distribution does not
differ significantly between signal and background, so
no cut on this quantaty can be done.

\begin{table}[htbp]
    \centering
    \caption{Cut-flow table for the low $\Delta M$ case.
      The first column shows the cut definition, and the following columns shows the
      number of events passing all cuts up to, and including, the current row.
      The ``overlay-only cuts'' are described in detail in the text.
      The model point is  M$_{\widetilde{\tau}}$ = 245\,GeV $\Delta M$ = 10\,GeV, and the numbers correspond to an integrated luminosity of 1.6\,ab$^{-1}$ at 500\,GeV for 
    polarisation $\mathcal{P}_{+-}$.}
    \label{tab:cutsflow_dm10}
    \begin{tabular}{lS[table-format=4.1]S[table-format=4.1]S[table-format=4.1]S[table-format=4.1]S[table-format=4.1]S[table-format=4.1]}
      \hline\hline
      \addlinespace[2pt]
      Cut &\alc{c}{Signal}& \alc{c}{$e^+e^- \rightarrow $}     &\alc{c}{$e^+e^- \rightarrow $}  &\alc{c}{$e^+e^- \rightarrow $} &\alc{c}{$\gamma\gamma (\gamma e) \rightarrow $}&\alc{c}{All other}  \\[2pt]
          &               & \alc{c}{$\tau\tau/\mu\mu ~\nu\nu$} &\alc{c}{$ l l ~\nu\nu$}        & \alc{c}{$\tau\tau/\mu\mu $}   &\alc{c}{2(3) fermions}                         &                    \\[2pt]
          &               &                                    &\alc{c}{($\ge 1~l$ not $\tau$)}& \alc{c}{(no $\nu$'s)}         &                                               &                    \\[2pt]
     \hline
     \addlinespace[1pt]     
      no cut                             &  672.8   &  134100 &  533900 & \alc{l}{4.377 \tento{6}}   &\alc{l}{2197 \tento{6}}   & \alc{l}{2616 \tento{6}}\\
      BeamCal veto                       &  672.1  &  133966 &  532800 & \alc{l}{4.368 \tento{6}}    &\alc{l}{1933 \tento{6}}   & \alc{l}{2603 \tento{6}}\\
      2 $\le N_{charged} \le$ 10         & 664.3 &  106309 &  402048 & \alc{l}{3.547 \tento{6}}        &\alc{l}{910.4 \tento{6}}   &\alc{l}{1720 \tento{6}}\\
      $ q \cos{ \theta_{jet}} > -1$     & 585.2 &   70645 &  216728 & \alc{l}{2.671 \tento{6}}        &\alc{l}{800.6 \tento{6}}  &\alc{l}{1504 \tento{6}}\\
      $|M_{vis} - M_Z| > 4$             & 585.2 &   66806 &  207120 & \alc{l}{2.555 \tento{6}}        &\alc{l}{795.2 \tento{6}}  &\alc{l}{1500 \tento{6}} \\
      $E_{miss} >2 M_{LSP}$              & 585.1 &    6208 &   854.1 &                  122748         &\alc{l}{536.8 \tento{6}}  &\alc{l}{1042 \tento{6}}\\
      $M_{vis} < \sqrt{s} - 2 M_{LSP}$   & 583.9 &    5655 &   722.0 &                  111070         &\alc{l}{519.0 \tento{6}}  &\alc{l}{969.9 \tento{6}}\\
      $P_{jet} < P_{max}$                & 514.6 &   3823 &    350.7 &                   69078         &\alc{l}{339.0 \tento{6}}  &\alc{l}{715.8 \tento{6}}\\
      $|\cos(\theta_{\sum \bar{p}})|<0.85$ & 437.2 &  1201 &   168.3 &                    14529        &\alc{l}{135.6 \tento{6}}  &  \alc{l}{54.32 \tento{6}}    \\
      $\tau$ decay modes               & 184.4 &   194.2 &    28.2 &                    4204         &\alc{l}{82.71 \tento{6}}  & \alc{l}{20.06 \tento{6}}\\
      $P_{T miss}$ > 4\,GeV               & 157.8 &   86.3 &      7.0 &                    1931       &\alc{l}{8.901 \tento{6}} &   427494     \\
      $\theta_{acop} \in$  [0.2,2.0]    & 136.2 &   43.7 &       3.9 &                  823.6        &       121625             &          118136  \\
      $\rho$ >  6 \,GeV                        &  34.8 &    3.7 &     0.6 &                     0.0         &         241.5            &         0.2 \\
      overlay-only cuts              &  20.1 &    1.0 &     0.2 &                     0.0         &         121.0            &          0.0 \\
     \addlinespace[2pt]
     \hline
      \addlinespace[2pt]

    \end{tabular}
  \end{table}
\begin{table}[htbp]
    \centering
    \caption{Final counts for the small $\Delta M$ case for all polarisation cases,
      with integrated luminosities as defined by the standard ILC running scenario. The model point is M$_{\widetilde{\tau}}$ = 245\,GeV $\Delta M$ = 10\,GeV.
For reference, the counts
      assuming unpolarised beams is also given.}

    \label{tab:finalcounts_pols_dm10}
    \begin{tabular}{lcS[table-format=4.1]S[table-format=4.1]S[table-format=4.1]S[table-format=4.1]S[table-format=4.1]S[table-format=4.1]}
      \hline\hline
      \addlinespace[2pt]
      Beam&Integrated &\alc{c}{Signal}& \alc{c}{$e^+e^- \rightarrow $}     &\alc{c}{$e^+e^- \rightarrow $}  &\alc{c}{$e^+e^- \rightarrow $} &\alc{c}{$\gamma\gamma (\gamma e) \rightarrow $}&\alc{c}{All other}  \\[2pt]
      polarisation    &Luminosity&               & \alc{c}{$\tau\tau/\mu\mu ~\nu\nu$} &\alc{c}{$ l l ~\nu\nu$}        & \alc{c}{$\tau\tau/\mu\mu $}   &\alc{c}{2(3) fermions}                         &                    \\[2pt]
          &   [ab$^{-1}$]            &                                    &\alc{c}{($\ge 1~l$ not $\tau$)}& \alc{c}{(no $\nu$'s)}         &                                               &                    \\[2pt]
      \hline
      \addlinespace[1pt] 
     \hline
      \addlinespace[1pt] 
      $\mathcal{P}_{+-}$  &1.6& 20.1   &    1.0   &       0.2  &                     0.0                &    121.0                        &         0.0 \\
      $\mathcal{P}_{-+}$  &1.6& 13.6  &    17.0   &       1.0 &                      0.0                &     65.0                   &         0.1  \\
      $\mathcal{P}_{++}$  &0.4&  3.0  &     0.5    &      0.0 &                     0.0                 &     17.43                   &         0.0  \\
      $\mathcal{P}_{--}$  &0.4&  2.2  &     2.3   &       0.0 &                     0.0                 &     11.02                    &         0.0  \\
     \hline
      Unpolarised        &4.0&  34.0   &   18.2 &         1.0 &                     0.0                &    214.9                     &         0.1   \\
      \hline
      \addlinespace[2pt]
      \addlinespace[2pt]
    \end{tabular}
  \end{table}

The $\gamma\gamma$ events still remaining at this stage are of two distinct types.
First, there are 
the ``overlay-only'' events, i.e.\ bunch crossings that
only contain one or several  low-$\it{P_{T}}$-hadron events discussed in Sec.~\ref{sec:ILC}.
Secondly, there are the  multi-peripheral events yielding $\tau$ pairs, i.e.\ the events of the
type illustrated by the diagram in Fig.~\ref{fig:redbckdiagrams}(b).

\begin{figure}[htbp]
    \centering
    \subcaptionbox{}{\includegraphics [width=0.4\textwidth]{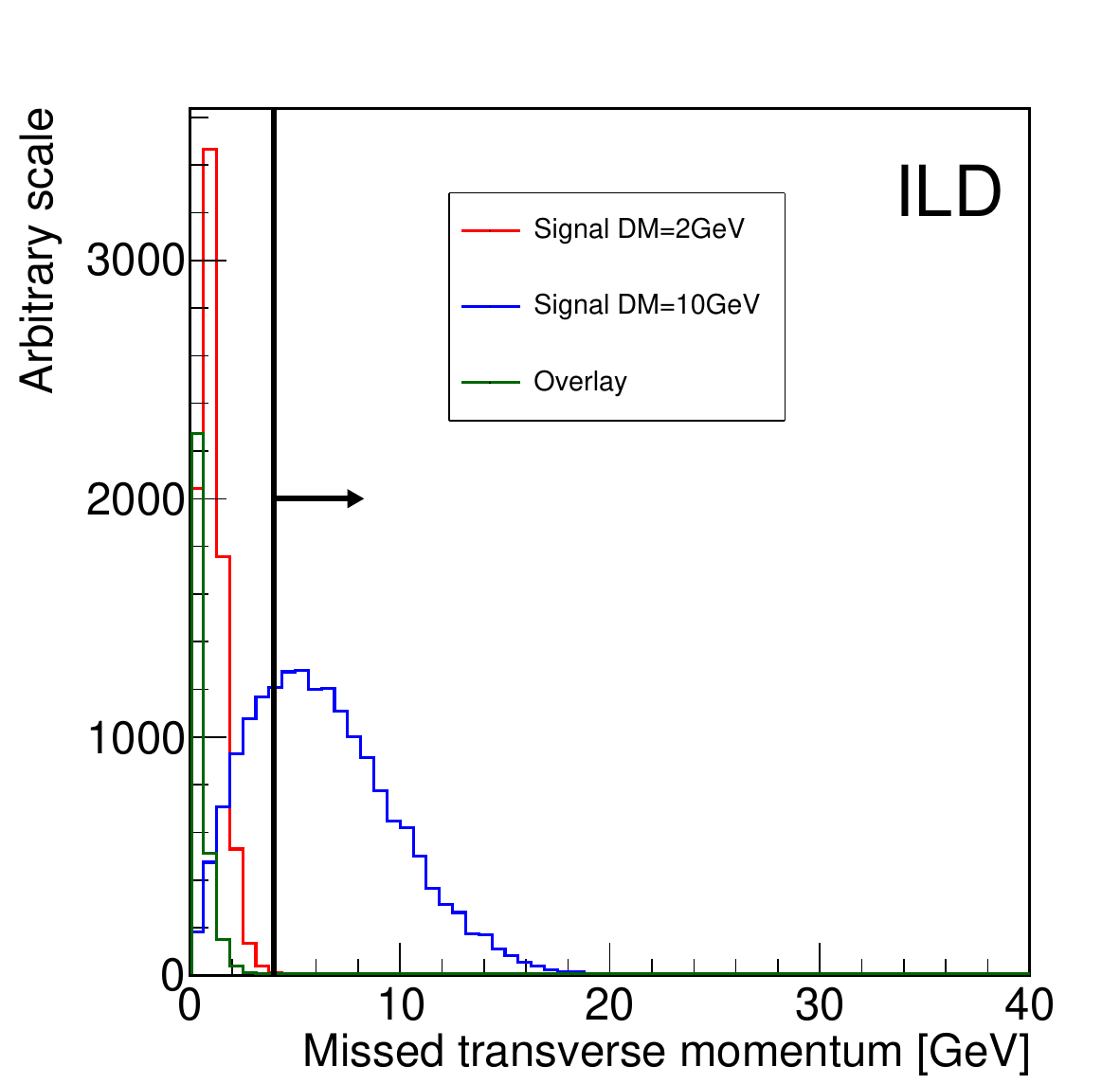}}
    \subcaptionbox{}{\includegraphics [width=0.4\textwidth]{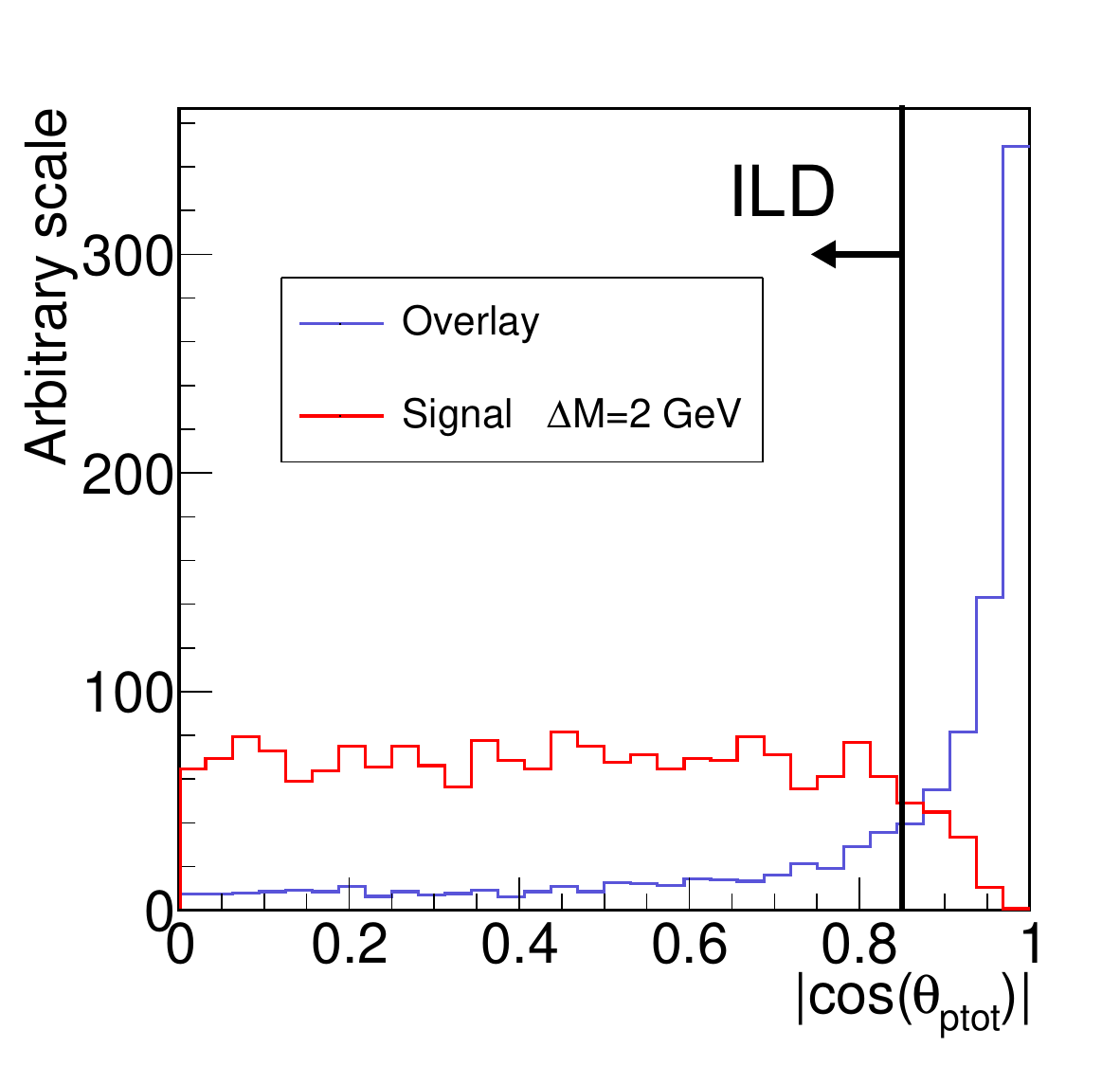}}
    \caption{(a) : Transverse momentum distribution for tracks in signal and overlay-only events. The arrow indicates the region where the events are accepted for the $\Delta M$ = 10\,GeV. There is no cut for $\Delta M$ = 2\,GeV.
      (b) Distributions of the polar angle of the total seen momenta for signal and overlay-only events.
      The arrow indicates the region for accepted events.}
    \label{fig:overlonly_missedpt_costhetaptot}
  \end{figure}

  The rejection factors
  against the ``overlay-only''  events,
  obtained applying the cuts described so far
for mass differences 2 and 10\,GeV are $2.6 \tento{-3}$ and $< 2.7 \tento{-6}$, respectively.
In both cases only $\gamma\gamma$ to low-$\it{P_{T}}$-hadron events survive the cuts, and in fact, no events
  of any type survives in the  $\Delta M$ = 10\,GeV  case (hence the rejection factor is only an upper limit in this case).
The large difference between the factor for both mass differences can be explained by the cut in the missing transverse 
momentum, as one can see in Fig.~\ref{fig:overlonly_missedpt_costhetaptot}(a).
Cuts affecting multiplicity and ${\tau}$-identification are similar in both cases.
For the case of $\Delta M$ = 2\,GeV an important contribution to the rejection comes from the cut in $|\cos(\theta_{\sum \bar{p}})|$,
as shown in Fig.~\ref{fig:overlonly_missedpt_costhetaptot}(b).
 
While the cuts are strong enough to only leave a handful of the simulated overlay-only events,
in both cases the rejection factors are not sufficient to reduce the overlay-only  events to an acceptable level,
for which a reduction factor of $\sim 10^{-9}$ is needed, see Sec.~\ref{sec:ILC}.
It is thus necessary to search
for additional cuts.
Moreover,
since the available Monte-Carlo statistics of overlay-only events has been
exhausted after applying the cuts already made,
a detailed optimisation of the cuts is  not possible.
Instead,
the approach was taken to determine sets of independent cuts,
and calculating the reduction factor for each of them separately.
The total factor  can be calculated by multiplying the rejection factors obtained
by applying the cuts individually.
The aim is to obtain a factor large enough that one can make certain that the
overlay-only background will be significantly lower than that from other sources,
and hence will not influence the final result.
The details of the procedure and verifications are given in Appendix~\ref{appendix:lowdm}.

Since no overlay-only events passed the cuts for mass-differences above 2\,GeV,
we first sub-divide the cuts in two groups, that are expected to be 
independent: on one hand the cuts on $\rho$ and $P_{T miss}$, on the other hand
the rest of the cuts.
By multiplying the reduction factors from each of these two groups of cuts,
we obtain a non-zero value. 

The first additional independent cut is 
based on vertex properties:
In the overlay-only events, the tracks are produced in $\rho-\rho$ scattering, and hence
originate at the main vertex. The tracks in the signal, in contrast, originate from the
$\tau$ decays, and are {\it not} from the main vertex, and in addition
in most cases there is only one charged track detected from each $\tau$ decay.
These properties are not used in any other cuts,
and cuts on them are very likely to be independent of the other cuts.
This cut applied alone reduces the overlay-only background by a factor $1.9 \tento{-2}$.

A second added cut
required to detect a photon at high angles to the beam. 
The low-$\it{P_{T}}$ hadrons rarely contain
photons at high angles, while the  $\widetilde{\tau}$ events quite often do, as discussed in
Sections~\ref{sec:ILC} and~\ref{sec:thestau}, respectively.
Hence requiring to detect a photon at high angles
further enhances the $\tau$ identification.
We obtain good performance of the photon cut 
down to the lowest $\Delta M$.
The reduction-factor of overlay-only background of this cut alone is $1.7\tento{-4}$ .
As explained in Appendix~\ref{appendix:lowdm},
the cut is also efficient in reducing the 
multi-peripheral $\gamma\gamma$ background to an acceptable level.
However, to achieve this reduction 
at $\Delta M$ = 2\,GeV, the requirement needs to be supplemented
by a combined cut on  $P_{T miss}$ and the absolute value of the momentum of the jet system,
namely that
$P_{T miss} > |\sum \bar{p}_{jet}|$.

For $\Delta M$ between 10 and 3\,GeV, the reduction factor obtained by
multiplying this factor with the one obtained by the subset of 
standard cuts 
(i.e.\ all cuts except
those on $P_{T miss}$ and  $\rho$), $5.0\tento{-3}$, 
reduces the overlay-only events by a factor of $8.5\tento{-7}$, and when multiplied
with the factor obtained from the  
vertex-condition, $1.9\tento{-2}$, 
it yields a reduction factor of $1.6\tento{-8 }$.
The final factor is then obtained by multiplying by the factor from the  $P_{T miss}$ and  $\rho$ cuts alone, $2\tento{-5}$,
to yield a final factor of  $2.7\tento{-13}$.
For the model point with $\Delta M$ = 2\,GeV, the standard cuts combined with the photon requirement 
and the vertex condition reduces the overlay-only events by
a factor of $8.4\tento{-9}$. When multiplied by the factor from the further combined $P_{T miss}$ and $|\sum \bar{p}_{jet}|$ cut,
the final reduction is $1.0\tento{-12}$.

To conclude,
we can assert that the overlay-only background can be reduced by factors well below the required 
level of $10^{-9}$ by these additional cuts.
In fact, at none of the  model-points studied, more than two such events are expected
in the full 4 ab$^{-1}$ sample, so that they will not influence the
expected limits.
The cuts are necessary to keep the overlay-only background under control,
but we note that when they are applied to the sample {\it without} overlay-only background,
both signal and background are reduced by approximately the same factor (see last two lines of 
Table~\ref{tab:cutsflow_dm10}),
so that the significance becomes lower when applying the cuts.
This indicates that it is probable that an analysis with computing resources large enough
to simulate $\mathcal{O}(10^{10})$ overlay-only events could further optimise the cuts.

\section{Calculating exclusion/discovery limits and determining the worst case\label{sec:limitcalc}}
  
\begin{figure}[htbp]
    \centering
     \subcaptionbox{}{\includegraphics [width=0.45\textwidth]{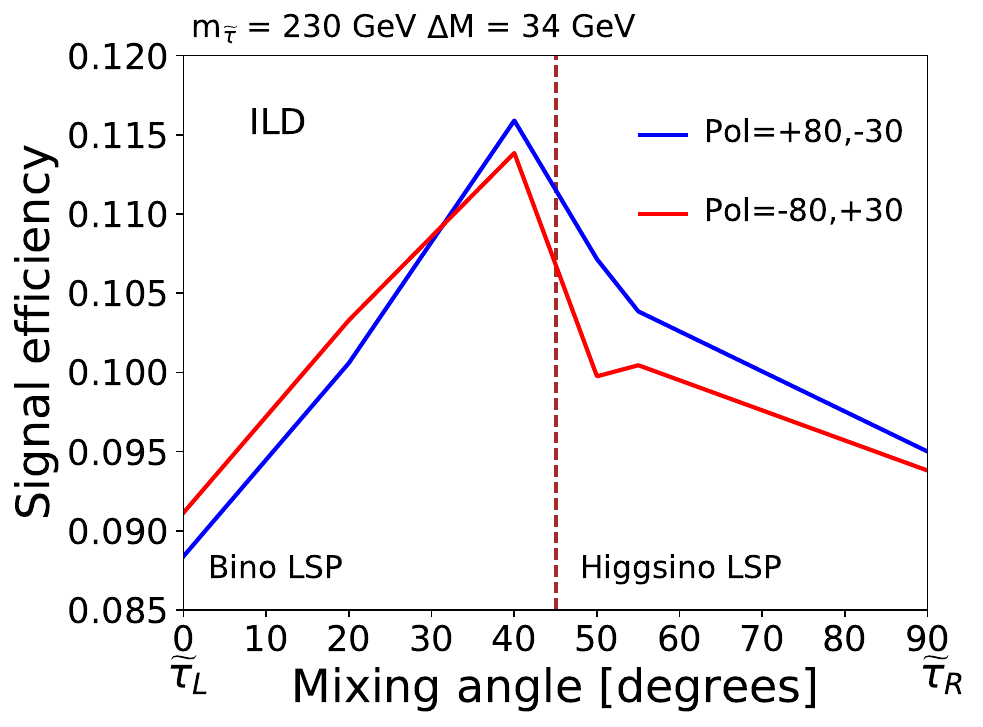}}
     \subcaptionbox{}{\includegraphics [width=0.45\textwidth]{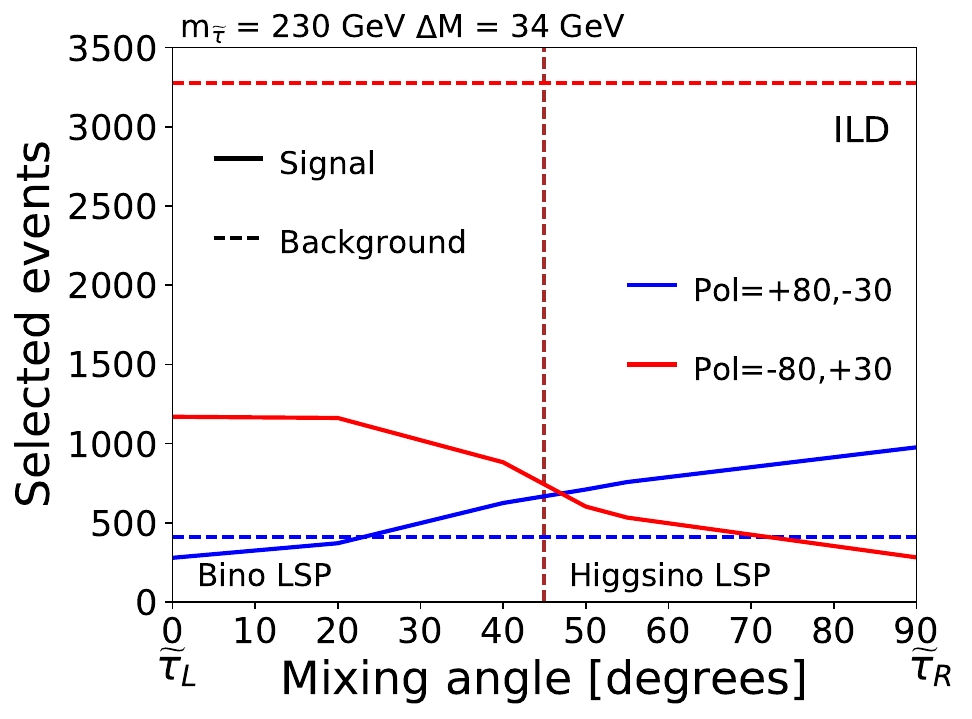}}
    \caption{(a) Signal efficiency as a function of the $\widetilde{\tau}$ mixing angle for both main ILC polarisations.
      The LSP composition was chosen giving the worst case for each mixing.
       (b) Number of signal and background selected events as a function of the $\widetilde{\tau}$ mixing angle for both main ILC polarisations. }
    \label{fig:signal_eff_events_mixings}
  \end{figure}

  Final significance values were computed by combining the contribution of
  all polarisations.
  The signal cross section depends on the beam polarisation.
  Even more so, the background levels are expected
  to be quite different for the different cases,
  mainly because the strongly polarisation dependent $e^+e^- \rightarrow W^+W^-$ process is one of the most
  important sources of background.
  Because of this the samples are 
  weighted by the likelihood ratio test statistic~\cite{Read:2000ru,Neyman:1933wgr}.
  In this study a pure event-counting is performed, thus the likelihood ratio statistic has the simple form
  \begin{equation}
    N_{\sigma} = \frac{\sum^{n_{samp}}_{i=1} s_i \ln{(1+s_i/b_i)}}{
      \sqrt{ \sum^{n_{samp}}_{i=1} n_i \left [\ln{(1+s_i/b_i)} \right]^2}}
    \label{eq:LR}
  \end{equation}
  where $s_i$ and $b_i$ are expected number of signal and background events in each sample,
  and $N_{\sigma}$ is the significance of the test, expressed as the equivalent number of standard deviations in
  the Gaussian limit.
 Depending on which hypothesis is tested (exclusion or discovery), $n_i$ is either $s_i +b_i$ (exclusion),
  or $b_i$ (discovery).
  Table~\ref{tab:polarisations} shows the number of signal and background events for a specific model point for
  the four ILC running polarisations and for unpolarised beams.
  The difference in the number of signal events arises from the dependence of the
  cross section on the polarisation, as well as from the effect on selection efficiency due to the 
  polarisation of the $\tau$ coming from the $\widetilde{\tau}$, see Fig.~\ref{fig:signal_eff_events_mixings}.
  The dependence of the
  cross section on the polarisation is the main factor for the difference in $WW$ events, $e^+e^- \rightarrow\tau\nu\tau\nu$.
  One can see that the signal-to-background ratio is clearly enhanced for the case of mainly right-handed electrons,
  left-handed positrons.

  Taking the definition of expected exclusion at 95$\%$ CL$_s$ as $N_\sigma > 2$ (cf.\ Eq.~\ref{eq:LR})\footnote{Note that Eq.~\ref{eq:LR} trivially states that $N_\sigma = S/\sqrt{S+B}$ (or $N_\sigma = S/\sqrt{B}$) in the case that
    there is only one sample.},
  it can be seen that combining the four polarisation samples with the likelihood ratio statistic does strengthen the limit,
  even though the $\mathcal{P}_{-+}$ and, even more so, the $\mathcal{P}_{++}$ and  $\mathcal{P}_{--}$ samples give much weaker limits than the  $\mathcal{P}_{+-}$ one, cf.\ Table~\ref{tab:polarisations}.

\begin{figure}[htbp]
    \centering
    \includegraphics [width=0.45\textwidth]{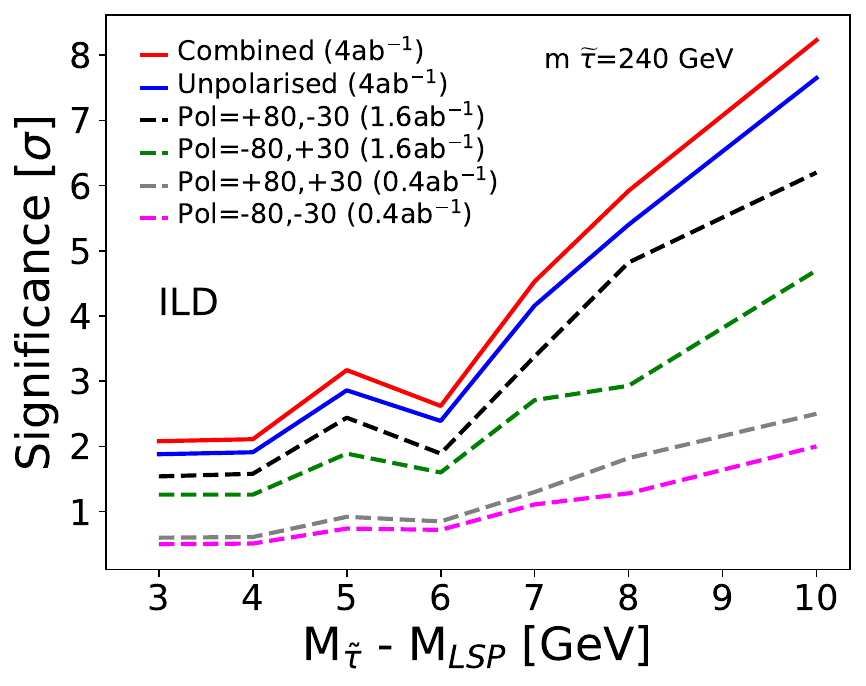}
    \caption{Significance for exclusion as a function of the mass difference for unpolarised beams, the different standard ILC polarisation settings and their combination using the likelihood ratio statistic.
    }
    \label{fig:sigma_different_polarisations}
  \end{figure}

  It is also shown that unpolarised beams would allow neither exclusion nor discovery, even though the sample corresponds to
  the same integrated luminosity as the sum of the polarised samples.
  This is partly because the ILC has {\it both} beams polarised, meaning that {\it more
    than half} of the collisions in the $\mathcal{P}_{-+}$ and $\mathcal{P}_{+-}$ samples
  are between opposite polarised particles, which is necessary to allow for s-channel processes,
  such as $\widetilde{\tau}$ pair production,
  but also because of the possibility to combine samples with different beam polarisations in an optimal way
  is not available for unpolarised beams\footnote{Polarisation is not only important in the enhancement of the signal over background but also
  plays an important role in the parameter determination~\cite{Bechtle:2009em,Berggren:2015ar}.}.
Figure~\ref{fig:sigma_different_polarisations} shows the significances for the
exclusion hypothesis in Table~\ref{tab:polarisations} together with the
  ones corresponding to different mass differences.

    \begin{table}[htbp]
    \centering
    \caption{
      Remaining signal and background events after the application of the selection cuts for
      $M_{\widetilde{\tau}}$=240\,GeV and mass difference with the LSP of 4\,GeV. }
    \label{tab:polarisations}

    \begin{tabular}{lS[table-format=4.1]S[table-format=4.1]S[table-format=4.1]S[table-format=4.1]}
      \hline\hline
      \addlinespace[2pt]
      Polarisation &\alc{c}{Signal}& \alc{c}{$e^+e^- \rightarrow $}     &\alc{c}{$\gamma\gamma (\gamma e) \rightarrow $} & \alc{c}{N$_\sigma$}\\[2pt]
                   &               & \alc{c}{$\tau\tau/\mu\mu ~\nu\nu$} &\alc{c}{2(3) fermions}   & \alc{c}{(Exclusion)}\\[2pt]
      \hline
      \addlinespace[1pt]
      $\mathcal{P}_{+-}$   &  39.3 &    0.4 & 574.0 & 1.6 \\
      $\mathcal{P}_{-+}$   &  24.9 &    7.5 & 360.5 & 1.3\\
      $\mathcal{P}_{++}$   &   5.7 &    0.2 &  83.8 & 0.61 \\
      $\mathcal{P}_{--}$   &   4.1 &    1.0 &  59.5 & 0.51\\
      Combined            & \alc{c}{-}   & \alc{c}{-}  & \alc{c}{-}  &2.17\\
      Unpolarised        &  64.7 & 8.0  &  1078.0 & 1.9\\
     \hline
      \addlinespace[2pt]
    \end{tabular}
  \end{table}

 \begin{figure}[htbp]
    \centering
    \subcaptionbox{}{\includegraphics [width=0.45\textwidth]{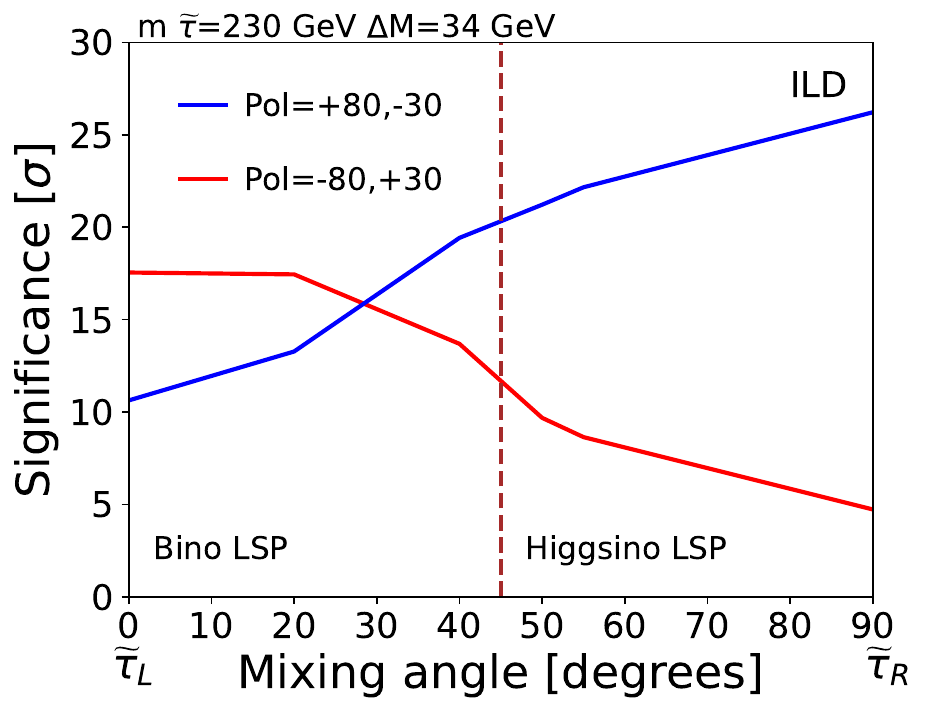}}
     \subcaptionbox{}{ \includegraphics [width=0.45\textwidth]{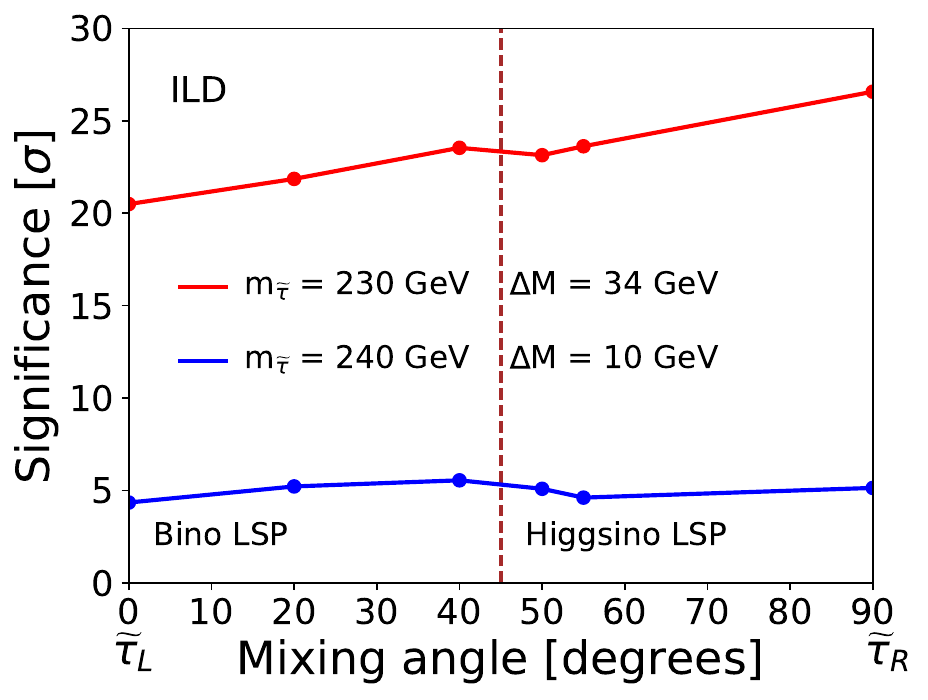}}
   \caption{(a) Signal exclusion significance as a function of the $\widetilde{\tau}$ mixing angle for both main ILC polarisations.
     (b) Combined signal exclusion significance of all polarisation settings of the ILC standard scenario using the likelihood ratio statistic.
  }
    \label{fig:sigmas_mixings_weighted}
  \end{figure}

  The significance for exclusion as a function of the mixing angle was computed for
each polarisation, as shown in Fig.~\ref{fig:sigmas_mixings_weighted}(a). 
Final significance values were computed adding the contribution of all
polarisations weighted by the likelihood ratio statistic.
The results are plotted in
Fig.~\ref{fig:sigmas_mixings_weighted}(b). 
One can see that rather uniform sensitivity to all mixing angles is obtained, and that for the 
smallest mass differences - the ones closest to the critical region - a mixing angle around 53$^\circ$ can indeed 
be considered as the
worst one,  and validates our choice of this mixing angle being the worst case
\footnote{
  Note that the estimation of the worst scenario, illustrated in
  Figs.~\ref{fig:signal_eff_events_mixings} and ~\ref{fig:sigmas_mixings_weighted}, was not done with the final cuts used in
the analysis. The final cuts were further optimised taking into account the whole parameter
 space. This does not affect the conclusion of this section.}.

\section{Exclusion/discovery reach\label{sec:limits}}

\subsection{Reach of ILC at 500\,GeV and existing limits\label{sec:ilclimits}}

\begin{figure}[htbp]
    \centering
     \subcaptionbox{}{\includegraphics [width=0.45\textwidth]{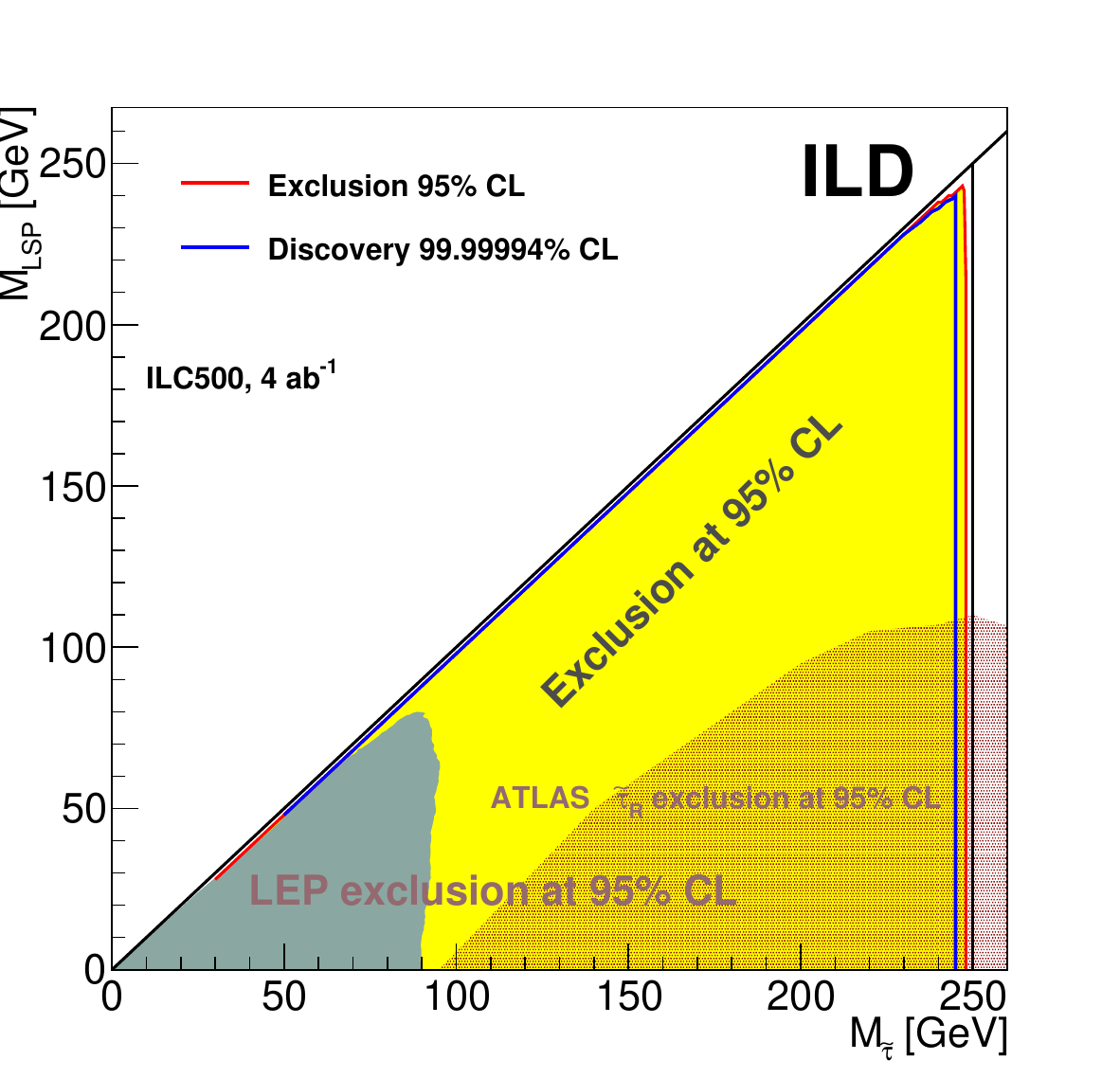}}
     \subcaptionbox{}{\includegraphics [width=0.45\textwidth]{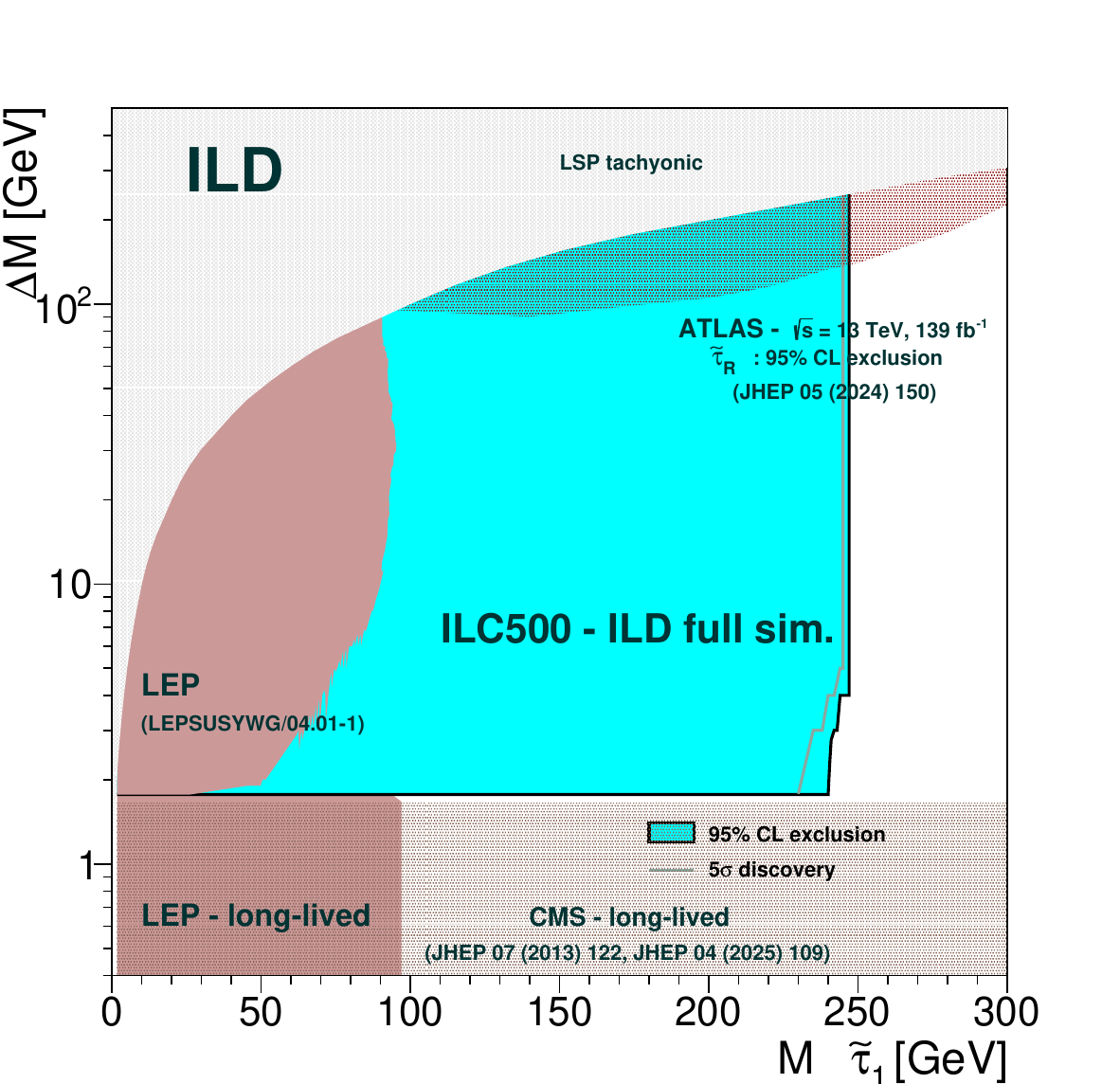}}
   \caption{
      Exclusion and discovery $\widetilde{\tau}$ limits from the current studies compared to the ones from LEP studies~\cite{LEPSUSYWG/04-01.1} and from ATLAS at LHC~\cite{ATLAS:2024fub}.
      The ILC region corresponds to the full standard data-set at $E_{CMS}$ = 500\,GeV, i.e.\ 1.6\,ab$^{-1}$ at 
each of the beam polarisations
$\mathcal{P}_{+-}$ and  $\mathcal{P}_{-+}$, 0.4\,ab$^{-1}$ at
$\mathcal{P}_{++}$ and  $\mathcal{P}_{--}$.
(a) In the $M_{\widetilde{\tau}}$-$M_{LSP}$ plane; (b)
in the $M_{\widetilde{\tau}}$-$\Delta  M$ plane.
In (b) also the region with mass differences below the mass of the ${\tau}$ is shown with the results from
LEP ~\cite{LEPSUSYWG/02-05.1,LEPSUSYWG/02-09.2} and from CMS at LHC~\cite{CMS:2013czn,CMS:2024nhn}.
}
    \label{exclusion_mstau_1}
  \end{figure}

  The exclusion and discovery reach extracted from this study is illustrated in Fig.~\ref{exclusion_mstau_1}.
In Fig.~\ref{exclusion_mstau_1}(a), the result is shown in the  $M_{\widetilde{\tau}}$-$M_{LSP}$ plane,
and in Fig.~\ref{exclusion_mstau_1}(b), in the  $M_{\widetilde{\tau}}$-$\Delta  M$ plane.
    The limits assume the lightest $\widetilde{\tau}$ to be the NLSP and the lightest neutralino the LSP,
    and are valid for any  $\widetilde{\tau}$ mixing angle.
  It is also relevant to compare these results with the current $\widetilde{\tau}$ limits coming from
  LEP and LHC. The LEP limits illustrated are also valid for any value the
  model parameters not explicitly shown.
 The limit from LHC shown are obtained by the ATLAS collaboration~\cite{ATLAS:2024fub},
  Since LHC limits are highly model-dependent, the comparison in
  this case have to be taken with care: here limits considering only the $\widetilde{\tau}_{R}$-pair production
  are shown, since, while still being optimistic, they are closest to the ones expected for the lightest
  $\widetilde{\tau}$ at minimal cross section.
  It should also be noted that the LHC limit only is an exclusion limit - no discovery potential is expected.

  \begin{figure}[htbp]
    \centering
     \subcaptionbox{}{\vspace*{0.65cm}\includegraphics [width=0.45\textwidth]{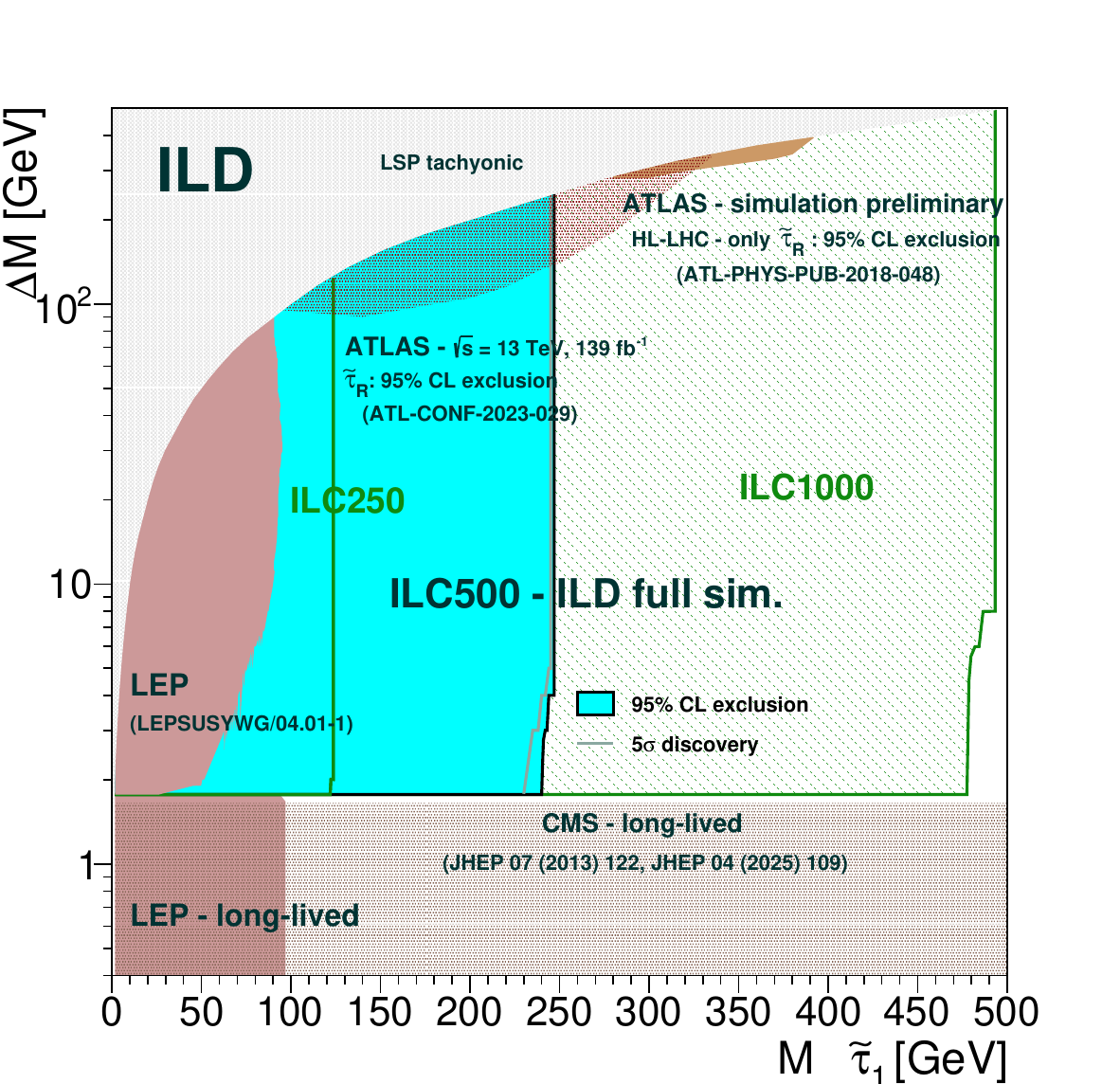}}
     \subcaptionbox{}{\includegraphics [width=0.5\textwidth]{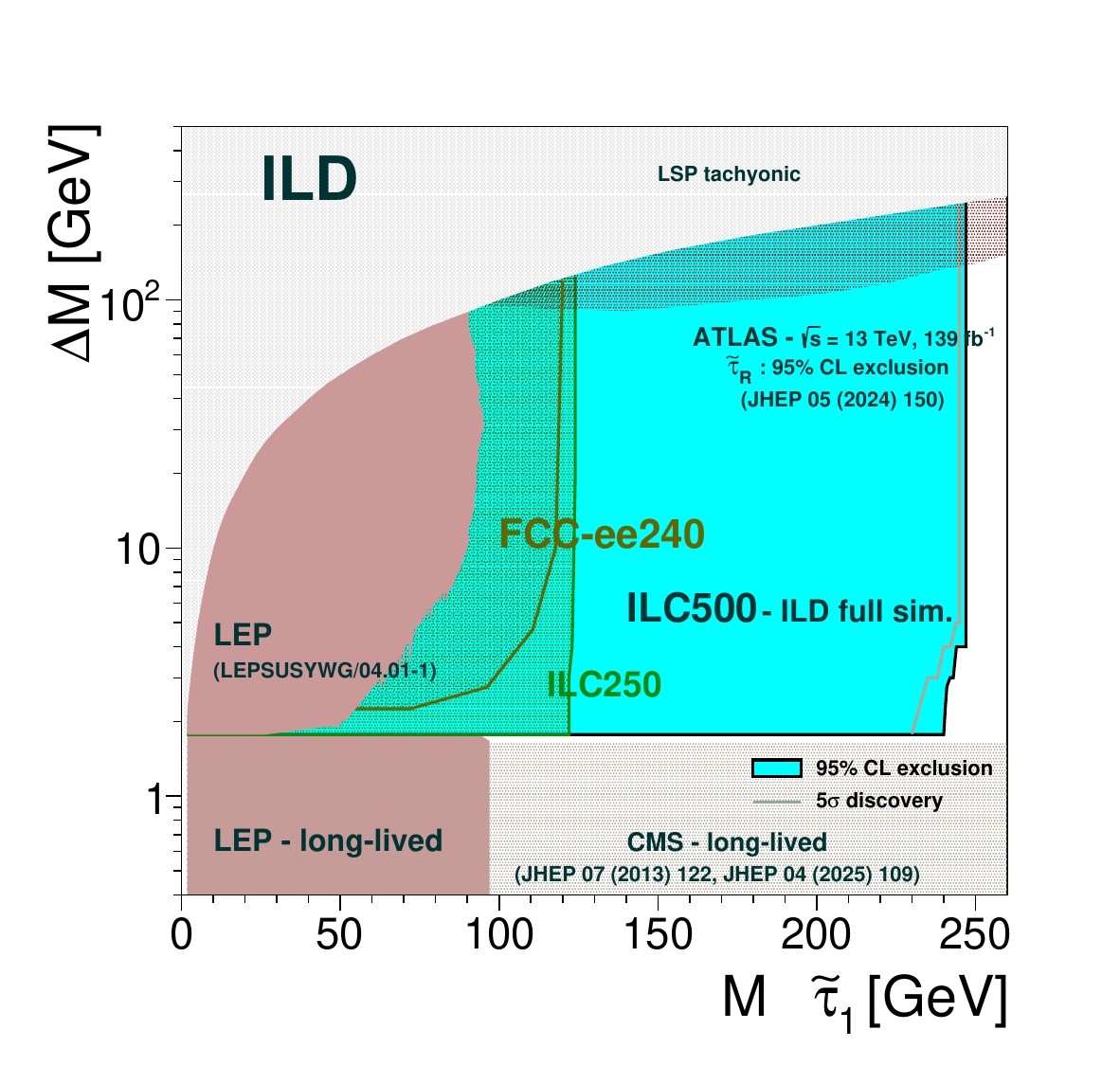}}
   \caption{
     $\widetilde{\tau}$ limits in the $M_{\widetilde{\tau}}$-$\Delta  M$ plane. 
  (a) ILC results from the current studies together with limits
  from LEP~\cite{LEPSUSYWG/04-01.1}, ATLAS at LHC~\cite{ATLAS:2024fub} 
  and projection for ATLAS at HL-LHC~\cite{ATLAS:2018diz}. The ILC region corresponds to the full standard data-set
      at $E_{CMS}$ = 500\,GeV, i.e.\ 1.6\,ab$^{-1}$ at each of the beam polarisations $\mathcal{P}_{+-}$ and  $\mathcal{P}_{-+}$,
 0.4\,ab$^{-1}$ at
 $\mathcal{P}_{++}$ and  $\mathcal{P}_{--}$.
       The shown LHC results and HL-LHC projection, only exclusion limits without discovery potential, consider the $\widetilde{\tau}_{R}$-pair production,
      which is expected to be the closest to the lightest $\widetilde{\tau}$ at minimal cross section. 
      The region with mass differences below the mass of the ${\tau}$ is also shown with results from LEP ~\cite{LEPSUSYWG/02-05.1,LEPSUSYWG/02-09.2} and from CMS at LHC~\cite{CMS:2013czn,CMS:2024nhn}.
      even if it is not covered by this study. 
      The extrapolation of
      the ILC current results to the ILC 250\,GeV and 1\,TeV running scenarios is also shown.
       (b) A zoom in of (a) with, in addition, the recast to FCCee at 240\,GeV, assuming combining results
        from four experiments, i.e.\ an integrated luminosity of 10.8\,ab$^{-1}$ and a systematic
          uncertainty of the $\gamma\gamma$ background of 15 \permil .
 }
    \label{exclusion_mstau_recasts}
  \end{figure}

The exclusion 
  region for mass differences below the mass of the $\tau$ is not included in the current study.
  For this region 
  results from LEP and from CMS at LHC~\cite{CMS:2013czn,CMS:2024nhn} are show,
  The LEP studies cover not only the region where the $\widetilde{\tau}_{1}$ travels through the
  detector without decaying but also the region with decays at a certain distance from the production
  vertex. In those regions acoplanar leptons, tracks with large impact parameters and kinked tracks
  are looked for, depending on the $\widetilde{\tau}_{1}$ lifetime~\cite{LEPSUSYWG/02-05.1,LEPSUSYWG/02-09.2}.
Studies of searches for other signals with these characteristics done within the ILD collaboration~\cite{Nakajima:2024hdm,Sasikumar:2020qxa}
indicate that there would be no issue to exclude or discover a $\widetilde{\tau}_{1}$ in this region up to quite close to
the kinematic limit.
  
\subsection{Extrapolations to other energies and other e$^+$e$^-$ colliders\label{sec:extrapolatedlimits}}
To extrapolate an analysis performed at one centre-of-mass energy to other energies,
an estimate of how background and signal are expected to change with energy is required.
This can be achieved as follows: 
For the background, the total measured energy scales up or down
    linearly with $E_{CM}$.
    Away from resonances, the angular distributions do not change with
    $E_{CM}$, so also components of momenta
    (e.g. transverse momenta)
    scales linearly with $E_{CM}$.
For the signal, the highest possible $P_T$ of any visible decay 
products of the  $\widetilde{\tau}$  is $ P_{max}$ (Eq.~\ref{eq:pmax}).
So, if one scales both $M_{\widetilde{\tau}}$  and $M_{LSP}$ by $E_{beam}$, both brackets in the equation remain
unchanged, so that $ P_{T~max}$ scales with $E_{beam}$, just like the background.
    The conclusion is that one expects S/B at one modified energy, $E_{CM~recast}$, to be the same as that at a
reference energy,
    $E_{CM~ref}$, if one scales the kinematic cuts and the SUSY masses with the
    ratio of the two, $E_{CM~recast}/E_{CM~ref}=R_{CM}$,
    while leaving angular cuts and cuts on decay topology unchanged.
This argument would be expected to hold well in our case, as long as the
the mass-difference is not too small, $\gtrsim$ 5\,GeV. 
At lower differences, 
details of the beam conditions, beam-spot size and  detector performance become more important,
as can be understood from the discussion in Sec.~\ref{sec:lowdmcuts}.
These words of caution are particularly relevant if the recast is done to
a {\it different} machine, with a {\it different} detector.  
    For the {\it significance}, S/$\sqrt{\mathrm{B}}$, at constant S/B, one notes that both 
background and signal
    production cross sections scales as 1/$s$ = 1/$E^2_{CM}$.
Since S/B remains unchanged if the scaling above is applied, the significance will scale 
as 1/$R_{CM}$, i.e.\ with the inverse of the ratio of the two centre-of-mass energies.
This is so also when threshold-effects are included, since the cross section  for pair production
scales with the ratio of the momentum to the energy of the particle
in the lab-frame, i.e. with $\beta$ to a power $n$
(n=1 for fermion pair-production, and n=3 for scalar pair-production),
and $\beta$ will be the same when $M_{\widetilde{\tau}}$  is scaled by $E_{beam}$,
as we suggest to do.
Thus, if the same amount of data is collected at two different beam-energies, 
the significance for a $\widetilde{\tau}$ signal with both $M_{\widetilde{\tau}}$  and $M_{LSP}$ as well as any kinematic cut scaled by $R_{CM}$, would change as  $1/R_{CM}$.
Furthermore, at linear  colliders, the luminosity increases linearly with $E_{CM}$, so one would expect that
the significance would acquire another factor $\sqrt{R_{CM}}$,
if the same amount of {\it time} is spent at the two energies.
Taken together, the total effect, at linear colliders, is expected to be that the 
significance of a signal at one energy should be equal to $1/\sqrt{R_{CM}}$ times that at the reference
energy, but at SUSY masses scaled by $R_{CM}$.
The detailed simulation study performed with ILD at ILC500 thus allows to estimate the prospects for ILD at
other ILC centre-of-mass energies.
Figure~\ref{exclusion_mstau_recasts}(a) shows the extrapolation of the ILC limits for the scenarios
with centre-of-mass energy 250\,GeV and 1\,TeV.

The detailed simulation study also allows to estimate the prospects for other detector concepts and colliders
by considering the impact of the particularities of ILD and ILC on the final result: 
Trivially, the beam energy of the collider provides
a kinematic limit to pair production of any new particle.
Larger data-sets at a given centre-of-mass energy could improve the discovery reach, but only in a limited region close to the kinematic limit.
For instance with $5$ ($10$) instead of $2$\,ab$^{-1}$ at $\sqrt{s}=250$\,GeV, the exclusion reach for $\Delta M=2$\,GeV would
increase from $M_{\widetilde{\tau}}<$ $112$\,GeV to $117$ ($117$)\,GeV, while the effect is negligible (less than $1$\,GeV) for $\Delta M=10$\,GeV.
However, any upgrade of the centre-of-mass energy of the collider will immediately outperform these small
improvements from more luminosity.

The most important feature of ILD used in this study is its excellent hermeticity, down to $6$\,mrad to the beam-axis. 
  In particular without the ability to veto forward-scattered high-energy electrons, positrons and photons, all kind of
  high-rate backgrounds from interactions of real or virtual photons are expected to explode.
  The impact of having to trigger the read-out for selected events of interest is hard to quantify in absence of a
  full understanding of accelerator-related backgrounds, including the substantial synchrotron radiation in circular colliders.
  However it is clear that the signature of low $\Delta M$ will suffer most:
  at LEP, the very soft $\tau$-decay products were not enough to trigger on. Instead, an additional ISR photon had to be used as
  trigger requirement, reducing the signal rate by about a factor 10, and causing the significant loss in sensitivity
  below $\Delta M \simeq 8$\,GeV, clearly visible in Fig.~\ref{exclusion_mstau_1}(b).
  Therefore we consider the ability of the linear colliders 
  to operate trigger-less as an asset when
  searching for soft or even unexpected signatures.
It is unclear if such trigger-less operation is possible at circular colliders, with a collision rate two orders of
magnitude larger.

\begin{figure}[htbp]
    \centering
    \includegraphics [width=0.45\textwidth]{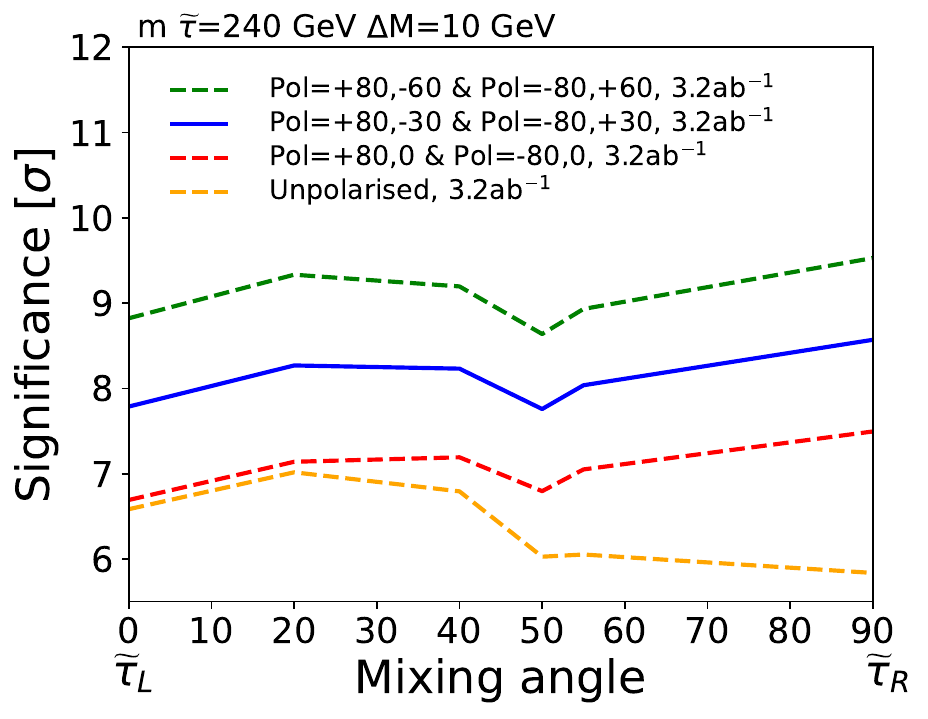}
    \caption{Comparison of the signal significance as a function of the mixing angle in the standard ILC conditions to the ones changing the ILC polarisations to the ones foreseen for CLIC or C$^3$, $P(e^{-},e^{+})=(+80\%, 0)$ and $P(e^{-},e^{+})=(-80\%, 0)$, and for the LCF at 550\,GeV, $P(e^{-},e^{+})=(+80\%, -60\%)$ and $P(e^{-},e^{+})=(-80\%, +60\%)$. For each case both polarisations were combined using the likelihood ratio statistic. The significances without polarisation are also shown.}
    \label{fig:significances_ilcclic}
  \end{figure}

For the other proposed {\it linear}  $e^+e^-$ colliders,  C$^3$  and  CLIC, the expected results would not be very different
from the ILC case.
However, neither of these foresee positron polarisation, so some loss of sensitivity is expected,
both from the lack of the enhancement of the effective luminosity that positron polarisation brings,
and from the the fact that the likelihood ratio weighting of samples becomes less powerful if only
one beam is polarised\footnote{This absence of beam polarisation of the positron beam neither change the worst case nor increase the dependence on the
mixing angle. This is also so if neither beam is polarised.}.
The absence of polarisation decreases the signal significance noticeably, as can be seen
in Fig.~\ref{fig:significances_ilcclic}, while an increase of the positron polarisation to 60\,\% as foreseen for LCF raises the significance visibly.
The different bunch-structures expected at these machines will require different optimisations
of the detector-system,
which, however, are not expected to have any important impact in the context of this analysis.
The most prominent example is probably the larger number of $e^+e^-$ pairs produced from beam-strahlung at the high-energy stages of CLIC.
However, past studies have shown that this can be mitigated very well with timing information~\cite{CLICdp:2018cto}.
A residual effect might be a somewhat worse hermeticity of the detector at CLIC.
On the other hand, the higher centre-of-mass energy foreseen at CLIC largely outweighs these disadvantages.
 
  \begin{figure}[htbp]
    \centering
    \subcaptionbox{$\Delta M = $ 3\,GeV}{\includegraphics [width=0.45\textwidth]{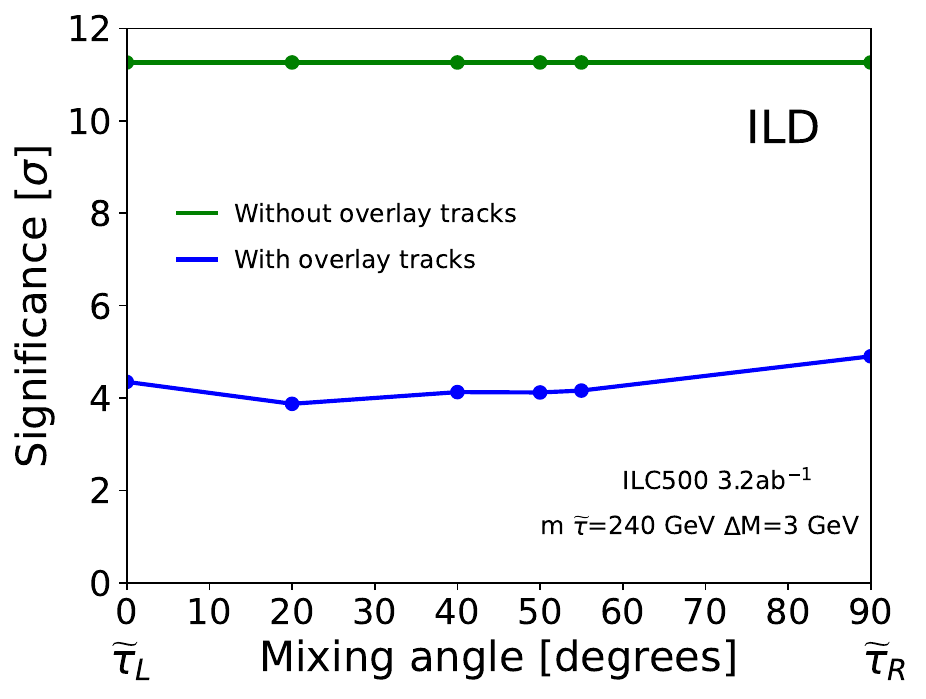}}
     \subcaptionbox{$\Delta M = $ 10\,GeV}{
     \includegraphics [width=0.45\textwidth]{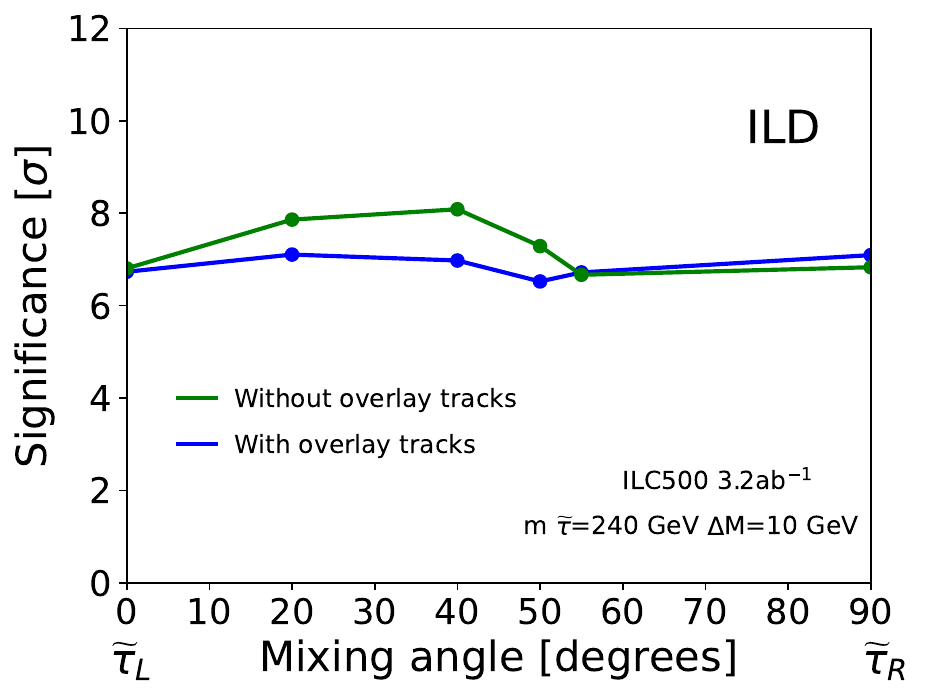}}
  
    \vspace*{0.4cm}
    \caption{Significance for a $\widetilde{\tau}$ with 
      $M_{\widetilde{\tau}} =$ 240\,GeV.  (a) shows $\Delta M = $ 3\,GeV, (b) shows $\Delta M = $ 10\,GeV.
        The plots combines the two main polarisation settings using the likelihood ratio
      statistic. Blue lines correspond to the case with all backgrounds, including overlay-on-physics tracks, 
     while the green curves correspond to the study where the overlay-on-physics tracks are not included. 
    }
    \label{nb_sigmas_fullsimu_sgv}
  \end{figure}

  For {\it circular}  $e^+e^-$ colliders,  FCCee  and  CepC,
  the differences are expected to be more prominent.
It is beyond the scope of this work to make a full study for FCCee-240 or CepC.
However, some well-founded conclusions can be drawn by extrapolating the
ILC-500 results to FCCee-240 conditions.
This includes both re-scaling the results to a lower $E_{CM}$,
taking the different beam conditions into account,
and evaluating the effect of the change in detector acceptance:
the machine-detector interface at FCCee does not allow for coverage lower than 50 mrad
to the beam,
compared to 6 mrad for ILC.
Unfortunately,
the effect of the possible need of a trigger cannot be evaluated at this stage.
As already pointed out,
it will be difficult to evaluate the results at low mass differences without
full detector and machine background simulation.
From the discussion above,
for an experiment at $E_{CM}$=240\,GeV, rather than 500\,GeV, i.e.\ with
$R_{CM}=$ 1/2.08,
the significance for a signal with SUSY masses 2.08 times smaller would be 2.08 times higher,
but only if all conditions are otherwise the same. 

Some of the effects of the various differences between the ILC and the FCCee conditions
can readily be found, by (hypothetically) changing the conditions for the ILC-500
analysis. By removing polarisation from the analysis,
the increase of
effective luminosity is lost, and the possibility to do likelihood ratio weighting
no longer exists, leading to a considerable loss in sensitivity, as discussed in Sec.~\ref{sec:limitcalc}.
Due to the much less strongly focused beams, 
the overlay-on-physics can be expected to be absent at FCCee,
both the low $P_T$ hadrons as well as the pairs background.
For the former, this is because the luminosity per bunch crossing is much lower, 
making it very improbable that such an event would happen in the same bunch crossing as a 
signal event, and for the latter because of the much smaller beam-beam interactions.
To take this into account, we also performed an analysis in which we didn't add overlay-on-physics
neither to the simulated signal nor to the background\footnote{Note, however, that there
is another, potentially bothering, background at each beam-crossing at FCCee, largely absent at the linear colliders, namely synchrotron
radiation.
This has not been considered in our study.}.
While the absence of overlay-on-physics under FCCee conditions is an advantage
at the very lowest mass-differences, this advantage is no longer present for  
$\Delta M$ = 8\,GeV or larger, see Fig.~\ref{nb_sigmas_fullsimu_sgv}.
For the overlay-only events, we assume that they can be kept under control also under FCCee
conditions.

To evaluate the effect of lower hermeticity of the detectors at FCCee,
we note that the background at $\Delta M \lesssim$ 10\,GeV is dominated by
the $\gamma\gamma$ background: For $M_{\widetilde{\tau}}$ = 245\,GeV and
$\Delta M$ = 10\,GeV, 215 such background events are expected in the ILD at the ILC
in the case of
unpolarised beams, compared to only 19 from all other sources.
(The significance of the signal at this point is just above 2 $\sigma$
with unpolarised beams).
We can therefore make an estimate of the increase in background from
modifying the acceptance of the forward calorimetry at generator level,
and for the $\gamma\gamma$ background only.
In doing so, we only included the background from virtual photons,
not from real ones, since the latter is expected to 
be much reduced due to the lower beam-beam effects at FCCee.
We find that at this model-point the background at such an hypothetical machine 
operating at 500 GeV, but with FCCee beam-conditions, would be  60 times higher
in a detector having FCCee acceptance than in a
detector with  ILC acceptance,
and
the significance would decrease from 2\,$\sigma$ to 0.3\,$\sigma$.
Alternatively, one could increase the $\rho$ cut such that one obtains the same
background level, but then one would need to increase the cut in $\rho$ by 95\,\%,
leading to a loss of 96\,\% of the signal, so an even more reduced significance.
According to the scaling with  $E_{CM}$ above, the same S/B would be expected for 
the signal point $M_{\widetilde{\tau}}$ = 118\,GeV and
$\Delta M$ = 4.8\,GeV at FCCee-240. Both signal and background cross sections would be
4.3 times higher, and the significance would be 2.08 times better, i.e.\ 0.6 $\sigma$.
To reach 2 $\sigma$, even at this more than two times lower  $\widetilde{\tau}$ mass,
the FCCee would need to collect 10 times more luminosity
than what is foreseen for the ILC at 500\,GeV, i.e.\ 40\,ab$^{-1}$, much more than the 10.8\,ab$^{-1}$ planned 
at 240\,GeV for FCCee with four experiments~\cite{FCC:2025uan}.
At lower $\widetilde{\tau}$ masses, the production cross section becomes much higher (due to the $\beta^3$
dependence),
and higher levels of background will be acceptable.
This allows to lower the cut in $\rho$,
yielding better efficiency at lower mass-differences, even though it will increase the background.
In our analysis at ILC, no particular optimisation of cuts was needed - the cuts optimised
for masses close to the kinematic limit were largely sufficient to ensure discovery down to $\Delta M = m_\tau$
also at lower $\widetilde{\tau}$ masses.
However, under FCCee conditions, the cut in $\rho$ that would naively yield the optimal sensitivity
would correspond to background levels approaching the millions of events.
In this case, the result would become dominated by systematics dominated - the $\gamma\gamma$ background would need to be know at the
per mil level, which is beyond the current state of the art~\cite{Sasikumar:2020qxa}.
As already pointed out, our recast method becomes quite un-reliable for small $\Delta M$
and in addition the generator-level estimate becomes less adequate.
Nevertheless,
if we follow the same recast-procedure as for $M_{\widetilde{\tau}}$ = 245\,GeV
for lower  $M_{\widetilde{\tau}}$,
but assuming a 15 \permil ~ systematic uncertainty on the level of $\gamma\gamma$ background,
we can estimate the reach down to masses already explored by the LEP experiments.
The result is shown in Fig.~\ref{exclusion_mstau_recasts}(b).
It can be seen that FCCee is not expected to be able to
exclude a $\widetilde{\tau}$ at all mass differences at any $\widetilde{\tau}$ mass.
In particular, the co-annihilation region, a region where SUSY could provide
an excellent dark-matter candidate,  would still be left partially un-explored,
and the LEPII limit would remain the only model-independent one.

We stress once again, that many aspects of the conditions at FCCee
are not included in this estimate, among others that the larger beam-spot, thicker beam-pipe and
vertex-detector and lower B-field will make the non-vertex track method less power-full,
that not only the coverage of electromagnetic calorimetry, but also hadronic calorimetry is weaker,
that we optimistically assumed that there is {\it no} background from real $\gamma$'s created from beam-beam
interactions, and that since the LumiCal in an FCCee detector is placed far in front of the
end-cap calorimeters, there is an acceptance-hole not only below it, but also above. 
Assessing the impact of these additional changes w.r.t.\ the ILD concept in its ILC version requires
a full Geant4-based simulation of the FCCee detector concepts,
which goes beyond the scope of this study, but should be pursued in the future.

\section{Summary and Conclusions\label{sec:concl}}
When searching for SUSY, the search for a $\widetilde{\tau}$ NLSP stands out in that it
represents the worst possible case:
Unlike other candidate  NLSPs, its standard model partner always decays
partially invisibly (to neutrinos) before it can be detected.
Being a third generation sfermion,
mixing between the weak hyper-charge states can be large.
This can reduce the production cross section,
and might yield $\tau$'s polarised in such a way that most of
their momentum is carried away by neutrinos after the decay.
Due to the sew-saw mechanism, the mixing also makes it likely that the
$\widetilde{\tau}$ is the lightest sfermion,
and can well be separated from the LSP by only a few GeV.
This taken together makes the $\widetilde{\tau}$ the probe by preference
to evaluate the SUSY exclusion or discovery potential of an experiment:
If a   $\widetilde{\tau}$ NLSP can be excluded or discovered,
any other candidate will, as well.

We have pointed out that the current projections of  $\widetilde{\tau}$ searches
at the HL-LHC estimate that there will be no potential to discover it at that
facility, 
and that the exclusion reach will be very limited.
More broadly, LHC and HL-LHC have little power to unambiguously exclude SUSY,
but does have good discovery potential for certain specific parts
of SUSY parameter-space.

Electron-positron colliders, such as the various proposed e$^+$e$^-$ Higgs factories,
in contrast,
do have excellent potential for loop-hole free exclusions,
as already demonstrated by LEPII. 
At such machines, 
the discovery reach is typically only a few GeV lower than the exclusion one,
contrary to the situation at hadron colliders,
where this difference is several hundreds of GeV.

It is sometimes claimed that the reach of SUSY searches at lepton colliders is
simply full coverage up to almost the kinematic limit.
But is this true when confronted to realistic experimental conditions?
To answer that question,
in the present study,
we have made a full simulation evaluation of the search for the $\widetilde{\tau}$
with the ILD concept at ILC500.
Not only do we include all SM backgrounds, fully simulated, but also
beam-induced backgrounds and low-$\it{P_{T}}$ hadrons from 
vector meson dominated (VDM) $\gamma\gamma$ interactions.
For the latter two,
we studied their effect both when such processes distorts the
reconstruction of hard interaction events, be it signal or SM events,
but also the signal-imitating aspects of such events in
bunch-crossings with no hard interaction.

We find,
for the first time in such a complete study,
that the answer to the question is ``yes'':
At large NLSP-LSP mass differences,
the exclusion limit is at 247\,GeV (3\,GeV below the kinematic limit),
and the discovery-reach is up to 245\,GeV, only 2\,GeV less than the exclusion reach.
At mass-difference less than 5\,GeV the limit becomes weaker,
but up to 240\,GeV, all mass-differences down to $M_\tau$ would be possible to exclude,
and up to 230\,GeV, any  mass-differences down to $M_\tau$ would be discoverable.

However,
this is the situation with ILD at ILC500,
and it should not be transposed to other conditions without caution.
We presented a general way to recast results at one energy to
another,
and applied that to evaluate the prospects for ILC250 and ILC1000.
Compared to other linear machines, ILC has an advantage in that
it is foreseen to have both beams polarised at ILC, unlike at C$^3$ or CLIC,
where only the electron-beam would be polarised.
Having both beams polarised increases the effective luminosity, i.e.\ the
fraction of the data-set where opposite polarised particles are colliding,
a pre-requisite for s-channel processes (such as $\widetilde{\tau}$-pair production) 
to be possible.
In addition,
it makes likelihood-ratio weighing of sub-samples more powerful.
Such weighing effectively gives more weight to the sub-samples
where the signal-to-background ratio is more favourable,
and with both beams polarised, the difference in  signal-to-background
between samples is larger,
adding even more strength to the weighing procedure.
Therefore, we expect the reach to be somewhat smaller for CLIC or C$^3$ than
for the ILC with the same energy and luminosity.
For CLIC at the highest energy, 3\,TeV, the beam-induced background will
be quite different, so the coverage at the lowest mass-differences,
where these backgrounds are of most concern,
would need a dedicated study.

Clearly,
the centre-of-mass energy of the accelerator is what defines the kinematic
limit, and therefore the linear Higgs factories, with their much higher
energy reach are preferred to circular ones for any SUSY search.
However,
this is not the only issue with the experimental environment at circular
Higgs factories.
Some of the differences are advantageous for this type of study, others not.
Among the advantages of circular colliders are the higher total integrated luminosity expected up to
the Higgs-factory stage,
and the lower per-bunch-crossing luminosity.
The latter almost completely avoids the risk to have overlaid parasitic
$\gamma\gamma$ interactions in the same bunch-crossing as a signal event.
It also reduces the total number of $\gamma\gamma$ interactions, since
the component of such interaction coming from real photons produced
by beam-beam interactions is largely reduced.
Neither of these effects, however, can compensate the increased reach in energy of
the linear Higgs factories, since these would be operating at a
higher centre-of-mass energy, 250\,GeV, rather than 240\,GeV.
The main disadvantage of the circular machines is their much
reduced hermeticity: Due to the need to have the final focusing system
quite close to the interaction point (inside the tracking volume),
the current design for FCCee foresees coverage down to 50 mrad to the
beam, compared to 6 mrad in the case of ILC or C$^3$.
A generator-level study indicated that this leads to an increase 
of the $\gamma\gamma$ background by factors between
30 and 60 (depending on the model-point), even when the above-mentioned
absence of the real photon induced part is taken into account.
Apart from obviously decreasing the sensitivity, it also leads to
the search becoming dominated by systematic uncertainties long before the full
data-set is exploited, and that the higher collected luminosity
will not extend the reach.
Due to this, we find that it is unlikely that FCCee would be 
able to fully exploit the model-space - at no $\widetilde{\tau}$ mass
would it be possible to exclude mass-differences to the LSP down
to the $\tau$ mass.

Further disadvantages at circular machines is that the continuous
operation (not allowing to power-off the detector between
bunch-trains, as can be done at the linear machines), and
the high levels of synchrotron radiation implies that both
the detector and the beam-pipe needs active cooling, making
them thicker.
This, together with the much larger beam-spot,
and lower detector-solenoid field (leading to worse momentum resolution)
will make exploitation of impact-parameters to separate
signal and background less effective.
The exact implications of this reduced impact-parameter
precision  must await a full simulation
study.

To conclude, 
we have shown that for the ILD concept at ILC500 indeed can
exclude or discover SUSY almost to the kinematic limit,
with no loop-holes, by finding that this is the case even
for the worst case, when the NLSP is the  $\widetilde{\tau}$,
and performing a full simulation study including all backgrounds,
including machine-related ones.
By recasting to other proposed linear colliders, we estimate
that the same would be true also there. 
The limits would, however, be slightly worse due to lack of positron
polarisation, and in the case of CLIC slightly worse hermeticity.
We also find that it is unlikely that FCCee would be able to do this:
Not only is the reach in energy modest at FCCee, also the much worse hermeticity 
will dramatically increase backgrounds at low mass-differences,
leading to less sensitivity, and systematics domination,
which
indicates that the lowest mass-difference, down to  $M_\tau$, will not be
completely covered.

\section{Acknowledgements\label{sec:ackn}}
  We would like to thank the LCC generator working group for producing the Monte Carlo samples used in this study.
  We also thankfully acknowledge the support by the the Deutsche Forschungsgemeinschaft (DFG, German Research Foundation) under 
  Germany's Excellence Strategy EXC 2121 ``Quantum Universe'' 390833306.
  This work has benefited from computing services provided by the German National Analysis Facility (NAF)~\cite{Haupt:2010zz}.
\appendix
\section{Appendix}\label{appendix:lowdm}
In this appendix we give details on the estimation of the total
reduction factor of the overlay-only background,
introduced in Sec.~\ref{sec:lowdmcuts}.
We discuss the two added cuts,
and describe how we conclude that it is  likely that the cuts are independent,
so that their combined effect is given by the product of each of them individually.
\subsection{The vertex condition}
The main vertex was fitted with the beam-spot as a constraint,
and any tracks not consistent with coming from the main interaction point
were used to find and fit secondary vertices.
Note that only {\it one} main vertex is fitted, but it is quite possible that there is
more than one  low-$\it{P_{T}}$-hadron event in the same bunch crossing.
This would result in a reconstructed event having one main vertex and one or more secondary vertices,
although the secondary vertices might also very well be consistent with the beam-spot.

\begin{figure}[htbp]
    \centering
    \subcaptionbox{Signal events ($\Delta M$ = 2\,GeV)}{\includegraphics [width=0.45\textwidth]{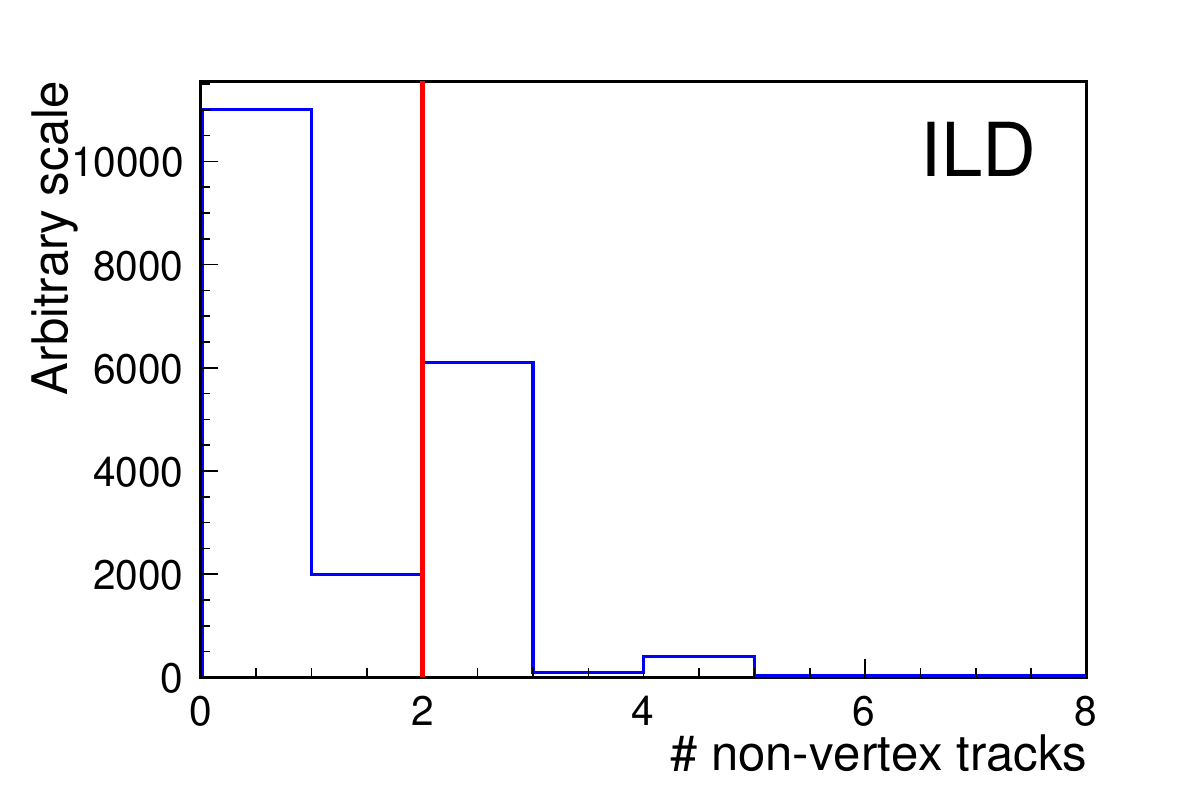}}
    \subcaptionbox{Signal events ($\Delta M$ = 10\,GeV)}{\includegraphics [width=0.45\textwidth]{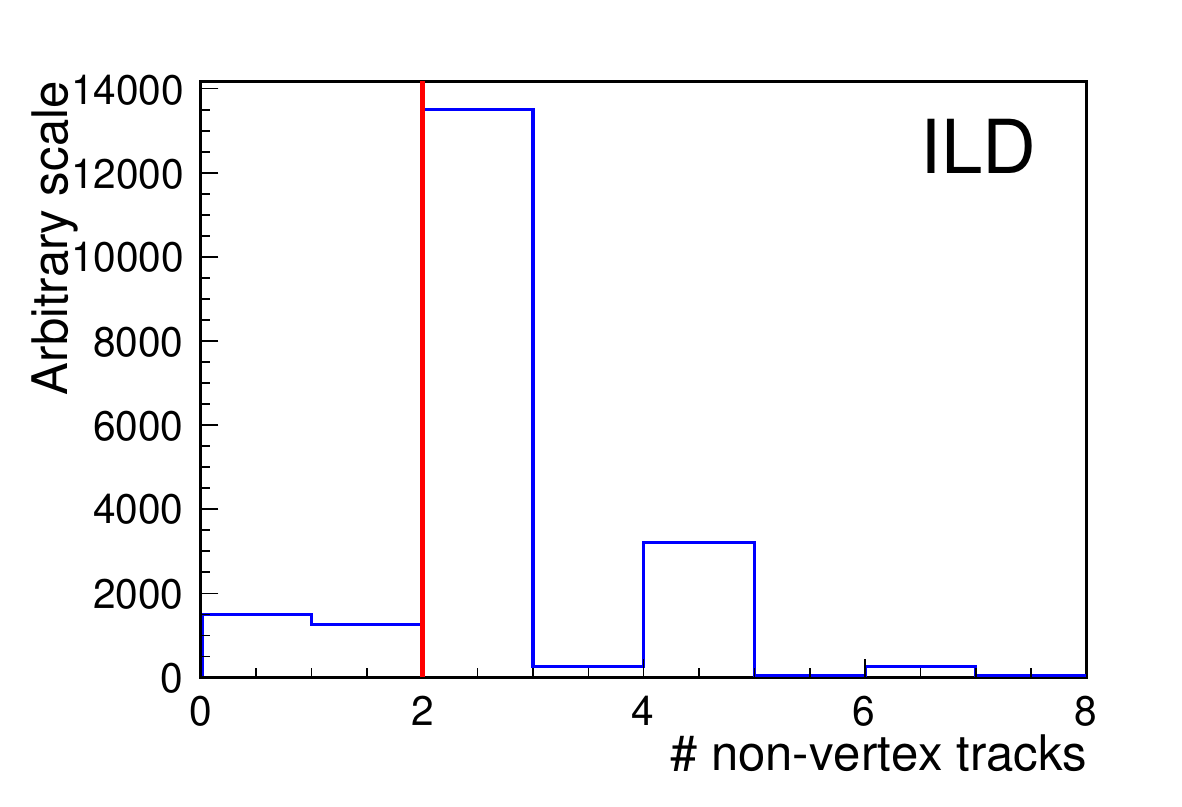}}
    \subcaptionbox{Overlay-only events (low-$\it{P_{T}}$ hadrons)}{\includegraphics [width=0.45\textwidth]{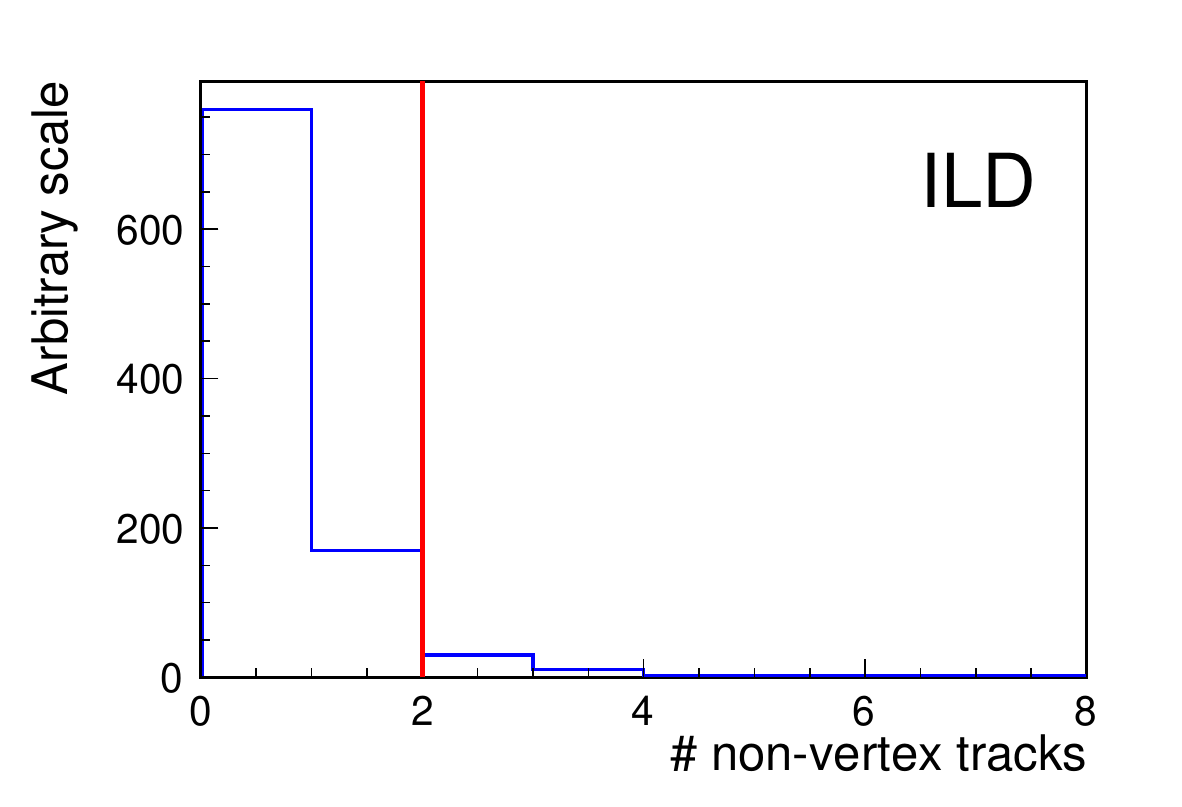}}
    \subcaptionbox{Overlay-only event (electrons)}{\includegraphics [width=0.45\textwidth]{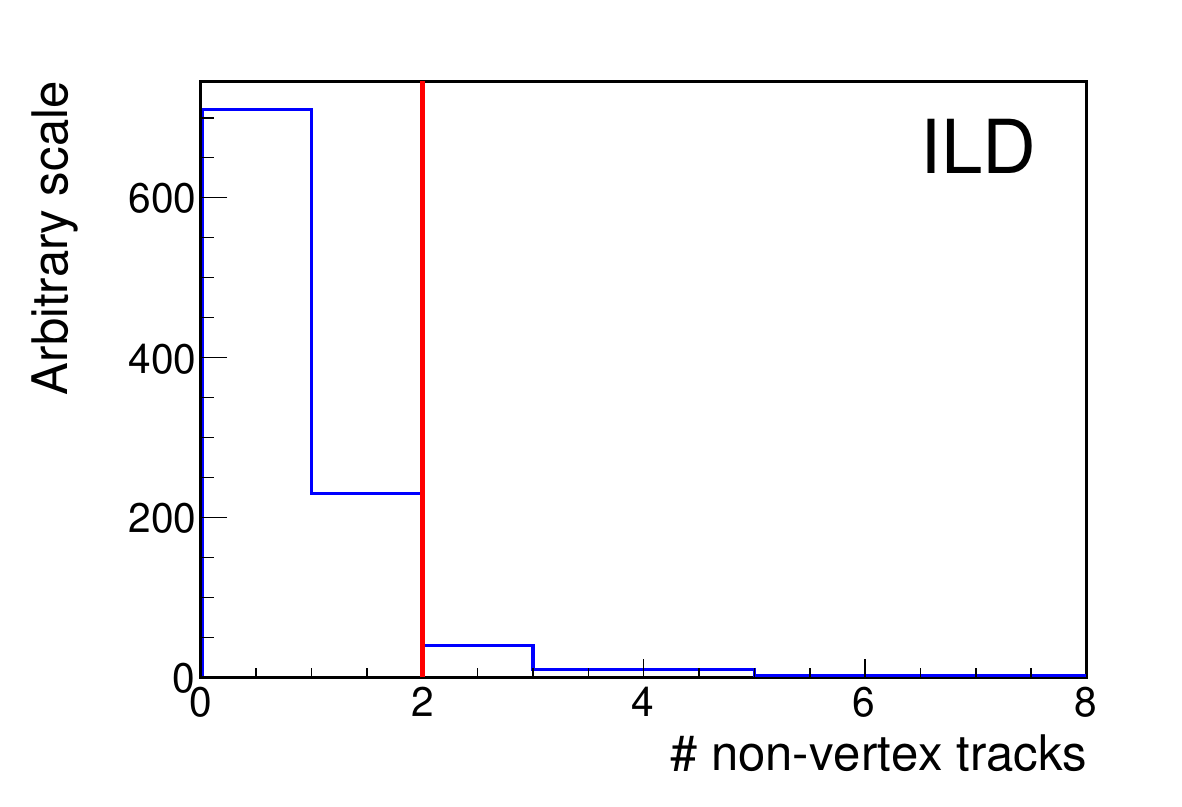}}
    \caption{Number of non-vertex tracks in signal and overlay-only events. The red lines show the lower cut for accepted events.}
    \label{fig:overlonly_vertex}
  \end{figure}

Both the fitted main and secondary vertices should contain at least two tracks.
Tracks in the event that could not be fitted neither to the main vertex, nor to any secondary vertex
are the  ``non-vertex'' tracks.
Figure~\ref{fig:overlonly_vertex} shows the number of
tracks not included in any vertex for overlay-only and signal events
for two different mass splittings, showing that the secondary multiplicity strongly depends on the 
momentum available, and the hence boost-factor for the tau-lepton.
The reduction of overlay-only events requiring only that the events have at least two ``non-vertex'' tracks
is  $1.9\tento{-2}$.

\subsection{The photon condition}
\begin{figure}[htbp]
  \centering
  \includegraphics [width=0.45\textwidth]{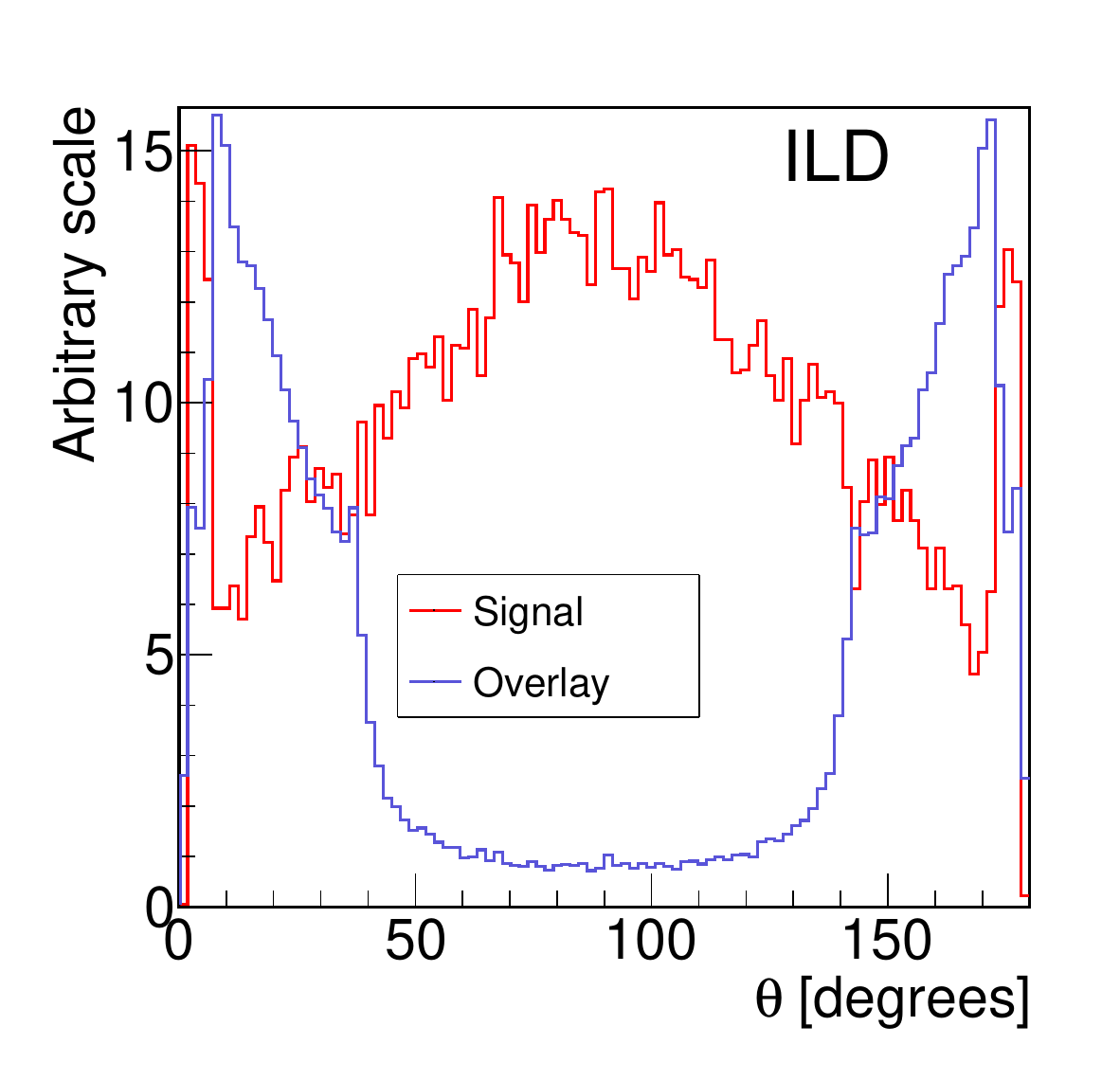}
  \caption{Distributions of the polar angle of photons in signal and overlay-only events. The signal
    distribution is for  M$_{\widetilde{\tau}}$ = 240\,GeV, $\Delta M$ = 2\,GeV }
  \label{fig:overlonly_thetaisr}
\end{figure}

As explained in Sec.~\ref{sec:lowdmcuts},
requiring to to detect a photon with sizeable angle to the beam-axis and
with sizeable energy, enhances to $\tau$ identification,
since photons are often present among the decay products of $\tau$'s.
In overlay-only events, on the other hand, 
photons with such properties are rare.  
The angle and photon energy was optimised for getting enough rejection without
removing all simulated overlay-only events.
The decrease in the signal efficiency when requiring a photon is about a factor two, regardless of the cut in $\theta$.
Figure~\ref{fig:overlonly_thetaisr} shows the polar angle distribution for photons
in overlay-only and in signal
events.
The cut in the angle was set to 30$^\circ$ to the beam axis; higher angles could also have been
used, with little loss in signal efficiency,
but in this case no events were left in the simulated  low-$\it{P_{T}}$ hadrons sample.
The energy cut was set to 2.2\,GeV.
This requirement alone reduces the overlay-only background by a factor $1.7\tento{-4}$.

We also note that this photon requirement in addition reduces the other important source
of background at low $\Delta M$,  the multi-peripheral $\gamma\gamma$ events.
The effect of enhancing the $\tau$ identification is not an effect that
   enhances the signal-to-background ratio in this case, because at this point in the cut-chain,
   the remaining background events of this type actually are almost entirely composed of $\gamma\gamma \rightarrow \tau^+\tau^-$
   events, so the enhanced $\tau$ identification has the same effect on signal as on background.
However, it still does, but for a different reason:
If the presence of an {\it ISR photon} is requested, the incoming electron or positron that
  emitted the ISR must have recoiled against the ISR.
  Since this background is a
  scattering process, not an annihilation one, the electron (positron) is still present in the final state.
  Therefore, if it is required to see a high transverse momentum ISR, the final state
  electron (positron) will have acquired a recoil transverse momentum big enough to be deflected into the BeamCal, and
  thus to have been rejected already at the pre-selection stage. 
  On the other hand, if the ISR was emitted from
  an electron or positron that was subsequently annihilated into a $Z$, as is the case for the signal process,
  the transverse momentum of the ISR is included
  in the decay products of the $Z$, and no signal is expected in the BeamCal.
  The only way a multi-peripheral $\gamma\gamma$ event would at the same time have a high $P_T$ ISR photon detected,
  but no signal in the BeamCal is if the fermion-antifermion pair in the final state accidentally would have acquired
  exactly a transverse momentum opposite to that of the ISR photon.
  In this case, the outgoing electron or positron would have been steered back close to the beam-axis.
  While this is a rare random occurrence,
  it can not be neglected due to the very high cross section of  multi-peripheral $\gamma\gamma$ events.
  However,
  in this configuration, the fermion-antifermion pair is recoiling against the ISR photon,
  which is not the case for the signal: In the signal, the {\it $\widetilde{\tau}$-pair} recoils against the ISR photon,
  and the $\tau$-pair from the $\widetilde{\tau}$ decays has no particular relation to the direction of the
  ISR photon, due to the invisible LSPs.
  At the lowest mass-difference,
  $\Delta M$ = 2\,GeV, the photon cut needs to be supplemented by the requirement that  $P_{T miss} > |\sum \bar{p}_{jet}|$,
  to reduce the  multi-peripheral $\gamma\gamma$ background to an acceptable level. 
  Using the total momentum of the jet system, rather than the transverse component of it in this comparison further enhances the
  performance of the cut. The background is usually at lower angles to the beam-axis and has closer-together jets than
  the signal, and  therefore the momentum of the jet system exceeds the transverse momentum more for background than for signal.

\subsection{Verification of independence}
  To verify if the cuts can be considered independent, we use the defining property of independent events: If $P(A \cap B) = P(A) \times P(B)$,
then $A$ and $B$ are independent.
Hence, by verifying if the fraction of events passing any combination of two cuts is compatible with the fraction
obtained by multiplying the fractions passing any one of the cuts,
we assume that the two cuts are indeed independent.
We also note that for three cuts to be mutually independent, it is a necessary condition that
they are pair-wise independent.
While it is not a sufficient condition, it remains a strong indication that they are
mutually independent,
and that
the product of the fractions of events passing any one of the three cuts 
is a good estimate
on the fraction that would pass all three at the same time.
To evaluate pair-wise factors, we found that we often needed to make the studied cut weaker than what was finally
applied to signal and the other backgrounds (where the MonteCarlo statistics is sufficient), in order 
to get any simulated overlay-only events passing the different pair-wise combinations.

At all mass-differences studied,
the fraction of overlay-only events that  pass the photon cut is $1.7 \pm 0.3 \tento{-4}$,
and $3.4 \pm 0.1 \tento{-3}$ pass a softened photon cut, where it is only
required to have detected a photon with energy above 2.2\,GeV, at whatever angle.
The fraction $1.9 \pm 0.03 \tento{-2}$ pass the vertex cut alone.

At the model-point with $\Delta M$ from 3 to 10\,GeV,
the fraction  of overlay-only events that passes the standard cuts excluding the cuts on $\rho$ and $P_{T miss}$ is
$5 \pm 0.1 \tento{-3}$.
No such events pass the  combined cuts on $\rho$ and $P_{T miss}$.
If these cuts are weakened to $\rho > 2$\,GeV and $P_{T miss} > 2$\,GeV,
a fraction $2.0 \pm 0.9 \tento{-5}$ of the  overlay-only events pass.
However, even this weaker cut leaves no events when combined with either
the weaker photon cut or with the standard cut.
On  the other hand, we find that 
if the cut on $\rho$ is removed, so that only the $P_{T miss}$ cut is applied, 
the fraction that passes it and the weaker photon cut at the same time is $2.4 \pm 0.9 \tento{-5}$.
The fraction passing the $P_{T miss}$ cut is $3.0 \pm 0.1 \tento{-3}$, so that 
the product of the individual factors is $1.02 \pm 0.05 \tento{-5}$.
The difference between the two is  $1.4 \pm 0.9   \tento{-5}$, i.e.\ $1.5 \sigma$ from zero.
Likewise, 
the fraction passing the standard cuts and the weaker photon cut  is $2.4 \pm 0.9 \tento{-5}$, the
product is  $1.7 \pm 0.8 \tento{-5}$, and the difference becomes $0.7 \pm 1.2   \tento{-5}$, 
i.e.\ $0.6 \sigma$ from zero.
However, no events pass the third combination - the standard cuts and the $P_{T miss}$ cut. 
The product of the two in this case is  $1.5 \pm 0.1 \tento{-5}$,
which is at the edge of what could have been observed. 
The vertex cut is expected to be independent of the other cuts by construction.
Assuming mutual independence,
the total reduction factor
by multiplying the individual factors for the vertex cut, the standard cuts and
the photon cut is 
$(1.9 \times 5.0 \times 1.7) \tento{-9} = 1.6 \tento{-8}$.
When this is multiplied with the factor obtained with the weaker combined $\rho$ and  $P_{T miss}$ cuts,
a final factor of  $2.7 \tento{-13}$ is obtained.
The actual factor should be even smaller, but since no events pass
the final  combined $\rho$ and  $P_{T miss}$ cuts, this could not be
evaluated.

At the model-point with $\Delta M$= 2\,GeV,
the fraction passing the standard cuts is $2.6 \pm 0.1 \tento{-3}$
The fraction passing these and the weaker photon cut is
 $1.2 \pm 0.7 \tento{-5}$, the product is  $0.88 \pm 0.5 \tento{-5}$,
and the difference between the two is $0.32 \pm 0.86   \tento{-5}$, i.e.\ $0.4 \sigma$ from zero.
The vertex cut is expected to be independent of the other cuts by construction.
Assuming mutual independence,
the total reduction factor
by multiplying the individual factors for the vertex cut, the standard cuts and
the photon cut is $(1.9 \times 2.6 \times 1.7) \tento{-9} = 8.4 \tento{-9}$
This factor is not quite sufficient to conclude that the overlay-only events
are under control.
However,
at $\Delta M$= 2\,GeV there is the further cut $P_{T miss} > |\sum \bar{p}_{jet}|$.
This cut alone reduces the overlay-only by a factor $1.2 \pm 0.2 \tento{-4}$.
Unfortunately,
no events pass this cut together with either of the weakened photon cut or
the standard cuts. In both cases, the product of the factors is indeed
less than what would allow any simulated event to pass.
We therefore assume that this factor can indeed be used to get the
final reduction factor of $1.0 \tento{-12}$

\printbibliography

\end{document}